# Asymptotic Exit Location Distributions in the Stochastic Exit Problem

**Robert S. Maier**[*]

rsm@math.arizona.edu

*Dept. of Mathematics*

*University of Arizona*

*Tucson, AZ 85721, USA*

**D. L. Stein**[†]

dls@ccit.arizona.edu

*Dept. of Physics*

*University of Arizona*

*Tucson, AZ 85721, USA*

[*]Partially supported by the National Science Foundation under grant NCR-90-16211.
[†]Partially supported by the U.S. Department of Energy under grant DE-FG03-93ER25155.

**Abstract**

Consider a two-dimensional continuous-time dynamical system, with an attracting fixed point $S$. If the deterministic dynamics are perturbed by white noise (random perturbations) of strength $\epsilon$, the system state will eventually leave the domain of attraction $\Omega$ of $S$. We analyse the case when, as $\epsilon \to 0$, the exit location on the boundary $\partial\Omega$ is increasingly concentrated near a saddle point $H$ of the deterministic dynamics. We show using formal methods that the asymptotic form of the exit location distribution on $\partial\Omega$ is generically non-Gaussian and asymmetric, and classify the possible limiting distributions. A key role is played by a parameter $\mu$, equal to the ratio $|\lambda_s(H)|/\lambda_u(H)$ of the stable and unstable eigenvalues of the linearized deterministic flow at $H$. If $\mu < 1$ then the exit location distribution is generically asymptotic as $\epsilon \to 0$ to a Weibull distribution with shape parameter $2/\mu$, on the $O(\epsilon^{\mu/2})$ lengthscale near $H$. If $\mu > 1$ it is generically asymptotic to a distribution on the $O(\epsilon^{1/2})$ lengthscale, whose moments we compute. The asymmetry of the asymptotic exit location distribution is attributable to the generic presence of a 'classically forbidden' region: a wedge-shaped subset of $\Omega$ with $H$ as vertex, which is reached from $S$, in the $\epsilon \to 0$ limit, only via 'bent' (non-smooth) fluctuational paths that first pass through the vicinity of $H$. We show that as a consequence the Wentzell-Freidlin quasipotential function $W$, which governs the frequency of fluctuations to the vicinity of any point $x$ in $\Omega$ and is the solution of a Hamilton-Jacobi equation, generically fails to be twice differentiable at $x = H$. This nondifferentiability implies that the classical Eyring formula for the small-$\epsilon$ exponential asymptotics of the mean first exit time, which includes a prefactor involving the Hessian of $W$ at $x = H$, is generically inapplicable. Our treatment employs both matched asymptotic expansions and probabilistic analysis. Besides relating our results to the work of others on the stochastic exit problem, we comment on their implication for the two-dimensional analogue of Ackerberg-O'Malley resonance.

# 1 Introduction

We consider the problem of noise-activated escape from a connected planar domain $\Omega \subset \mathbb{R}^2$ with smooth boundary, in the limit of weak noise. If $\boldsymbol{b} = (b^i)$, $i = 1, 2$ is a smooth vector field on a neighborhood of the closure $\bar{\Omega}$, we define the random process $\boldsymbol{x}_\epsilon(t) = (x^i_\epsilon(t))$, $i = 1, 2$ by the Itô stochastic differential equation

$$dx^i_\epsilon(t) = b^i(\boldsymbol{x}_\epsilon(t)) + \epsilon^{1/2} \sum_\alpha \sigma^i{}_\alpha(\boldsymbol{x}_\epsilon(t)) \, dw_\alpha(t) \tag{1}$$

and an appropriate initial condition. Here $w_\alpha(t)$, $\alpha = 1, 2$, are independent Wiener processes and $\boldsymbol{\sigma} = (\sigma^i{}_\alpha)$ is a 2-by-2 noise matrix, like $\boldsymbol{b}$ a function of position $\boldsymbol{x} = (x^i)$, $i = 1, 2$. The associated diffusion tensor $\boldsymbol{D} = (D^{ij})$ is defined by

$$D^{ij} = \sum_\alpha \sigma^i{}_\alpha \sigma^j{}_\alpha, \tag{2}$$

*i.e.*, $\boldsymbol{D} = \boldsymbol{\sigma}\boldsymbol{\sigma}^T$. We assume strict ellipticity, *i.e.*, that $\boldsymbol{D}$ is nonsingular on $\Omega$ and its boundary, and that its $\boldsymbol{x}$-dependence is smooth on a neighborhood of $\bar{\Omega}$. $\epsilon > 0$ is a noise strength parameter. The subscripts on $x^i_\epsilon$ and $\boldsymbol{x}_\epsilon$ emphasize the $\epsilon$-dependence of the random process, which may be viewed as a dynamical system stochastically perturbed by noise.

Of interest in applications is the case when $\Omega$ contains only a single stable fixed point $S$ of the drift field $\boldsymbol{b}$, and $S$ serves as an attractor for the whole of $\Omega$. If $\Omega$ is the entire domain of attraction of $S$, the boundary $\partial \Omega$ will not be attracted to $S$: it will be a separatrix between domains of attraction. This 'characteristic boundary case' is particularly important and difficult to study. We shall analyse this case, assuming that $\partial \Omega$ is a smooth characteristic curve of $\boldsymbol{b}$ containing fixed points (alternating saddle points and unstable fixed points, as in Figure 1). We allow $\Omega$ to be unbounded. If the initial condition is $\boldsymbol{x}_\epsilon(0) = S$ and $\tau_\epsilon$ is the first passage time from $S$ to the boundary (*i.e.*, the first exit time), we shall study the behavior of the exit location distribution $p_\epsilon(\boldsymbol{x}) \, d\boldsymbol{x} = \mathsf{P}\{\boldsymbol{x}_\epsilon(\tau_\epsilon) \in \boldsymbol{x} + d\boldsymbol{x}\}$ on $\partial \Omega$ in the $\epsilon \to 0$ limit, and the small-$\epsilon$ asymptotics of the mean first passage time (MFPT) $\mathsf{E}\tau_\epsilon$.

Exit problems of this sort, in more than one dimension, have a long history [65]. They arose originally in chemical physics [2, 4, 9, 29, 40], but occur in other fields of physics [64, 75] as well as in systems engineering [53, 76] and theoretical ecology [49, 57, 58]. In recent years two different approaches have been used: rigorous large deviations theory [14, 15, 26, 27] and formal but systematic asymptotic expansions [49, 61, 65, 74]. The rigorous approach yields comparatively weak but still very useful results. In particular much light is thrown on exit problems by the Wentzell-Freidlin quasipotential [27], or classical action function, $W : \bar{\Omega} \to \mathbb{R}^+$. $W(\boldsymbol{x})$ is best thought of as a measure of how difficult it is for the process $\boldsymbol{x}_\epsilon(t)$ to reach the point $\boldsymbol{x}$; as $\epsilon \to 0$ the frequency of excursions to the vicinity of $\boldsymbol{x}$ will be suppressed exponentially, with rate constant $W(\boldsymbol{x})$. As we review in Section 4, $W(\boldsymbol{x})$ has an interpretation in terms of certain 'classical' trajectories extending from $S$ to $\boldsymbol{x}$, interpreted as the most probable (as $\epsilon \to 0$) fluctuational paths from $S$ to $\boldsymbol{x}$ [49, 56].

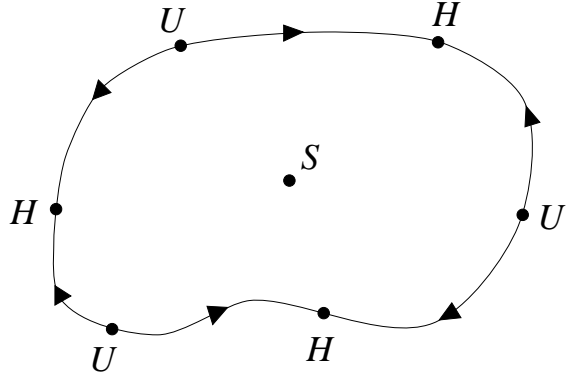

Figure 1: The structure of the drift field $b$ on $\Omega$, when $\Omega$ is bounded. The boundary $\partial\Omega$, being a separatrix, is not attracted to the stable fixed point $S$, though all points in $\Omega$ are attracted to $S$. Saddle points ($H$) and unstable fixed points ($U$) alternate around the boundary.

Normally one expects that $W$ will attain a minimum on $\partial\Omega$ at one of the fixed points, in particular at a saddle point. This can be shown to imply that as $\epsilon \to 0$, the exit location on $\partial\Omega$ converges in probability to the saddle point. Actually the behavior of $W$ on $\partial\Omega$, and its consequences, are not yet fully understood. (For some partial results on the smoothness of $W$, see Day and Darden [19] and Day [18].) Recent treatments [16, 54] indicate that as $\epsilon \to 0$ it is possible for the exit location to converge to an *unstable* fixed point on $\partial\Omega$, due to local constancy of $W$ on the boundary. In this paper we consider only models displaying the conventional behavior: as $\epsilon \to 0$ the exit location should converge to some distinguished saddle point $H$ on $\partial\Omega$, due to $W$ on $\partial\Omega$ having a unique global minimum at $H$.

The method of formal asymptotic expansions has provided evidence for an unusual phenomenon: *skewing* of the exit location distribution in the vicinity of the saddle point $H$. What is meant by skewing is that as $\epsilon \to 0$, the exit location converges to $H$ but its distribution $p_\epsilon(\boldsymbol{x})\,d\boldsymbol{x}$ on $\partial\Omega$ is not asymptotic to a Gaussian centered on $H$. (Skewing was discovered by Bobrovsky and Schuss [7]; for recent rigorous work, see Bobrovsky and Zeitouni [8] and Day [17].) This phenomenon has until now remained unclarified, though examples of skewed asymptotic exit location distributions have been given by Maier and Stein [55]. Using formal methods we show in this paper that skewing is a generic phenomenon. We also derive a general result, analogous to the central limit theorem, which characterizes skewing: *As $\epsilon \to 0$, in any generic stochastic exit model (with characteristic boundary) of the above sort, the exit location distribution, on an appropriate $\epsilon$-dependent lengthscale near $H$, will be asymptotic to a non-Gaussian distribution that belongs to one of two well-defined classes*. Which asymptotic distribution occurs is largely determined by the behavior of the model near $H$ (*i.e.*, the ratio $\mu = |\lambda_s(H)|/\lambda_u(H)$ of the stable and unstable eigenvalues of the linearization of $\boldsymbol{b}$ at $H$, which we assume to be nonsingular, and to a lesser extent the value of $\boldsymbol{D}(H)$). One of the two classes is the class of Weibull distributions, which is familiar from statistics [3]. The asymptotic exit location distributions in the second class are more complicated, and we do not derive explicit expressions for them here. We do however provide an algorithm for computing their moments, in terms of the correlation functions of a conditioned three-dimensional Bessel process.

Much previous work, in particular on physical applications [32], has dealt with a special case: when the mean drift $b^i$ equals $-D^{ij}\partial_j\Phi$, for some smooth potential function $\Phi$. (Here $\partial_j \equiv \partial/\partial x^j$; summation over repeated Roman indices is assumed henceforth.) In this case the point $H$ will simply be a saddle point of $\Phi$. This gradient (or 'conservative') case finds many applications in physics, but from a mathematical point of view is highly nongeneric. $W$ can be solved for exactly (it equals $2\Phi$), and one can show formally that in the limit of weak noise the distribution of the exit location on $\partial\Omega$ is asymptotic to a Gaussian centered on $H$, with $O(\epsilon^{1/2})$ standard deviation. It turns out that generic, nongradient drift fields, which find application in nonequilibrium statistical mechanics and stochastic modelling generally, are qualitatively different.

In the light of the large deviations approach, the fact that generic drift fields give rise to non-Gaussian asymptotic exit location distributions is surprising. One might expect that on $\partial\Omega$, $W$ in general attains a quadratic minimum at $H$. If $\partial^2 W/\partial s^2(s=0)$ equals $\sigma^{-2}$ ($s$ being the arc length along $\partial\Omega$, measured from $H$), the abovementioned exponential suppression as $\epsilon \to 0$ would presumably give rise to a factor $\exp(-s^2/2\sigma^2\epsilon)$ in the exit location density $p_\epsilon(s)$. This is precisely a Gaussian centered on $H$, on the $O(\epsilon^{1/2})$ lengthscale. The generic case would therefore seem to resemble the gradient case.

It is tempting to ascribe the discrepancy between this prediction and our results to what theoretical physicists would call a 'prefactor effect', *i.e.*, the presence of subdominant (as $\epsilon \to 0$) terms in the exit location density that are not included in the usual exponential factor arising from large deviations theory. Surprisingly, this is not the case. The prediction assumed that if $W$ has a minimum at $H$, it has a quadratic minimum there. We shall see that generically, $W$ is *not even twice differentiable* at $H$; strikingly, $\partial^2 W/\partial s^2$ is generically discontinuous at $s = 0$. This discontinuity causes the exit location density $p_\epsilon(s)$, even if (in the small-$\epsilon$ limit) it is localized on the $O(\epsilon^{1/2})$ lengthscale near $H$, to have different Gaussian falloff rates as $s/\epsilon^{1/2} \to +\infty$ and $s/\epsilon^{1/2} \to -\infty$. This is the true cause of skewing. The abovementioned limiting Weibull distributions, which are one-sided, arise when either $\partial^2 W/\partial s^2(0+)$ or $\partial^2 W/\partial s^2(0-)$ equals zero (a generic occurrence, when $\mu < 1$); on account of the corresponding zero Gaussian falloff rate, they are localized on a larger lengthscale than $O(\epsilon^{1/2})$. The appropriate lengthscale turns out to be $O(\epsilon^{\mu/2})$.

The cause of the generic discontinuity in $\partial^2 W/\partial s^2$ at $s = 0$ is explained in Section 5; it has an easy (indeed pictorial) interpretation. The reader may wish to glance ahead at Figure 3, which displays the typical behavior of the piecewise classical trajectories giving rise to $W(\boldsymbol{x})$, for $\boldsymbol{x}$ in the vicinity of $H$. The difference between trajectories incident on boundary points with $s > 0$ and those incident on boundary points with $s < 0$ is apparent. There is a wedge-shaped 'classically forbidden' region emanating from $H$, which is reached only by piecewise classical (rather than classical) trajectories. This region includes all the boundary points to one side of $H$.

In Sections 2, 3 and 4 we explain the aspects of our approach that make contact with previous treatments: the reduction of the exit location problem to the study of the principal eigenfunction of the forward Kolmogorov equation, and the approximation of this eigenfunction by a characteristic (ray) expansion [49]. The rays employed will be the classical trajectories of Wentzell-Freidlin theory, and the same action function $W$ will appear. We obtain a system of ordinary differential equations which must be integrated along the rays [49, 55, 56]. Our equations extend those obtained by Matkowsky, Schuss and Tier [60, 61]; being ordinary rather than partial differential equations, they are more suited to numerical computation. A key role is played by a Riccati equation for the matrix of second derivatives $\boldsymbol{Z} = (\partial_i \partial_j W)$, $i, j = 1, 2$. We comment on the fact that our equations are covariant: they transform systematically under changes of coordinates, and have a geometric (coordinate-free) interpretation.

In Section 6 we consider the nongeneric case when $(\partial^2 W/\partial s^2)(0+)$ and $(\partial^2 W/\partial s^2)(0-)$ are equal, which gives rise to a Gaussian asymptotic exit location distribution. In Sections 7 and 8 we turn to the generic case, and analyse models with $\mu < 1$ and $\mu > 1$. We study the fine behavior of the process $\boldsymbol{x}_\epsilon(t)$ and the function $W$ in the vicinity of $H$, and by a combination of matched asymptotic expansions (involving the construction of a novel 'inner' approximation in a boundary layer near $\partial\Omega$) and probabilistic analysis, derive the possible asymptotic exit location distributions on the appropriate lengthscales. We also extend earlier work by showing how our system of coupled ordinary differential equations may be employed to compute the small-$\epsilon$ asymptotics of the MFPT $\mathsf{E}\tau_\epsilon$. It is known that $\mathsf{E}\tau_\epsilon$ grows exponentially as $\epsilon \to 0$, with rate constant $W(H)$. But in the case of a characteristic boundary, the numerical computation of the pre-exponential factor in the MFPT asymptotics is more complicated than previous treatments [60, 61, 65, 69, 74] have suggested. Until now it has been generally believed that the pre-exponential factor may be computed from the Hessian matrix of $W$ at the saddle point $H$. But generically this matrix of second derivatives does not exist! We show how this is due to the ray expansion becoming singular at $H$, and how previous treatments may be modified to take this phenomenon into account.

In Section 9 we comment on the implications of our results for a related area: the solution of multidimensional singularly perturbed elliptic boundary value problems. In one-dimensional singularly perturbed boundary value problems with turning points, the phenomenon of Ackerberg-O'Malley resonance [1, 66] is well understood [42, 43], but the corresponding multidimensional phenomenon has been little explored. We are able to show that at least one case of two-dimensional Ackerberg-O'Malley resonance, which can be reduced to a stochastic exit problem on $\Omega$ ($\partial\Omega$ being characteristic, and containing fixed points) has unusual features. On account of skewing near the dominant fixed point and an anomalous exit location lengthscale, a singular perturbation expansion for the matching 'outer' solution in the body of $\Omega$ must in general employ *irrational* powers of the perturbation strength $\epsilon$. This is strikingly different from the case of a non-characteristic boundary, where an expansion in powers of $\epsilon$ suffices [60, 61]. It also differs from the corresponding one-dimensional problem (on an interval whose endpoints are turning points), where subdominant terms in the outer expansion are exponentially suppressed. Our results indicate that multidimensional Ackerberg-O'Malley resonance is likely to be more complicated than one-dimensional resonance. Our overall conclusions appear in Section 10.

## 2  Mathematical Preliminaries

The random process $\boldsymbol{x}_\epsilon(t)$, $t \geq 0$ has generator

$$\mathcal{L}_\epsilon = -(\epsilon/2) D^{ij} \partial_i \partial_j - b^i \partial_i \tag{3}$$

with formal adjoint $\mathcal{L}_\epsilon^*$ defined by

$$\mathcal{L}_\epsilon^* \rho = -(\epsilon/2)\partial_i\partial_j[D^{ij}\rho] + \partial_i[b^i\rho]. \tag{4}$$

The density $\rho(\boldsymbol{x},t)$ of the probability distribution $\rho(\boldsymbol{x},t)\,d\boldsymbol{x} = \mathsf{P}\{\boldsymbol{x}_\epsilon(t) \in \boldsymbol{x} + d\boldsymbol{x}\}$ will satisfy $\dot\rho = -\mathcal{L}_\epsilon^*\rho$, the forward Kolmogorov (or Fokker-Planck) equation. So the spectral theory of the operators $\mathcal{L}_\epsilon$ and $\mathcal{L}_\epsilon^*$ is of interest. If they are equipped with Dirichlet (absorbing) boundary conditions on $\partial\Omega$ and $\Omega$ is bounded, it follows by standard methods [39] that they have pure point spectrum, with smooth eigenfunctions, and that the principal eigenvalues of $\mathcal{L}_\epsilon$ and $\mathcal{L}_\epsilon^*$ (the ones with minimum real part) are real. Also, the corresponding principal eigenfunctions may be taken to be real and positive on $\Omega$. If $\Omega$ is unbounded we assume these properties continue to hold; they will hold if the stochastic model is sufficiently stable at infinity. We write $\lambda_\epsilon^{(0)}, \lambda_\epsilon^{(1)}, \lambda_\epsilon^{(2)}, \ldots$ for the eigenvalues of $\mathcal{L}_\epsilon$, ordered by increasing real part, and $u_\epsilon^0, u_\epsilon^1, u_\epsilon^2, \ldots$ for the corresponding eigenfunctions. Since $\lambda_\epsilon^{(0)}$ is real, the eigenvalues of $\mathcal{L}_\epsilon^*$ will be $\lambda_\epsilon^{(0)}, \lambda_\epsilon^{(1)*}, \lambda_\epsilon^{(2)*}, \ldots$; we denote the corresponding eigenfunctions by $v_\epsilon^0, v_\epsilon^1, v_\epsilon^2, \ldots$ If $\langle\cdot,\cdot\rangle$ denotes the usual $L^2$ inner product on $\Omega$, we choose the $u_\epsilon^n$ and $v_\epsilon^n$ to satisfy $\langle u_\epsilon^m, v_\epsilon^n\rangle = \delta_{mn}$.

It is noteworthy that if $b^i = -D^{ij}\partial_j\Phi$ for some function $\Phi$ on $\Omega$, then the operator $\mathcal{L}_\epsilon$ is related by a similarity transformation, known as the Liouville transformation, to a formally self-adjoint operator; equivalently, $\mathcal{L}_\epsilon$ has a domain of self-adjointness. Consequently in the gradient case, the eigenvalues $\lambda_\epsilon^{(n)}$ will be real for all $\epsilon > 0$. The Liouville transformation is widely used in the physics literature [10, 63]. But although the gradient case occurs frequently in applications (for example, in the modelling of overdamped [inertialess] motion in a conservative force field, by means of a Smoluchowski equation), it is not generic.

As $\epsilon \to 0$ the operators $\mathcal{L}_\epsilon$ and $\mathcal{L}_\epsilon^*$ degenerate to first-order operators $\mathcal{L}_0$ and $\mathcal{L}_0^*$, so analysis of the $\epsilon \to 0$ limit is a singular perturbation problem. Formally one expects that $\lambda_\epsilon^{(n)} \to \lambda_0^{(n)}$ and $u_\epsilon^n \to u_0^n$, where $\lambda_0^{(n)}$ and $u_0^n$ satisfy the first-order differential equation $\mathcal{L}_0 u_0^n = \lambda_0^{(n)} u_0^n$. It will not in general be possible for the function $u_0^n$ to satisfy the Dirichlet boundary conditions on $\partial\Omega$, so a boundary layer near $\partial\Omega$ of thickness tending to zero as $\epsilon \to 0$ is expected to be present. The convergence $u_\epsilon^n \to u_0^n$ will be uniform only on compact subsets of $\Omega$.

The formal limiting eigenvalues $\lambda_0^{(n)}$ are easily worked out. Let $\boldsymbol{B} = (B^i{}_j)$ be the linearization of the drift field $b^i$ at $\boldsymbol{x} = S$, i.e., $B^i{}_j = \partial_j b^i(S)$; consider the case when $\boldsymbol{B}$ is negative definite.

Approximation of $\mathcal{L}_0 u_0^n = \lambda_0^{(n)} u_0^n$ near $S$, using $b^i(\boldsymbol{x}) \approx B^i{}_j(x^j - S^j)$, gives

$$-B^i{}_j(x^j - S^j)\partial_i u_0^n = \lambda_0^{(n)} u_0^n. \tag{5}$$

In this linear approximation the $u_0^n$ will be homogeneous polynomials in the normalized coordinates $x^i - S^i$; $u_0^0$ will be constant and the $u_0^n$, $n \geq 1$, will be of higher degree. For each choice of positive degree the system (5) may be solved for the polynomial approximate eigenfunctions and their eigenvalues. In the gradient case, for example, the eigenvalues turn out to be of the form $n_1|b^{(1)}| + n_2|b^{(2)}|$, $n_i \in \mathbb{N}$, for $b^{(1)}, b^{(2)}$ the two (negative) eigenvalues of $B$.

The fact that $u_0^0$ is always a constant function obviously holds beyond the linear approximation; the corresponding limiting eigenvalue $\lambda_0^{(0)}$ equals zero. But the principal eigenfunction $v_\epsilon^0$ of $\mathcal{L}_\epsilon^*$ is less well-behaved than $u_\epsilon^0$ as $\epsilon \to 0$. An analysis similar to the above indicates that the $v_\epsilon^n$, for all $n$, are localized on the $O(\epsilon^{1/2})$ lengthscale near $S$. Their limits as $\epsilon \to 0$ are accordingly singular. The asymptotics of the principal eigenfunction $v_\epsilon^0$ will be studied in detail in Sections 3, 6, and 7.

It is easier to justify rigorously this formal analysis, and control the $\epsilon \to 0$ limit of $\lambda_\epsilon^{(0)}$ and $u_\epsilon^0$, when the boundary of $\Omega$ is non-characteristic: the integral curves of $\boldsymbol{b}$ cross $\partial\Omega$ at all points, so that $\partial\Omega$ as well as $\Omega$ is attracted to $S$. In this case Friedman [28] proved that $\lambda_\epsilon^{(0)} \to 0$ exponentially, and bounded it above and below; also Devinatz and Friedman [21] proved that $u_\epsilon^0$ converges to a constant uniformly on compact subsets of $\Omega$. In the one-dimensional case asymptotics of higher order can be obtained; see de Groen [20]. The behavior of the higher ($n \geq 1$) eigenmodes seems not to have been rigorously investigated, but exponential convergence of $\lambda_\epsilon^{(n)}$ to $\lambda_0^{(n)}$ is expected as $\epsilon \to 0$.

In the case of a characteristic boundary there are some rigorous results of Eizenberg and Kifer [26] on the $\epsilon \to 0$ asymptotics of $\lambda_\epsilon^{(0)}$, but they do not apply when the boundary contains fixed points. Most rigorous work on the characteristic boundary case is probabilistic rather than analytic, so we remind the reader of the connections between the operators $\mathcal{L}_\epsilon$, $\mathcal{L}_\epsilon^*$ and the stochastic exit problem. Since the forward Kolmogorov equation $\dot{\rho} = -\mathcal{L}_\epsilon^* \rho$ may be written as a continuity equation $\dot{\rho} + \partial_i J^i = 0$, with the probability current

$$J^i = -(\epsilon/2)\partial_j[D^{ij}\rho] + b^i\rho, \tag{6}$$

the influx of probability into $\partial\Omega$ (on which $\rho = 0$ on account of absorption) has density $-(\epsilon/2)\partial_j[D^{ij}\rho](\boldsymbol{x})$ at point $\boldsymbol{x} \in \partial\Omega$. Accordingly the exit location density $p_\epsilon$, as a function of $\boldsymbol{x} \in \partial\Omega$, will satisfy

$$p_\epsilon(\boldsymbol{x}) = \int_0^\infty J^i(\boldsymbol{x},t) n_i(\boldsymbol{x}) \, dt = -(\epsilon/2) n_i(\boldsymbol{x}) \partial_j \left\{ D^{ij} \int_0^\infty \rho(\cdot,t) \, dt \right\}(\boldsymbol{x}) \tag{7}$$

where $n_i(\boldsymbol{x})$ is the outward normal to $\partial\Omega$ at $\boldsymbol{x}$. In the special case when $\boldsymbol{x}_\epsilon(0) = \boldsymbol{y}$ this yields

$$\begin{aligned} p_{\epsilon,\boldsymbol{y}}(\boldsymbol{x}) &= -(\epsilon/2) n_i(\boldsymbol{x}) \partial_j \left\{ D^{ij} \int_0^\infty [\exp(-\mathcal{L}_\epsilon^* t) \delta_{\boldsymbol{y}}] \, dt \right\}(\boldsymbol{x}) \\ &= -(\epsilon/2) n_i(\boldsymbol{x}) \partial_j \{ D^{ij} [(\mathcal{L}_\epsilon^*)^{-1} \delta_{\boldsymbol{y}}] \}(\boldsymbol{x}) \end{aligned} \tag{8}$$

the second subscript on $p_\epsilon$ now denoting the initial condition. Here $\delta_{\boldsymbol{y}}$ is a Dirac delta function. But we have

$$[(\mathcal{L}_\epsilon^*)^{-1} \delta_{\boldsymbol{y}}](\boldsymbol{x}) = \sum_{n=0}^\infty \left( \lambda_\epsilon^{(n)*} \right)^{-1} u_\epsilon^n(\boldsymbol{y}) v_\epsilon^n(\boldsymbol{x}), \tag{9}$$

since the right-hand side is the integral kernel of $(\mathcal{L}_\epsilon^*)^{-1}$. Substitution into (8) yields

$$p_{\epsilon,\boldsymbol{y}}(\boldsymbol{x}) = -(\epsilon/2) n_i(\boldsymbol{x}) \sum_{n=0}^\infty \left( \lambda_\epsilon^{(n)*} \right)^{-1} u_\epsilon^n(\boldsymbol{y}) \partial_i [D^{ij} v_\epsilon^n](\boldsymbol{x}). \tag{10}$$

Equation (10) is the point of contact between the analytic and probabilistic approaches to the stochastic exit problem. In the analytic approach asymptotic expansions, either formal or rigorously justified, are constructed for the eigenvalues and eigenfunctions. In the probabilistic approach the $\epsilon \to 0$ asymptotics of the exit location measure $p_{\epsilon,\boldsymbol{y}}(\boldsymbol{x}) \, d\boldsymbol{x}$ and the corresponding MFPT $\mathsf{E}_{\boldsymbol{y}} \tau_\epsilon$ are studied directly.

The rigorous probabilistic approach was pioneered by Wentzell and Freidlin; in the case of non-characteristic boundary their results are as follows [27]. If the classical action function $W$ attains a global minimum on $\partial\Omega$ at some point $\boldsymbol{x}^*$, for any $\boldsymbol{y} \in \Omega$ the first hitting location on $\partial\Omega$ is as $\epsilon \to 0$ increasingly concentrated near $\boldsymbol{x}^*$, i.e., $p_{\epsilon,\boldsymbol{y}} \to \delta_{\boldsymbol{x}^*}$ as $\epsilon \to 0$. Moreover $\mathsf{E}_{\boldsymbol{y}} \tau_\epsilon \sim \exp(W(\boldsymbol{x}^*)/\epsilon)$, $\epsilon \to 0$; the action $W(\boldsymbol{x}^*)$ is interpreted as the asymptotic exponential growth rate of the MFPT. These results have been extended to the case of a characteristic boundary $\partial\Omega$ by Day [15].

The lack of $\boldsymbol{y}$-dependence in the $\epsilon \to 0$ limit has a probabilistic interpretation: irrespective of the choice of starting point $\boldsymbol{y} \in \Omega$, for sufficiently small $\epsilon$ a typical sample path of the process $\boldsymbol{x}_\epsilon(t)$ tends to flow toward the attracting point $S$, and remain near $S$ for a substantial (exponentially long) time, before experiencing a fluctuation large enough to drive it to $\partial\Omega$. (With high probability this will be along a Wentzell-Freidlin classical trajectory.) The lack of $y$-dependence also has an interpretation in terms of the series expansion (10). Since $\lambda_\epsilon^{(0)}$ decays exponentially as $\epsilon \to 0$ and the other $\lambda_\epsilon^{(n)}$ are expected to converge to nonzero values, up to exponentially small relative errors it should suffice to keep only the $n = 0$ term in (10). That is,

$$p_{\epsilon,y}(x) \sim -(\epsilon/2)\left(\lambda_\epsilon^{(0)}\right)^{-1} u_\epsilon^0(y)\, n_i(x)\partial_j[D^{ij} v_\epsilon^0](x), \qquad \epsilon \to 0. \tag{11}$$

The asymptotic constancy of $u_\epsilon^0$ on $\Omega$, from this point of view, is the cause of the asymptotic $y$-independence of the measure $p_{\epsilon,y}(x)\,dx$ on $\partial\Omega$. Alternatively the boundary layer displayed by $u_\epsilon^0$ near $\partial\Omega$ may be viewed as a consequence of the fact that when the starting point $y$ is sufficiently near $\partial\Omega$, exit from $\Omega$ is likely to occur immediately rather than after a long sojourn in the vicinity of $S$.

The asymptotic validity of the reduction from (10) to (11) has not been rigorously proved, but is assumed (often implicitly) in most formal treatments of the exit problem. A related assumption, which we shall also make, is that up to exponentially small relative errors

$$\mathsf{E}_y\,\tau_\epsilon \sim \left(\lambda_\epsilon^{(0)}\right)^{-1}, \qquad \epsilon \to 0 \tag{12}$$

for any $y \in \Omega$, the subscript $y$ on the expectation denoting the initial condition for the process. This is similar to assuming that as $\epsilon \to 0$ the integral kernel of the evolution operator $\exp(-\mathcal{L}_\epsilon^* t)$, which equals the infinite sum

$$\sum_{n=0}^{\infty} \exp\left(-\lambda_\epsilon^{(n)*} t\right)\, u_\epsilon^n(y) v_\epsilon^n(x), \tag{13}$$

may be approximated to high accuracy at large $t$ by the $n = 0$ term. If so the distribution of $\tau_\epsilon$ for small $\epsilon$ will be essentially exponential, with expectation $\sim \left(\lambda_\epsilon^{(0)}\right)^{-1}$ as in (12). The assumption that only the $n = 0$ term in (13) is significant at large $t$ is justified by the fact that if the higher eigenvalues $\lambda_\epsilon^{(n)}, n \geq 1$ tend to nonzero limits as $\epsilon \to 0$, the $n \geq 1$ terms, considered individually, should unlike the $n = 0$ term decay on an $O(1)$ timescale rather than an exponentially long timescale. As $\epsilon \to 0$ there occurs a *separation of timescales*.

The assumption (12) ties asymptotic expansions and Wentzell-Freidlin theory firmly together: it permits the identification of $W(x^*)$ with the asymptotic exponential decay rate of $\lambda_\epsilon^{(0)}$ as $\epsilon \to 0$. Note that whether or not this assumption is made, for any $\epsilon > 0$ the eigenvalue $\lambda_\epsilon^{(0)}$ always has a rigorous interpretation as the exponential decay rate of $\mathsf{P}\{\tau_\epsilon > t\}$ as $t \to \infty$ [39]. Similarly for any $\epsilon > 0$, the eigenfunction $v_\epsilon^0$ of $\mathcal{L}_\epsilon^*$ corresponding to $\lambda_\epsilon^{(0)}$ has an interpretation as a *quasistationary density*. As $t \to \infty$, the $n = 0$ term in (13) will dominate, and irrespective of the choice of initial condition $\rho(\cdot, 0)$,

$$\rho(\cdot, t) = \exp(-\mathcal{L}_\epsilon^* t)\rho(\cdot, 0) \tag{14}$$

will be asymptotically proportional to $v_\epsilon^0$.

At this point we introduce the singularly perturbed boundary value problem mentioned in Section 1. Consider the problem

$$\mathcal{L}_\epsilon u_\epsilon = 0 \text{ in } \Omega, \, u_\epsilon = f \text{ on } \partial\Omega, \tag{15}$$

in which $f : \partial\Omega \to \mathbb{R}$ is a specified smooth function. It is easily checked that by Green's Theorem, for any $\epsilon > 0$ the problem has solution

$$u_\epsilon(\boldsymbol{y}) = -(\epsilon/2) \int_{\boldsymbol{x} \in \partial\Omega} f(\boldsymbol{x}) \partial_j \{[D^{ij}(\mathcal{L}_\epsilon^*)^{-1}\delta_{\boldsymbol{y}}]\}(\boldsymbol{x}) n_i(\boldsymbol{x}) \, d\boldsymbol{x}. \tag{16}$$

That is, by (8),

$$u_\epsilon(\boldsymbol{y}) = \int_{\boldsymbol{x} \in \partial\Omega} f(\boldsymbol{x}) p_{\epsilon,\boldsymbol{y}}(\boldsymbol{x}) \, d\boldsymbol{x} = \mathsf{E}_{\boldsymbol{y}} f(\boldsymbol{x}_\epsilon(\tau_\epsilon)). \tag{17}$$

In other words $u_\epsilon(\boldsymbol{y})$ equals the expected value of $f$ at the first hitting point $\boldsymbol{x}$ of the random process $\boldsymbol{x}_\epsilon(t)$ on $\partial\Omega$, provided that the starting point $\boldsymbol{x}_\epsilon(0)$ equals $\boldsymbol{y}$. This is an analytic rederivation of a well-known fact from the theory of random processes. In particular if $p_{\epsilon,\boldsymbol{y}} \to \delta_{\boldsymbol{x}^*}$, as will be the case if $W$ attains a minimum on $\partial\Omega$ at $\boldsymbol{x}^*$, $u_\epsilon$ on $\Omega$ will tend as $\epsilon \to 0$ to a constant, namely $f(\boldsymbol{x}^*)$. And by (11) and (17), it will be possible to work out the small-$\epsilon$ asymptotics of $u_\epsilon$ on $\Omega$ from the small-$\epsilon$ asymptotics of $v_\epsilon^0$ near $\partial\Omega$.

There are a handful of rigorous results on the boundary value problem which suggest that the asymptotic validity of the reduction from (10) to (11) is correct. Kamin [38] showed in the case of non-characteristic boundary that even if $W$ does not attain a unique minimum on $\partial\Omega$, the solution $u_\epsilon$ 'levels' as $\epsilon \to 0$, *i.e.*, tends to a constant. Eizenberg [25] has shown that this levelling occurs exponentially rapidly, with rate constant that may be taken arbitrarily close to the infimum of $W$ on $\partial\Omega$. This is consistent with the $n \geq 1$ terms in (10) being smaller than the $n = 0$ term by an exponentially small $O(\lambda_\epsilon^{(0)})$ factor. It is also known that $u_\epsilon$ will have a boundary layer near $\partial\Omega$, of width $O(\epsilon)$ if the boundary is non-characteristic and of width $O(\epsilon^{1/2})$ if the boundary is characteristic [33–35, 37]. This boundary layer may be viewed as arising from the boundary layer of the dominant eigenfunction $u_\epsilon^0$. The boundary value problem will be studied further in Section 9.

## 3 Matched Asymptotic Expansions

It is clear from the last section that the weak-noise ($\epsilon \to 0$) asymptotics of the stochastic exit problem are determined by the quasistationary density, *i.e.*, the principal eigenfunction $v_\epsilon^0$ of the forward Kolmogorov operator $\mathcal{L}_\epsilon^*$. The principal eigenfunction $u_\epsilon^0$ of the backward operator $\mathcal{L}_\epsilon$ is expected to level exponentially rapidly as $\epsilon \to 0$, so for all starting points $\boldsymbol{y} \in \Omega$ the expression (11) for the $\epsilon \to 0$ asymptotics of the density $p_{\epsilon,\boldsymbol{y}}$ of the exit location measure on $\partial\Omega$ may be written

$$p_{\epsilon,\boldsymbol{y}}(\boldsymbol{x}) \propto n_i(\boldsymbol{x})\partial_j[D^{ij}v_\epsilon^0](\boldsymbol{x}), \qquad \epsilon \to 0, \tag{18}$$

where $n_i(\boldsymbol{x})$ is the outward normal to $\partial\Omega$ at $\boldsymbol{x}$. The constant of proportionality here, which is independent of the starting point $\boldsymbol{y}$, is fixed by the normalization condition

$$\int_{\boldsymbol{x}\in\Omega} p_{\epsilon,\boldsymbol{y}}(\boldsymbol{x})\,d\boldsymbol{x} = 1. \tag{19}$$

We shall take $\boldsymbol{y} = S$ henceforth, though any other $\boldsymbol{y} \in \Omega$ could be chosen, and shall often drop the subscript. The interpretation of the asymptotic $\boldsymbol{y}$-independence is as discussed in the last section.

$p_{\epsilon,S}$ is by (18) asymptotic to the absorption location density of the eigenmode $v_\epsilon^0$ on $\partial\Omega$; also, the corresponding eigenvalue $\lambda_\epsilon^{(0)}$ of $\mathcal{L}_\epsilon^*$ may be viewed as the absorption rate of the mode. If $J_0^i$ is the probability current arising from $v_0^\epsilon$, *i.e.*,

$$J_0^i = -(\epsilon/2)\partial_j[D^{ij}v_\epsilon^0] + b_i v_\epsilon^0 \tag{20}$$

then

$$\lambda_\epsilon^{(0)} = \int_{\boldsymbol{x}\in\partial\Omega} J_0^i(\boldsymbol{x})n_i(\boldsymbol{x})\,d\boldsymbol{x} \tag{21}$$

provided that $v_\epsilon^0$ is normalized to total unit mass. If this normalization condition does not hold,

(21) must be replaced by

$$\lambda_\epsilon^{(0)} = \frac{\int_{\boldsymbol{x}\in\partial\Omega} J_0^i(\boldsymbol{x}) n_i(\boldsymbol{x})\, d\boldsymbol{x}}{\int_{\boldsymbol{x}\in\Omega} v_\epsilon^0(\boldsymbol{x})\, d\boldsymbol{x}}. \tag{22}$$

Since we are assuming that (up to exponentially small relative errors) $\mathsf{E}_S \tau_\epsilon \sim \left(\lambda_\epsilon^{(0)}\right)^{-1}$, the MFPT asymptotics are determined by the asymptotics of $v_\epsilon^0$.

We shall employ a now standard method of matched asymptotic expansions [52, 54–56, 65] to approximate $v_\epsilon^0$ as $\epsilon \to 0$, and use (18) and (22) to compute the asymptotics of $p_{\epsilon,S}$ and $\lambda_\epsilon^{(0)}$. Our treatment is facilitated by the fact that $\lambda_\epsilon^{(0)} \to \lambda_0^{(0)} = 0$ exponentially rapidly; up to exponentially small relative errors, it suffices when approximating $v_\epsilon^0$ in various regions of $\Omega$ to view it as a solution of $\mathcal{L}_\epsilon^* v_\epsilon^0 = 0$, rather than of $\mathcal{L}_\epsilon^* v_\epsilon^0 = \lambda_\epsilon^{(0)} v_\epsilon^0$. The alternative approach of approximating $\lambda_\epsilon^{(0)}$ and $v_\epsilon^0$ simultaneously is also possible, and has been applied to several models [11, 51].

The following three asymptotic regions may be distinguished. Recall that we are considering the case when $W$ attains a global minimum on $\partial\Omega$ at a distinguished saddle point $H$, so the limiting exit location $\boldsymbol{x}^*$ of the last section equals $H$.

- A local (stable) region of size $O(\epsilon^{1/2})$ centered on $S$.

- A region including all of $\Omega$ except the local region near $S$, and the boundary layer of width $O(\epsilon^{1/2})$ near the characteristic boundary $\partial\Omega$. This region contains the fluctuational paths from the vicinity of $S$ to the vicinity of $H$.

- A boundary region, of size as yet unspecified but centered on $H$, and lying within the boundary layer.

We now explain the corresponding asymptotic approximations for $v_\epsilon^0$: a Gaussian approximation near $S$, an outer approximation in the body of $\Omega$, and an inner approximation in the boundary region near $H$. The last, which we treat in Sections 5, 6 and 7, is where our treatment differs most radically from previous work; it will determine the asymptotics of the exit location density $p_{\epsilon,S}$.

# 4 The Outer Approximation

The outer approximation to the quasistationary density $v_\epsilon^0$ is a WKB expansion; equivalently, a characteristic (ray) expansion. We write

$$v_\epsilon^0(\boldsymbol{x}) \sim K(\boldsymbol{x}) \exp\left(-W(\boldsymbol{x})/\epsilon\right), \qquad \epsilon \to 0, \tag{23}$$

for certain functions $W : \bar{\Omega} \to \mathbb{R}$ and $K : \bar{\Omega} \to \mathbb{R}$ normalized so that $W(S) = 0$ and $K(S) = 1$, whose smoothness properties are as yet unspecified. We could instead attempt an approximation of higher order [68], with $K(\boldsymbol{x})$ in (23) replaced by $K_0(\boldsymbol{x}) + \epsilon K_1(\boldsymbol{x}) + \cdots$. However we shall see in Section 9 that in the case of a characteristic boundary there may be difficulties with matching such outer expansions (in integer powers of $\epsilon$) to the inner approximation in the boundary region.

Substituting the approximation (23) into the approximate forward Kolmogorov equation $\mathcal{L}_\epsilon^* v_\epsilon^0 = 0$, and collating the coefficients of powers of $\epsilon$, yields equations for $W$ and $K$:

$$H(x^i, \partial_i W) = 0 \tag{24}$$

$$\left[\frac{\partial H}{\partial p^i}(x^i, \partial_i W)\right] \partial_i K = -\left\{\frac{\partial^2 H}{\partial x^i \partial p_i}(x^i, \partial_i W) + \frac{1}{2}\frac{\partial^2 W}{\partial x^i \partial x^j}(\boldsymbol{x})\frac{\partial^2 H}{\partial p_i \partial p_j}(x^i, \partial_i W)\right\} K. \tag{25}$$

Here $H : \bar{\Omega} \times \mathbb{R}^2 \to \mathbb{R}$ is the Hamiltonian (or energy) function

$$H(x^i, p_i) = \frac{1}{2}D^{ij}(\boldsymbol{x})p_i p_j + b^i(\boldsymbol{x})p_i, \tag{26}$$

and the eikonal equation (24) is the corresponding zero-energy Hamilton-Jacobi equation. For consistency, we have also used this Hamiltonian in the transport equation (25). The presence of a Hamilton-Jacobi equation suggests a classical-mechanical interpretation. In classical mechanics the Hamiltonian (26) would determine the motion of a particle on $\Omega$; $\bar{\Omega} \times \mathbb{R}^2$ would be interpreted as phase space, and $\boldsymbol{p} = (p_i)$, $i = 1, 2$, as the momentum of the particle. But this Hamiltonian is precisely the Wentzell-Freidlin Hamiltonian governing the large fluctuations of the process $\boldsymbol{x}_\epsilon(t)$ away from $S$ [27]. Recall that if

$$L(x^i, \dot{x}^i) = \frac{1}{2}D_{ij}(\boldsymbol{x})(\dot{x}^i - b^i(\boldsymbol{x}))(\dot{x}^j - b^j(\boldsymbol{x})) \tag{27}$$

is the Lagrangian canonically conjugate to $H(x^i, p_i)$, with the covariant tensor field $D_{ij}$ defined by $D_{ij}D^{jk} = \delta_i{}^k$, then the Wentzell-Freidlin classical action function $W : \bar{\Omega} \to \mathbb{R}$ is defined by

$$W(\boldsymbol{x}) = \inf_{\substack{T>0 \\ \boldsymbol{q}:[0,T]\to\bar{\Omega} \\ \boldsymbol{q}(0)=S,\ \boldsymbol{q}(T)=\boldsymbol{x}}} \int_0^T L(\boldsymbol{q},\dot{\boldsymbol{q}})\,dt. \tag{28}$$

That is, in Wentzell-Freidlin theory $W(\boldsymbol{x})$ is computed as an infimum over trajectories from $S$ to $\boldsymbol{x}$, and is a classical action function in the sense of classical mechanics. The Wentzell-Freidlin $W$ will necessarily satisfy the Hamilton-Jacobi equation (24), so we may identify it with the function $W$ in the outer approximation (23).

By the calculus of variations, if the infimum in (28) is achieved by a trajectory extending from $S$ to $\boldsymbol{x}$, all portions of the trajectory that lie in $\Omega$ (rather than $\partial\Omega$) must consist of least-action (zero-energy) classical trajectories [19]. Classical trajectories $\boldsymbol{x}(\cdot)$ are the trajectories in $\Omega$ determined by $H$ (or $L$); they satisfy the Euler-Lagrange equation

$$\frac{d}{dt}\left(\frac{\partial L}{\partial \dot{x}^i}\right) - \frac{\partial L}{\partial x^i} = 0. \tag{29}$$

Equivalently, if a classical trajectory is viewed as a pair of functions $(\boldsymbol{x}(\cdot), \boldsymbol{p}(\cdot))$, specifying position and momentum (*i.e.*, location in $\Omega \times \mathbb{R}^2$) as a function of time, it must satisfy Hamilton's equations. Hamilton's equations are of the form

$$\frac{d}{dt}\begin{pmatrix} x^k \\ p_k \end{pmatrix} = \begin{pmatrix} \boldsymbol{0} & \boldsymbol{I} \\ -\boldsymbol{I} & \boldsymbol{0} \end{pmatrix} \begin{pmatrix} \frac{\partial H}{\partial x^k} \\ \frac{\partial H}{\partial p_k} \end{pmatrix}, \tag{30}$$

and express the fact that $t \mapsto (\boldsymbol{x}(t), \boldsymbol{p}(t))$ must be an integral curve of a vector field on $\Omega \times \mathbb{R}^2$ determined by $H$. For the Hamiltonian (26) they reduce to

$$\dot{x}^k = \frac{\partial H}{\partial p_k} = D^{jk}(\boldsymbol{x})p_j + b^k(\boldsymbol{x}) \tag{31}$$

$$\dot{p}_k = -\frac{\partial H}{\partial x^k} = \partial_k D^{ij}(\boldsymbol{x})p_i p_j + \partial_k b^i(\boldsymbol{x})\,p_i. \tag{32}$$

Notice that at all points along the trajectory, momentum and velocity uniquely determine each other; by (31),

$$p_i = D_{ij}(\boldsymbol{x})[\dot{x}^j - b^j(\boldsymbol{x})]. \tag{33}$$

This is familiar from classical mechanics.

If the least-action trajectory from $S$ to $\boldsymbol{x}$ exists and is unique, it is interpreted in Wentzell-Freidlin theory as the most probable (in the $\epsilon \to 0$ limit) fluctuational path from $S$ to $\boldsymbol{x}$: the one whose cost is least. Since $p_i = \partial_i W$ at all points along the trajectory, $W(\boldsymbol{x})$ (the cost of the trajectory) can be computed from

$$W(\boldsymbol{x}) = \int_S^{\boldsymbol{x}} p_i \, dx^i, \tag{34}$$

the line integral being taken along the trajectory. Equivalently, $W$ satisfies

$$\dot{W} = p_i \dot{x}^i = p_i [D^{ij}(\boldsymbol{x}) p_j + b^i(\boldsymbol{x})], \tag{35}$$

an ordinary differential equation which may be integrated along the corresponding trajectory in phase space. Note however that the infimum in (28) will not actually be achieved at finite transit time. It is readily verified for the Hamiltonian (26) that zero-energy trajectories emanating from a fixed point of the drift field $\boldsymbol{b}$, such as $S$, will have *infinite transit time*. The formally most probable fluctuational paths require an infinite amount of time to emerge from $S$, and are more naturally parametrized by $t \in (-\infty, T]$. We discuss the physical consequences of this phenomenon elsewhere [56]. The integration of (35), as well as that of (31) and (32), must begin at $t = -\infty$.

For each such path $\lim_{t \to -\infty} \boldsymbol{x}(t) = S$, and by (33), $\lim_{t \to -\infty} \boldsymbol{p}(t) = 0$ as well. But by examination $(S, 0)$ is a fixed point (of hyperbolic type) of the deterministic flow on the phase space $\Omega \times \mathbb{R}^2$ defined by Hamilton's equations. So in the language of dynamical systems, the most probable fluctuational paths of Wentzell-Freidlin theory may be viewed as forming the *unstable manifold* of the point $(S, 0) \in \bar{\Omega} \times \mathbb{R}^2$. This manifold, which we shall denote $\mathcal{M}^u_{(S,0)}$, is 2-dimensional; it is coordinatized (at least in a neighborhood of $(S, 0)$) by (i) the choice of outgoing trajectory, and (ii) the arc length along the trajectory. $\mathcal{M}^u_{(S,0)}$ is a *Lagrangian manifold*: it is invariant under the flow. The stable manifold $\mathcal{M}^s_{(S,0)}$ of $(S, 0)$ in $\Omega \times \mathbb{R}^2$ is clearly the manifold $\boldsymbol{p} \equiv \boldsymbol{0}$, which by (31) and (32) is also invariant under the flow. It comprises 'cost-free' ($\Delta W = 0$) trajectories that satisfy $\dot{x}^i = b^i(\boldsymbol{x})$, and simply follow the mean drift toward $S$.

The analogy with Hamiltonian dynamics can be pushed much further. The action $W$ as com-puted by (34) is a single-valued function of position on $\mathcal{M}^u_{(S,0)}$, and $p_i$ equals $\partial_i W(\boldsymbol{x})$ at every point $(\boldsymbol{x}, \boldsymbol{p}) \in \mathcal{M}^u_{(S,0)}$. But it does not follow that $W$ may be viewed as a single-valued function on $\Omega$. This is because the unstable manifold may fold back on itself, and the projection $(\boldsymbol{x}, \boldsymbol{p}) \mapsto \boldsymbol{x}$ from $\mathcal{M}^u_{(S,0)}$ to $\bar{\Omega}$ may not be one-to-one. Equivalently the zero-energy classical trajectories emanating from $S$ may cross each other, creating *caustics* [12, 14, 24, 36, 54–56]; these caustics are the projections onto $\bar{\Omega}$ of the folds of $\mathcal{M}^u_{(S,0)}$. We shall not pursue the phenomenon of caustics at any length; unless otherwise stated we assume that the solution of the Hamilton-Jacobi equation is single-valued on $\bar{\Omega}$. In this case the map $\boldsymbol{x} \mapsto \boldsymbol{p}(\boldsymbol{x})$ will be single-valued. The projection map from $\mathcal{M}^u_{(S,0)}$ to $\bar{\Omega}$ may also fail to be onto. If this occurs, $W(\boldsymbol{x})$, for at least some $\boldsymbol{x} \in \bar{\Omega}$, cannot be viewed as the action of a classical trajectory from $S$ to $\boldsymbol{x}$ which lies entirely in $\Omega$; the action-minimizing trajectory, if it exists, must pass through $\partial\Omega$ on its way from $S$ to $\boldsymbol{x}$. We defer discussion of this possibility to the next section.

If we wish to compute the value of $W$ at some point in $\Omega$ other than $S$, we can integrate Hamilton's equations from $(S, 0)$, continually updating $\boldsymbol{x}$ and $\boldsymbol{p}$, until the specified point is reached. The ordinary differential equation (35), which may be integrated simultaneously, will yield the value of $W$ at the endpoint. However the numerical computation of $K$, which has no elementary classical-mechanical interpretation, is more subtle. Since $\dot{x}^i = \partial H / \partial p_i$, $(\partial H / \partial p_i)\partial_i K$ equals $\dot{K}$, the time derivative of $K$ along the trajectory emerging from $S$. So (25) becomes

$$\dot{K} = -\left\{ \frac{\partial^2 H}{\partial x^i \partial p_i} + \frac{1}{2} \frac{\partial^2 W}{\partial x^i \partial x^j} \frac{\partial^2 H}{\partial p_i \partial p_j} \right\} K. \tag{36}$$

This equation cannot be integrated numerically unless the Hessian matrix $\partial_i \partial_j W$ is known at all points along the trajectory. Fortunately $\partial_i \partial_j W$ itself satisfies an ordinary differential equation, obtained as follows. Differentiating the Hamilton-Jacobi equation $H = 0$ twice with respect to position on $\mathcal{M}^u_{(S,0)}$ yields

$$\left( \frac{\partial}{\partial x^i} + \frac{\partial p_k}{\partial x_i} \frac{\partial}{\partial p_k} \right) \left( \frac{\partial}{\partial x^j} + \frac{\partial p_l}{\partial x^j} \frac{\partial}{\partial p_l} \right) H = 0. \tag{37}$$

By rearranging terms, and using Hamilton's equations and $\partial_i p_j = \partial_i \partial_j W$, we obtain

$$\dot{W}_{,ij} = -\frac{\partial^2 H}{\partial p_k \partial p_l} W_{,ik} W_{,jl} - \frac{\partial^2 H}{\partial x^j \partial p_k} W_{,ik} - \frac{\partial^2 H}{\partial x^i \partial p_k} W_{,jk} - \frac{\partial^2 H}{\partial x^i \partial x^j}. \tag{38}$$

(For notational convenience $W_{,ij}$ signifies $\partial_i \partial_j W$ henceforth; we also write $b^i{}_{,j}$ for $\partial_j b^i$, *etc.*)

This version of the inhomogeneous Riccati equation for the Hessian $W_{,ij}$ is due to the present authors [55], though Ludwig apparently made use of an equivalent equation in the 1970's [49]. In any event, equation (38) may be integrated to yield $W_{,ij}$ at all points along an outgoing Wentzell-Freidlin trajectory. In numerical work the system (31), (32), (35), (36), (38) of coupled ordinary differential equations would be integrated simultaneously to produce $W$ and $K$.

At any $\boldsymbol{x} \in \Omega$, $W_{,ij}(\boldsymbol{x}) = \partial_j p_i(\boldsymbol{x})$ specifies a 2-dimensional subspace of $\mathbb{R}^4$, the tangent space $T_{(\boldsymbol{x},\boldsymbol{p}(\boldsymbol{x}))}\mathcal{M}^u_{(S,0)}$ to the unstable manifold $\mathcal{M}^u_{(S,0)} \subset \bar{\Omega} \times \mathbb{R}^2$ at $(\boldsymbol{x}, \boldsymbol{p}(\boldsymbol{x}))$. The Riccati equation (38) can be viewed as determining a trajectory through the (compact) Grassman manifold of such 2-dimensional subspaces. This sort of interpretation is standard in the theory of matrix Riccati equations [71], and facilitates the integration of (38) through points where $W_{,ij}$ diverges. Such divergences occur however only when the classical trajectory encounters a caustic [24]. This is because $W_{,ij}$ diverges only at points $\boldsymbol{x}$ where the tangent space $T_{(\boldsymbol{x},\boldsymbol{p}(\boldsymbol{x}))}\mathcal{M}^u_{(S,0)}$ 'turns vertical' [22]. We shall not pursue the consequences of caustics further here.

Much more could be said about the geometric interpretation of the above system of equations, which is ultimately made possible by the symplectic structure of classical mechanics on $\Omega \times \mathbb{R}^2$. We confine ourselves here to pointing out the differential-geometric (coordinate-free) interpretation of the Wentzell-Freidlin fluctuational paths on $\Omega$. The contravariant tensor field $D^{ij}$ is naturally viewed as a Riemannian metric on $\Omega$, and may be used to raise and lower indices. If connection coefficients (Christoffel symbols) $\Gamma^i{}_{jk}$ are defined by

$$\Gamma^i{}_{jk} = \frac{1}{2}D^{il}(D_{lj,k} + D_{lk,j} - D_{kj,l}) \tag{39}$$

in the usual way, the covariant derivative $u^i{}_{;j}$ of a vector field $u^i$ on $\Omega$ will be given by

$$u^i{}_{;j} = u^i{}_{,j} + \Gamma^i{}_{jk}u^k. \tag{40}$$

It is easy to check that if the mean drift $\boldsymbol{b} = 0$, the Euler-Lagrange equations (29) for the velocity field $\dot{x}^i(\cdot)$ of the fluctuational paths on $\Omega$ reduce to the single covariant equation $\dot{x}^i{}_{;j}\dot{x}^j = 0$. This is a statement of *covariant constancy* of the velocity of each such path along itself: the most probable fluctuational paths in this case are simply the geodesics of $D^{ij}$ which emanate from the point $S$. If $\boldsymbol{b} \neq 0$ the situation is more complicated; computation yields

$$(\dot{x}^i - b^i)_{;j}\dot{x}^j + (\dot{x}^j - b^j)b_j{}^{;i} = 0. \tag{41}$$

In a similar way, if $\boldsymbol{b} = 0$ the Riccati equation (38), if viewed as applying to a function $W$ defined on $\Omega$ rather than on $\mathcal{M}^u_{(S,0)}$, simplifies to yield a covariant equation closely related to the equations of geodesic deviation on $\Omega$. Taking $\boldsymbol{b} \neq 0$ yields a generalization. A fuller discussion of the differential-geometric interpretation, which owes much to work of Graham [31], may appear elsewhere.

## 5 The Approximations Near the Fixed Points

The outer approximation of the last section to the principal eigenfunction $v^0_\epsilon$ of $\mathcal{L}^*_\epsilon$ must be supplemented by an inner approximation in a boundary region centered on $H$, and a Gaussian approximation in the stable region of size $O(\epsilon^{1/2})$ near $S$. To a certain extent it is possible to treat the approximations near $H$ and $S$ in parallel.

This is possible because $(H, 0)$, as well as $(S, 0)$, is a fixed point (of hyperbolic type) of the deterministic flow on the phase space $\bar{\Omega} \times \mathbb{R}^2$ specified by Hamilton's equations (30), and the Wentzell-Freidlin Hamiltonian (26). In fact to any fixed point of the drift field $\boldsymbol{b}$ on $\Omega$ (or $\bar{\Omega}$) there corresponds a fixed point of the flow on $\Omega \times \mathbb{R}^2$ (or $\bar{\Omega} \times \mathbb{R}^2$), at zero momentum. By examination, zero-energy classical trajectories which are incident on such points in phase space, as well as those which emanate from them, will have infinite transit time. As a consequence the zero-energy trajectories incident on $(H, 0)$ may be viewed as forming the stable manifold of $(H, 0)$ in $\bar{\Omega} \times \mathbb{R}^2$, which we shall denote $\mathcal{M}^s_{(H,0)}$.

We are assuming that the classical action $W$, computed from the variational definition (28), attains a minimum on $\partial \Omega$ at $H$. We shall also assume that the infimum in the expression (28) for $W(H)$ is achieved; in particular, that $W(H)$ is the action of a unique zero-energy classical trajectory $\boldsymbol{q}^* : \mathbb{R} \to \Omega$ which emanates from $S$ at time $t = -\infty$ and is incident on $H$ at $t = \infty$. This trajectory $\boldsymbol{q}^*$, which is the formally most probable fluctuational path from $S$ to $\partial \Omega$ as $\epsilon \to 0$, is called the *most probable exit path* (MPEP). It may be viewed as a trajectory in phase space, in which case it necessarily lies in both the unstable manifold $\mathcal{M}^u_{(S,0)}$ and the stable manifold $\mathcal{M}^s_{(H,0)}$. The intersection $\mathcal{M}^u_{(S,0)} \cap \mathcal{M}^s_{(H,0)}$ of these two 2-dimensional manifolds will consist of zero-energy trajectories from $(S, 0)$ to $(H, 0)$. Examples are known in which the intersection consists of more than a single trajectory [55], but generically we expect that there is a unique zero-energy trajectory from $S$ to $H$ with minimum action.

The behavior of $W$ near $S$ and $H$ will constrain the possible approximations to $v^0_\epsilon$ near $S$ and $H$.

Since at all points on $\mathcal{M}^u_{(S,0)}$ momentum $p_i$ equals $\partial_i W$ and the MPEP approaches $(\boldsymbol{x}, \boldsymbol{p}) = (S, 0)$ as $t \to -\infty$ and $(H, 0)$ as $t \to \infty$, we expect that $S$ and $H$ are critical points of $W$ on $\bar{\Omega}$. So to leading order, a quadratic approximation to $W$ would seemingly be appropriate near both $S$ and $H$. We shall shortly see that generically, the behavior of $W$ near $H$ (though not near $S$) is more complicated. But an exploration of the possible quadratic approximations to $W$ will prove useful nonetheless.

The assumption of quadratic behavior near the fixed points has the following consequences. As $t \to -\infty$ or $t \to \infty$ the left-hand side of the matrix Riccati equation (38) will tend to zero, yielding the *algebraic* Riccati equation

$$\frac{\partial^2 H}{\partial p_k \partial p_l} W_{,ik} W_{,jl} + \frac{\partial^2 H}{\partial x^j \partial p_k} W_{,ik} + \frac{\partial^2 H}{\partial x^i \partial p_k} W_{,jk} + \frac{\partial^2 H}{\partial x^i \partial x^j} = 0 \tag{42}$$

for the Hessian matrix $W_{,ij} = W_{,ij}(\mathfrak{p})$ of second derivatives of $W$ at the fixed point $\mathfrak{p}$, $\mathfrak{p} = S, H$. Here the partial derivatives of the Hamiltonian must be evaluated at $(\boldsymbol{x}, \boldsymbol{p}) = (\mathfrak{p}, 0)$. Substituting the explicit form (26) of the Wentzell-Freidlin Hamiltonian $H(\cdot, \cdot)$ into (42) yields the matrix equation

$$W_{,ij} D^{jk} W_{,kl} + W_{,ij} B^j{}_l + B^j{}_i W_{,jl} = 0 \tag{43}$$

for $W_{,ij}(\mathfrak{p})$. Here we have written $\boldsymbol{B} = (B^i{}_j)$ for the linearized drift field $b^i{}_{,j}(\mathfrak{p})$ at $\mathfrak{p}$, and $D^{ij}$ signifies the local diffusivity tensor $D^{ij}(\mathfrak{p})$. Recall that by assumption the matrix $D^{ij}$ is a positive definite quadratic form at both $S$ and $H$. We are also assuming that the matrix $B^i{}_j$ is nonsingular at both $S$ and $H$; this is the case for any generic drift field $\boldsymbol{b}$. Note that equation (43) could be derived directly from the forward Kolmogorov equation [49, 57], without using the matrix Riccati equation (38).

In matrix form equation (43) reads

$$\boldsymbol{Z} \boldsymbol{D} \boldsymbol{Z} + \boldsymbol{Z} \boldsymbol{B} + \boldsymbol{B}^t \boldsymbol{Z} = \boldsymbol{0} \tag{44}$$

where $\boldsymbol{Z} = \boldsymbol{Z}(\mathfrak{p}) = (W_{,ij}(\mathfrak{p}))$ is the 2-by-2 Hessian matrix of $W$ at $\mathfrak{p}$. Since $W$ is computed from zero-energy trajectories emanating from $S$ at $t = -\infty$, the matrix $W_{,ij}(\mathfrak{p}) = \partial p_i / \partial x^j(\mathfrak{p})$ specifies a 2-dimensional subspace of $\mathbb{R}^4$, the tangent space $T_{(\mathfrak{p},0)} \mathcal{M}^u_{(S,0)}$ of the unstable manifold $\mathcal{M}^u_{(S,0)}$ at $(\mathfrak{p}, 0)$. By assumption $\mathcal{M}^u_{(S,0)}$ extends from $S$ to $H$ along the MPEP $\boldsymbol{q}^*$, so $\mathfrak{p} = H$ as well as $\mathfrak{p} = S$ is meaningful here.

We now digress to discuss the solution of the algebraic Riccati equations for the 2-by-2 matrices $Z(S)$ and $Z(H)$. The analysis of the Riccati equation (44) is facilitated by a one-to-one correspondence between solutions $Z$ (in general complex) and certain linear subspaces of $\mathbb{C}^4$. This correspondence can be explained without reference to classical dynamics; it was in fact first explored in the context of optimal control theory [44, 70]. At each of the two fixed points $\mathfrak{p}$, define a 4-by-4 matrix $T = T(\mathfrak{p})$ by

$$T = \begin{pmatrix} B & D \\ 0 & -B^t \end{pmatrix} \tag{45}$$

$T$ may be written as

$$\begin{pmatrix} \frac{\partial^2 H}{\partial p_i \partial x^j} & \frac{\partial^2 H}{\partial p_i \partial p_j} \\ -\frac{\partial^2 H}{\partial x^i \partial x^j} & -\frac{\partial^2 H}{\partial x^i \partial p_j} \end{pmatrix} = \begin{pmatrix} 0 & I \\ -I & 0 \end{pmatrix} \begin{pmatrix} \frac{\partial^2 H}{\partial x^i \partial x^j} & \frac{\partial^2 H}{\partial x^i \partial p_j} \\ \frac{\partial^2 H}{\partial p_i \partial x^j} & \frac{\partial^2 H}{\partial p_i \partial p_j} \end{pmatrix} \tag{46}$$

to reveal its symplectic structure; the representation (46) will be useful later. It is easily checked that a 2-by-2 (complex) matrix $Z$ is a solution of the Riccati equation (44) if and only if the column space (over $\mathbb{C}$) of the 4-by-2 matrix

$$R = \begin{pmatrix} I \\ Z \end{pmatrix}, \tag{47}$$

which is a 2-dimensional subspace of $\mathbb{C}^4$, is $T$-invariant, *i.e.*, if and only if $TR = RL$ for some 2-by-2 complex matrix $L$. A 2-dimensional subspace of $\mathbb{C}^4$ which can be viewed as the column space of such a 4-by-2 matrix $R$ (equivalently, a 2-dimensional subspace which is complementary to the subspace spanned by $(0, 0, 1, 0)$ and $(0, 0, 0, 1)$) is known as a *graph subspace*. The one-to-one correspondence between solutions and subspaces facilitates the solving of the algebraic Riccati equation: one need only enumerate the 2-dimensional $T$-invariant subspaces of $\mathbb{C}^4$, and determine which of them are graph subspaces. Each such graph subspace yields a solution $Z$, and these are the only solutions.

The situation here is slightly more complicated in that for $Z(\mathfrak{p})$ we are interested only in solution matrices $Z$ which are real symmetric. Gohberg, Lancaster and Rodman [30, 44] and Shayman [70] show in arbitrary dimensionality ('2' and '4' replaced by $n$ and $2n$ respectively, and $T$ a $2n$-by-$2n$ real matrix) that if $D > 0$ every $n$-dimensional $T$-invariant subspace of $\mathbb{C}^{2n}$ is a graph subspace, and that Hermitian solutions $Z$ of (44) will correspond to $n$-dimensional $T$-invariant subspaces $\mathcal{N} \subset \mathbb{C}^{2n}$ which are $J$-*neutral* in the sense that

$$\boldsymbol{v}^\dagger \boldsymbol{J} \boldsymbol{v} = 0 \tag{48}$$

for all vectors $\boldsymbol{v} \in \mathcal{N}$. Here

$$\boldsymbol{J} = \begin{pmatrix} 0 & \boldsymbol{I} \\ -\boldsymbol{I} & 0 \end{pmatrix} \tag{49}$$

is the fundamental symplectic form. The problem of finding real symmetric solutions $Z$ thus reduces to constructing, from the spectral subspaces of $T$, $n$-dimensional $J$-neutral $T$-invariant subspaces whose corresponding matrices $Z$ are real.

This construction is much easier than it sounds. Since $T$ is real, its eigenvalue set will be symmetric about the real axis, and since its eigenvalues are those of $B$ and $-B^t$ the eigenvalue set will be symmetric about the imaginary axis as well. If all eigenvalues of $T$ have unit multiplicity Lancaster and Rodman [44] show that there are $2^k$ possible subspaces $\mathcal{N}$ satisfying the conditions, where $k$ is the cardinality of $\Lambda$: the set of eigenvalues $\lambda$ of $T$ satisfying $\operatorname{Re}\lambda > 0$, $\operatorname{Im}\lambda \geq 0$. If no eigenvalue of $T$ is pure imaginary, the possible $\mathcal{N}$ are formed from the $2^k$ subsets $\Lambda' \subset \Lambda$ as follows: each is the spectral subspace of $T$ corresponding to

1. the eigenvalues in $\Lambda'$,

2. their complex conjugates, if distinct,

3. the negatives of the eigenvalues in $\Lambda \setminus \Lambda'$, and

4. their complex conjugates, if distinct.

If $T$ has any pure imaginary eigenvalues, the construction of the $2^k$ possible $\mathcal{N}$ will be slightly different; each $\mathcal{N}$ must be expanded to include a unique subspace constructed from their eigenvectors [70]. If any eigenvalue of $T$ has nontrivial multiplicity the set of possible subspaces $\mathcal{N}$ will no longer have cardinality $2^k$. Nontrivial *geometric* multiplicity will engender an infinite number of subspaces $\mathcal{N}$ (and real symmetric solutions $Z$), though nontrivial *algebraic* multiplicity will yield only a finite (increased) number of real symmetric solutions [30, 44].

Let us consider the implications of these results for the algebraic Riccati equations arising in the stochastic exit problem, which have $n = 2$. The eigenvalues of $T$ are those of $B$, together with their negatives. But we are assuming that at both $\mathfrak{p} = S$ and $\mathfrak{p} = H$, $B$ is nonsingular. Since $S$ is stable, either $B(S)$ has two distinct eigenvalues $\lambda_{s1}(S)$, $\lambda_{s2}(S)$ with negative real parts (both pure real, or $\lambda_{s1}(S) = \lambda_{s2}(S)^*$), or a single negative eigenvalue with nontrivial multiplicity (either algebraic or geometric). And since the smooth boundary $\partial\Omega$ is characteristic, $B(H)$ must have one negative eigenvalue $\lambda_s(H) < 0$ (whose eigenvector points along $\partial\Omega$) and one positive eigenvalue $\lambda_u(H) > 0$ (whose eigenvector points into $\Omega$). At neither fixed point $\mathfrak{p}$ does $B(\mathfrak{p})$ or $T(\mathfrak{p})$ have any pure imaginary eigenvalues.

We see, accordingly, that at $\mathfrak{p} = H$ there are normally $k = 2$ distinct (real) eigenvalues of $T(\mathfrak{p})$ satisfying $\operatorname{Re}\lambda > 0$, $\operatorname{Im}\lambda \geq 0$; namely, $-\lambda_s(H)$ and $\lambda_u(H)$. This will be the case unless $-\lambda_s(H) = \lambda_u(H)$; equivalently, unless $\operatorname{tr} B(H) = b^i{}_{,i}(H) = 0$. *We now explicitly rule out this possibility*: we assume that $-\lambda_s(H) \neq \lambda_u(H)$. This is not a major restriction, since drift fields $b$ satisfying $\partial_i b^i(H) = 0$ are nongeneric. Recall that we have already defined the parameter

$$\mu \stackrel{\text{def}}{=} |\lambda_s(H)|/\lambda_u(H). \tag{50}$$

The nongeneric case $\mu = 1$ will be seen below to be a difficult 'boundary' case, intermediate between the radically different cases $\mu < 1$ and $\mu > 1$ (the subjects of Sections 7 and 8 respectively). The difficulty with $\mu = 1$ is closely connected to the existence, by the results of Gohberg, Lancaster, Rodman, and Shayman, of an infinite number of possible solutions $Z(H)$. This indeterminacy does not exist when $\mu \neq 1$; there are precisely $2^k = 4$ real symmetric solutions of the algebraic Riccati equation (44) at $\mathfrak{p} = H$. We shall examine these four solutions shortly.

We first dispose of the case $\mathfrak{p} = S$, which is comparatively straightforward. Here $k = 2$ except when $B(S)$ has only a single (negative) eigenvalue, so one expects four solutions for $Z(S)$ except in cases of nontrivial multiplicity. However it is easy to see that irrespective of the multiplicity, there will be only a single solution $Z(S)$ of full rank. This is because if $Z$ is of full rank, $Z^{-1}$ exists, and multiplying (44) left and right by $Z^{-1}$ yields

$$D + BZ^{-1} + Z^{-1}B^t = 0. \tag{51}$$

This is a system of three linear equations for the three independent elements of $Z^{-1}$. A bit of linear algebra shows that in the 2-by-2 case, the stability of $B = B(S)$ will guarantee the existence of a unique solution $Z^{-1}$ and hence a unique nonsingular $Z(S)$.

At $\mathfrak{p} = S$ the solutions to (44) of less than full rank, which include $Z = \mathbf{0}$, can be ruled out on physical grounds; $S$ must be a local minimum of $W$, and the cost $W(\boldsymbol{x})$ of fluctuations from $S$ to $\boldsymbol{x}$ must increase quadratically as $\boldsymbol{x} - S$ increases. So the correct quadratic behavior of $W$ near $S$ is specified by the nonsingular solution. The nonsingular Hessian matrix $W_{,ij}(S)$ immediately yields an approximation to the eigenfunction $v_\epsilon^0$ in the stable region of size $O(\epsilon^{1/2})$ near $S$. Since

$$v_\epsilon^0(\boldsymbol{x}) \sim K(\boldsymbol{x}) \exp\left(-W(\boldsymbol{x})/\epsilon\right) \tag{52}$$

is the outer approximation in the body of $\Omega$, the natural approximation to $v_\epsilon^0$ near $S$ is

$$v_\epsilon^0(\boldsymbol{x}) \sim \exp\left[-(x^i - S^i) W_{,ij}(S)(x^j - S^j)/2\epsilon\right]. \tag{53}$$

This bivariate Gaussian approximation will match to the outer approximation if $K$ is normalized so that $K(S) = 1$. Equivalently we could normalize the Gaussian so that it has total mass unity, and choose $K(S)$ accordingly. In any event the covariance matrix $\boldsymbol{Z}(S)^{-1}$ of the Gaussian, which governs the small fluctuations of the process $\boldsymbol{x}_\epsilon(t)$ on the $O(\epsilon^{1/2})$ lengthscale near $S$, can be computed numerically from (51); this has been known since the work of Ludwig [49]. On the $O(\epsilon^{1/2})$ lengthscale there are no well-defined most probable fluctuational paths: the process resembles a two-dimensional Ornstein-Uhlenbeck process.

We now return to an analysis of the quadratic (or putatively quadratic) behavior of $W$ near the saddle point $H$. A correct analysis depends crucially on the classical-mechanical interpretation of the matrix $\boldsymbol{T}$. It follows from the representation (46) that linearizing Hamilton's equations (30) near $(\boldsymbol{x}, \boldsymbol{p}) = (\mathfrak{p}, \mathbf{0})$ yields

$$\frac{d}{dt}\begin{pmatrix} \delta\boldsymbol{x} \\ \delta\boldsymbol{p} \end{pmatrix} = \boldsymbol{T}(\mathfrak{p})\begin{pmatrix} \delta\boldsymbol{x} \\ \delta\boldsymbol{p} \end{pmatrix} \tag{54}$$

where $(\delta\boldsymbol{x}, \delta\boldsymbol{p}) = (\boldsymbol{x}, \boldsymbol{p}) - (\mathfrak{p}, \mathbf{0})$. In other words, $\boldsymbol{T}(\mathfrak{p})$ specifies the linearization of the Hamiltonian flow near the fixed point $(\mathfrak{p}, \mathbf{0})$ in phase space.

So the eigenvalues and eigenvectors of $\boldsymbol{T}(H)$ will have an interpretation in terms of the flow of zero-energy classical trajectories in the vicinity of $H$. We introduce some notation: let the eigenvectors of the linearized drift $\boldsymbol{B}(H)$ with eigenvalues $\lambda_s(H) < 0$ and $\lambda_u(H) > 0$ be denoted $\boldsymbol{e}_s$ and $\boldsymbol{e}_u$. By the definition (45), the eigenvectors of $\boldsymbol{T}(H)$ with eigenvalues $\lambda_s(H)$ and $\lambda_u(H)$ will be $(\bm{e}_s, 0)$ and $(\bm{e}_u, 0)$. The eigenvectors of $\bm{T}(H)$ with eigenvalues $-\lambda_s(H)$ and $-\lambda_u(H)$ will be denoted $(\tilde{\bm{e}}_u, \tilde{\bm{g}}_u)$ and $(\tilde{\bm{e}}_s, \tilde{\bm{g}}_s)$; the interchange of subscripts is justified because $-\lambda_s(H) > 0$ and $-\lambda_u(H) < 0$, indicating instability and stability respectively. Since $(\bm{e}_s, 0)$ and $(\tilde{\bm{e}}_s, \tilde{\bm{g}}_s)$ are stable directions in phase space, the tangent space $T_{(H,0)}\mathcal{M}^s_{(H,0)}$ of the stable manifold $\mathcal{M}^s_{(H,0)}$ at $(H, 0)$ will be their linear span (over $\mathbb{R}$); similarly $T_{(H,0)}\mathcal{M}^u_{(H,0)}$ will be the linear span of $(\bm{e}_u, 0)$ and $(\tilde{\bm{e}}_u, \tilde{\bm{g}}_u)$.

If $W$ is determined in the vicinity of $H$ by the zero-energy classical trajectories emanating from $S$ and extending along $\mathcal{M}^u_{(S,0)}$, $W_{,ij}(H) = \partial_j p_i(H)$ may be interpreted as specifying the tangent space $T_{(H,0)}\mathcal{M}^u_{(S,0)}$ of $\mathcal{M}^u_{(S,0)}$ at $(H, 0)$. The four possible choices for the Hessian $W_{,ij}(H)$ found by solving the algebraic Riccati equation correspond to four possible choices for this 2-dimensional subspace of $\mathbb{R}^4$. Each such corresponding subspace is spanned by the vectors $(1, 0, W_{,11}, W_{,21})$ and $(0, 1, W_{,12}, W_{,22})$. In other words $T_{(H,0)}\mathcal{M}^u_{(S,0)}$ is the column space (over $\mathbb{R}$) of the 4-by-2 matrix

$$\bm{R}(H) = \begin{pmatrix} \bm{I} \\ \bm{Z}(H) \end{pmatrix}. \tag{55}$$

This is precisely the restriction to $\mathbb{R}^4$ of the subspace $\mathcal{N}$ used in the construction of $\bm{Z}(H)$. We found four (and only four) $\bm{T}(H)$-invariant $\bm{J}$-neutral 2-dimensional subspaces $\mathcal{N} \subset \mathbb{C}^4$ which yield real symmetric matrices $\bm{Z}(H)$, and we now see that their restrictions to $\mathbb{R}^4$ may be interpreted as the four possible choices for the tangent space $T_{(H,0)}\mathcal{M}^u_{(S,0)}$.

The interpretation of the four subspaces $\mathcal{N}$ as spectral subspaces of $\bm{T}(H)$ allows us to interpret the four possibilities for $T_{(H,0)}\mathcal{M}^u_{(S,0)}$ in terms of classical dynamics. There are $k = 2$ positive eigenvalues of $\bm{T}(H)$, $\lambda_u(H)$ and $-\lambda_s(H)$. By the above construction, the four subspaces $\mathcal{N}$ will be the four spectral subspaces corresponding to the two eigenvalues $\pm \lambda_u(H)$ and $\pm \lambda_s(H)$. The corresponding possibilities for $T_{(H,0)}\mathcal{M}^u_{(S,0)}$ are immediately expressible in terms of the eigenvectors of $\bm{T}(H)$. In particular the choice $(+, -)$ for the $\pm$ signs multiplying the two eigenvalues $\lambda_u(H)$ and $\lambda_s(H)$ yields $T_{(H,0)}\mathcal{M}^u_{(H,0)}$, and $(-, +)$ yields $T_{(H,0)}\mathcal{M}^s_{(H,0)}$. The two remaining possibilities are $(+, +)$, which yields the linear span of $(\bm{e}_u, 0)$ and $(\bm{e}_s, 0)$, and $(-, -)$, which yields the linear span of $(\tilde{\bm{e}}_u, \tilde{\bm{g}}_u)$ and $(\tilde{\bm{e}}_s, \tilde{\bm{g}}_s)$. It is readily verified that the $(-, -)$ subspace corresponds to a matrix $\bm{Z}(H)$ of full rank; the other three subspaces correspond to possible choices for $\bm{Z}(H)$ which are of less than full rank. The $(+, +)$ subspace corresponds to the solution $\bm{Z}(H) = \bm{0}$.

For any stochastic model of the form that we are considering, the $t \to \infty$ limit of $W_{,ij}$ along the MPEP $\bm{q}^*$ should exist, and equal one of the four possibilities for $W_{,ij}(H)$; equivalently, the tangent space $T_{(\boldsymbol{x},\boldsymbol{p})}\mathcal{M}^u_{(S,0)}$ should converge to one of the corresponding possibilities for $T_{(H,0)}\mathcal{M}^u_{(S,0)}$ as $(\boldsymbol{x},\boldsymbol{p}) \to (H,0)$ along the MPEP. The $(+,-)$ possibility, *i.e.*, $T_{(H,0)}\mathcal{M}^u_{(H,0)}$, can be ruled out immediately. Since $\boldsymbol{q}^*$ lies in $\mathcal{M}^u_{(S,0)}$ and by assumption $\boldsymbol{q}^*(t) \to H$ as $t \to \infty$, $T_{(H,0)}\mathcal{M}^u_{(S,0)}$ must include a stable direction. But $\lambda_u(H)$ and $-\lambda_s(H)$ are both positive, so $T_{(H,0)}\mathcal{M}^u_{(H,0)}$ cannot serve. The $(-,+)$ possibility, *i.e.*, $T_{(H,0)}\mathcal{M}^s_{(H,0)}$, can also be ruled out. Since $-\lambda_u(H)$ and $\lambda_s(H)$ are both negative, if $T_{(H,0)}\mathcal{M}^u_{(S,0)}$ equalled $T_{(H,0)}\mathcal{M}^s_{(H,0)}$ there would be an infinite number of zero-energy classical trajectories extending from $(S,0)$ to $(H,0)$, all with the same action. Such a situation would be highly nongeneric.

There remain the $(+,+)$ and $(-,-)$ possibilities for $T_{(H,0)}\mathcal{M}^u_{(S,0)}$. $\mathcal{N}_{(+,+)}$ and $\mathcal{N}_{(-,-)}$, as we shall call them, are 2-dimensional $\boldsymbol{T}(H)$-invariant subspaces of $\mathbb{R}^4$, spectral subspaces corresponding respectively to $\lambda_u(H)$, $\lambda_s(H)$, and to $-\lambda_u(H)$, $-\lambda_s(H)$. Since the stable eigenvalues of $\boldsymbol{T}(H)$ restricted to $\mathcal{N}_{(+,+)}$ and $\mathcal{N}_{(-,-)}$ are $\lambda_s(H)$ and $-\lambda_u(H)$ respectively, the MPEP $\boldsymbol{q}^*$ will approach $H$ exponentially, as $\exp(-|\lambda_s(H)|t)$ if $T_{(H,0)}\mathcal{M}^u_{(S,0)} = \mathcal{N}_{(+,+)}$ and as $\exp(-\lambda_u(H)t)$ if $T_{(H,0)}\mathcal{M}^u_{(S,0)} = \mathcal{N}_{(-,-)}$. In principle either $\mathcal{N}_{(+,+)}$ or $\mathcal{N}_{(-,-)}$ could serve as $T_{(H,0)}\mathcal{M}^u_{(S,0)}$, but the relative magnitude of $|\lambda_s(H)|$ and $\lambda_u(H)$ (*i.e.*, the parameter $\mu$, or the sign of $b^i{}_{,i}(H)$) turns out to determine which is generic.

To see why this is the case, note that the MPEP, viewed as a trajectory in $\mathcal{M}^u_{(S,0)} \subset \Omega \times \mathbb{R}^2$, will approach $(H,0)$ in a manner specified by the linearized Hamilton's equations (54), and that the negative eigenvalues of the linearized drift matrix $\boldsymbol{T}(H)$ in (54) are $\lambda_s(H)$ and $-\lambda_u(H)$. Generically, $\boldsymbol{q}^*$ will approach $(H,0)$ as $t \to \infty$ along the *less* contractive direction in phase space; the stochastic model would have to be carefully 'tuned' to arrange for the incoming MPEP to approach along the *more* contractive direction in (54). If $\mu < 1$, $-\lambda_u(H) < \lambda_s(H) < 0$ and $\lambda_s(H)$ is less contractive; in this case the MPEP will generically approach $H$ as $\exp(-|\lambda_s(H)|t)$, so if $\mu < 1$ then $T_{(H,0)}\mathcal{M}^u_{(S,0)} = \mathcal{N}_{(+,+)}$ will be generic. Similarly if $\mu > 1$, $\lambda_s(H) < -\lambda_u(H) < 0$ and the MPEP will generically approach $H$ as $\exp(-\lambda_u(H)t)$, so if $\mu > 1$ then $T_{(H,0)}\mathcal{M}^u_{(S,0)} = \mathcal{N}_{(-,-)}$ will be generic. Consequently, in phase space the MPEP will generically approach the fixed point $(H,0)$ along the tangent vector $(\boldsymbol{e}_s,0)$ if $b^i{}_{,i}(H) > 0$, and along $(\tilde{\boldsymbol{e}}_s,\tilde{\boldsymbol{g}}_s)$ if $b^i{}_{,i}(H) < 0$.

We now consider the extent to which the behavior of $W$ near $H$ can be quadratic. A great deal of light is thrown on this question by a sketch of the pattern of zero-energy classical trajectories emanating from $S$, when prolonged to the vicinity of $H$. We may regard these trajectories, which include the MPEP $\boldsymbol{q}^*$, as lying in the manifold $\mathcal{M}^u_{(S,0)}$. But in a sufficiently small neighborhood of $H$, the flow of these trajectories on $\mathcal{M}^u_{(S,0)}$ should be given to high accuracy by the the corresponding flow on the tangent space $T_{(H,0)}\mathcal{M}^u_{(S,0)}$. If $\mu < 1$, this tangent space is generically $\mathcal{N}_{(+,+)}$, the linear span of the stable vector $(\boldsymbol{e}_s, 0)$ and the unstable vector $(\boldsymbol{e}_u, 0)$; if $\mu > 1$, it is generically $\mathcal{N}_{(-,-)}$, the linear span of the stable vector $(\tilde{\boldsymbol{e}}_s, \tilde{\boldsymbol{g}}_s)$ and the unstable vector $(\tilde{\boldsymbol{e}}_u, \tilde{\boldsymbol{g}}_s)$. It follows that the zero-energy classical trajectories emanating from $S$, if projected 'down' from $\mathcal{M}^u_{(S,0)}$ to $\Omega$ by the map $(\boldsymbol{x},\boldsymbol{p}) \mapsto \boldsymbol{x}$, will generically trace out in the immediate vicinity of $H$ the hyperbolic patterns shown in Figures 2(a) and 2(b). The principal axes of the hyperbolæ are $\boldsymbol{e}_s$ and $\boldsymbol{e}_u$ if $\mu < 1$, and $\tilde{\boldsymbol{e}}_s$ and $\tilde{\boldsymbol{e}}_u$ if $\mu > 1$. By convention, we assume that the $\boldsymbol{e}_s$ ray is vertical, and that the MPEP approaches $H$ from the first quadrant.

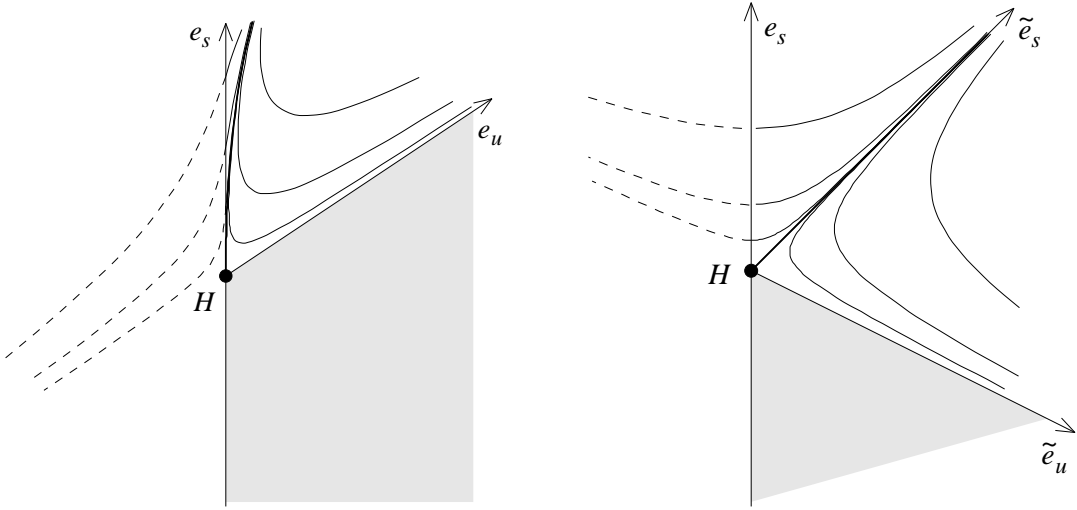

Figure 2: The generic appearance of the flow near $H$ of the zero-energy classical trajectories emanating from $S$; equivalently, the Hamiltonian flow on the manifold $\mathcal{M}^u_{(S,0)}$ projected 'down' to configuration space by the map $(\boldsymbol{x},\boldsymbol{p}) \mapsto \boldsymbol{x}$. Part (a) of the figure illustrates the case $\mu < 1$, *i.e.*, $\partial_i b^i(H) > 0$; part (b), the case $\mu > 1$, *i.e.*, $\partial_i b^i(H) < 0$. In both cases the MPEP [most probable exit path] is the solid curve incident on $H$. The wedge-shaped shaded regions are classically forbidden; also, the dashed trajectories (which extend beyond the $\boldsymbol{e}_s$ ray, which lies in $\partial\Omega$) are unphysical.

Figure 2(a) reveals a phenomenon not previously suspected: since the stable eigenvector $\boldsymbol{e}_s$ of $\boldsymbol{B}(H)$ lies in $\partial\Omega$, the MPEP $\boldsymbol{q}^*$ will generically be tangent to $\partial\Omega$ for stochastic models with $\mu < 1$. This prediction of a 'grazing MPEP' has been confirmed by numerical studies. Also, since the $(+,+)$ solution for $\boldsymbol{Z}(H)$ is the 2-by-2 zero matrix, when $\mu < 1$ one expects that that $W_{,ij}(\boldsymbol{x}) \to 0$ as $\boldsymbol{x} \to H$ along the MPEP. This too has been confirmed by numerical studies. The consequent 'flat' behavior of $W$ near the approaching MPEP has an appealing physical explanation: near $H$ the most probable fluctuational paths (perturbations of the grazing MPEP) come near to following the drift field $\boldsymbol{b}$, so very little additional action (cost) is built up as $H$ is approached.

We have not commented yet on the most visible feature of Figures 2(a) and 2(b), which is highly disconcerting: for both $\mu < 1$ and $\mu > 1$ there is a wedge-shaped region near $H$, beyond the boundary ray $e_u$ and the boundary ray $\tilde{e}_u$ respectively, which is not reached by any of the fluctuational paths. This wedge, the generic presence of which has been confirmed numerically, is 'classically forbidden' in that it cannot be reached by any zero-energy classical trajectory emanating from $S$. Unless the boundary ray lies in $\partial\Omega$, the wedge has nonempty interior. The generic presence of an unreachable region has a very simple interpretation: *Generically, the projection map $(\bm{x}, \bm{p}) \mapsto \bm{x}$ from $\mathcal{M}^u_{(S,0)}$ to $\bar{\Omega}$ is not onto; its range does not include all of $\Omega$.* The possibility that the projection map might fail to be onto on account of folds in the manifold $\mathcal{M}^u_{(S,0)}$ was mentioned in Section 4, but that this failure is generic is a new result.

We expect that $W(\bm{x})$, for any $\bm{x}$ in the wedge, is the action of an action-minimizing trajectory which passes through $\partial\Omega$ on its way from $S$ to $\bm{x}$ [19]. Only the portions of the trajectory that lie in $\Omega$ (rather than $\partial\Omega$) will be classical, *i.e.*, will be solutions of the Euler-Lagrange equation. The trajectory will be *piecewise classical*: it will have at least one 'corner,' or bend. The fact that generically, such fluctuational paths need to be considered for at least some endpoints $\bm{x}$ in any neighborhood of $H$ has not previously been realized.

By a careful optimization over piecewise classical trajectories it is possible to work out, for $\bm{x}$ in the wedge, the leading dependence of $W(\bm{x})$ on $\delta\bm{x} = \bm{x} - H$ in the vicinity of $H$. But we may economize on effort by applying our previous results. Suppose for the moment that to leading order $W(\bm{x})$ in the wedge behaves quadratically as $\bm{x} \to H$. Then the quadratic dependence will be specified by a limiting Hessian matrix $\hat{\bm{Z}}(H) = (\hat{W}_{,ij}(H))$, which must satisfy the algebraic Riccati equation (44). There are exactly four possibilities for $\hat{\bm{Z}}(H)$; equivalently, four possibilities for the corresponding graph subspace $\hat{\mathcal{N}}(H) \subset \mathbb{R}^4$. But $T_{(H,0)}\mathcal{M}^u_{(H,0)}$, *i.e.*, $\mathrm{sp}\{(\bm{e}_u, 0), (\tilde{\bm{e}}_u, \tilde{\bm{g}}_u)\}$, is the only one of the four possible subspaces on which the linearized Hamiltonian flow (45) has *positive* eigenvalues: $\lambda_u(H)$ and $-\lambda_s(H)$. The other three have at least one stable direction, and would correspond to a pattern of action-minimizing trajectories in the wedge which converge on $H$, rather than spreading outward from $H$. This is impossible, so we must have $\hat{\mathcal{N}}(H) = T_{(H,0)}\mathcal{M}^u_{(H,0)}$. If $W$ in the wedge behaves quadratically near $H$ then the limiting Hessian matrix $\hat{\bm{Z}}(H)$ will necessarily be the rank-1 2-by-2 matrix corresponding to $T_{(H,0)}\mathcal{M}^u_{(H,0)}$, and will differ from $\bm{Z}(H)$.

The extent to which the action-minimizing trajectories in $\Omega \times \mathbb{R}^2$ giving rise to this $\hat{\bm{Z}}(H)$ are only piecewise classical follows from the fact that they must emanate from $(H, 0)$, and thereafter (near $H$) lie in the subspace of $\mathbb{R}^4$ spanned by $(\bm{e}_u, 0)$ and $(\tilde{\bm{e}}_u, \tilde{\bm{g}}_u)$, the two expanding eigenvectors

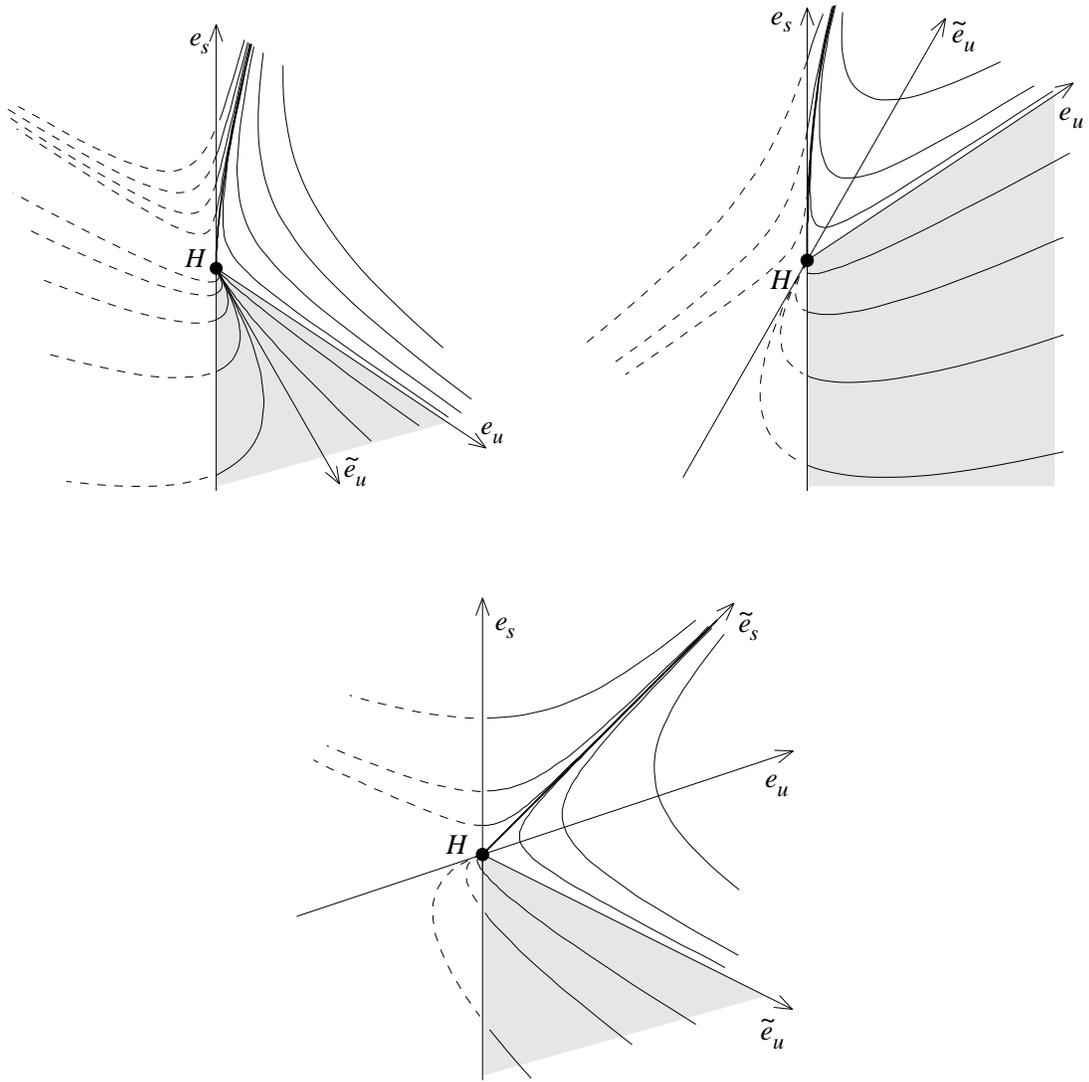

Figure 3: The generic appearance of the flow near $H$ of the most probable fluctuational paths (*i.e.*, the action-minimizing trajectories) emanating from $S$. The $e_s$ ray lies in $\partial\Omega$. Parts (a) and (b) of the figure illustrate Subcases A and B of the case $\mu < 1$; they differ in the relative placement of the $\tilde{e}_u$ and $e_u$ rays. Part (c) illustrates the case $\mu > 1$. In all three parts the MPEP [most probable exit path] is the solid curve incident on $H$. The other trajectories include zero-energy classical trajectories of the sort shown in Figure 2, and 'bent' trajectories which are prolongations (piecewise classical rather than classical) of the MPEP. The classically forbidden wedge-shaped shaded regions are reached (as $\epsilon \to 0$) only *via* bent trajectories. In part (a) these trajectories have only one bend, at $H$. But in parts (b) and (c) the dashed portions of the trajectories emanating from $H$ are unphysical, and the trajectories entering the shaded regions have two bends: they extend along $\partial\Omega$ before reentering $\Omega$.

of $T(H)$. Near $H$ each such trajectory, when parametrized by $t > -\infty$, should be of the form

$$t \mapsto (H, 0) + C_u \exp\left(\lambda_u(H)t\right)(e_u, 0) + \tilde{C}_u \exp\left(|\lambda_s(H)|t\right)(\tilde{e}_u, \tilde{g}_u), \tag{56}$$

for $C_u$, $\tilde{C}_u$ two constants that determine the trajectory. In general these action-minimizing trajectories will be tangent (as $t \to -\infty$, or as the trajectories emerge from $H$) to whichever is the *less* expansive of the two expanding eigenvectors of $T(H)$. If $\mu < 1$ this is $(\tilde{e}_u, \tilde{g}_u)$; if $\mu > 1$ it is $(e_u, 0)$. This tangency condition determines the behavior near $H$ of the velocity field $\dot{x}(\cdot)$ of the action-minimizing trajectories in the wedge. If $\mu < 1$ the portions of the trajectories emanating from $H$ should be tangent to $\tilde{e}_u$. If $\mu > 1$ they should be tangent to $e_u$.

Figure 3 is an extended version of Figure 2, which includes the rays $\tilde{e}_u$ and $e_u$, and the additional trajectories (emanating from $H$, and tangent to them) predicted by the assumption of quadratic behavior of $W$ in the wedge. The $\mu < 1$ case has two subcases, shown in Figures 3(a) and 3(b). We shall refer to them as Subcase A and Subcase B; they differ (as regards the projected eigenvectors of $T(H)$) in only one essential way: in the *relative* placement of the rays $\tilde{e}_u$ and $e_u$. We shall show in the next section that when $\mu > 1$, the $e_u$ ray necessarily lies between the $\tilde{e}_s$ and $\tilde{e}_u$ rays. This situation is shown in Figure 3(c); when $\mu > 1$, there are no subcases.

Subcase A of the case $\mu < 1$ is the most straightforward. It is clear from a glance at Figure 3(a) that in this subcase, the action-minimizing trajectory from $S$ to any point in the wedge is a prolongation of the MPEP: it extends from $S$ to $H$, experiences a discontinuous change in direction, and then enters the wedge. Interestingly, points on $\partial\Omega$ that are on the 'wedge side' of $H$ are reached only *viâ* trajectories that make a second passage through $\Omega$.

We stress that these 'bent' trajectories have a probabilistic interpretation. By Wentzell-Freidlin theory, any point $x$ in the interior of the wedge is (as $\epsilon \to 0$) preferentially reached during large fluctuations away from $S$ by a bent trajectory. That is, when $\mu < 1$ the fluctuation will in Subcase A preferentially drive the random process $x_\epsilon(t)$ to the vicinity of $H$ via the MPEP $q^*$, before the wedge is traversed and the vicinity of $x$ is finally reached. Although Figure 3(a) displays only the portion of the wedge in the vicinity of $H$ (*i.e.*, the wedge as computed in the linear approximation), studies of particular models show that it may extend a considerable distance from $H$. The numerical computation of $W(x)$, for $x$ in the wedge, requires an integration along the appropriate bent trajectory terminating at $x$.

There is a problem extending these conclusions to Subcase B, and to the case $\mu > 1$. It is clear

| Case [Subcase] | $Z(H)$ | $\hat{Z}(H)$ | $\mathcal{N}(H)$ | $\hat{\mathcal{N}}(H)$ |
|---|---|---|---|---|
| $\mu < 1$, *i.e.*, $\partial_i b^i(H) > 0$ [A] | 0 | rank-1 | $\mathcal{N}_{(+,+)} = \mathrm{sp}\{(e_s, 0), (e_u, 0)\}$ | $T_{(H,0)}\mathcal{M}^u_{(H,0)}$ |
| $\mu < 1$, *i.e.*, $\partial_i b^i(H) > 0$ [B] | 0 | — | $\mathcal{N}_{(+,+)} = \mathrm{sp}\{(e_s, 0), (e_u, 0)\}$ | — |
| $\mu > 1$, *i.e.*, $\partial_i b^i(H) < 0$ | rank-2 | — | $\mathcal{N}_{(-,-)} = \mathrm{sp}\{(\tilde{e}_s, \tilde{g}_s), (\tilde{e}_u, \tilde{g}_u)\}$ | — |

Table 1: The generic limiting behavior (as $x \to H$) of the Hessian matrix $Z(x) = (W_{,ij}(x))$, and of the corresponding graph subspace $\mathcal{N}(x) \subset \mathbb{R}^4$. $Z(x) \to Z(H)$ if $x \to H$ from outside the wedge, *i.e.*, along the manifold $\mathcal{M}^u_{(S,0)}$. Similarly $\mathcal{N}(x) \to \mathcal{N}(H)$. In Subcase A of the case $\mu < 1$, $Z(x) \to \hat{Z}(H)$ as $x \to H$ from within the wedge; similarly $\mathcal{N}(x) \to \hat{\mathcal{N}}(H)$. Each of $\mathcal{N}(H)$ and $\hat{\mathcal{N}}(H)$, if defined, is the linear span of a pair of eigenvectors of the linearized Hamiltonian flow $T(H)$, since the tangent space $T_{(H,0)}\mathcal{M}^u_{(H,0)}$ equals $\mathrm{sp}\{(e_u, 0), (\tilde{e}_u, \tilde{g}_u)\}$. In Subcase B and when $\mu > 1$ the behavior of the action in the wedge is not quadratic near $H$.

in Figures 3(b) and 3(c) that classical trajectories which are of the form

$$t \mapsto H + C_u \exp\left(\lambda_u(H)t\right) e_u + \tilde{C}_u \exp\left(|\lambda_s(H)|t\right) \tilde{e}_u, \tag{57}$$

and which are tangent to the $\tilde{e}_u$ ray (resp. the $e_u$ ray), are necessarily *unphysical*. They are the dashed trajectories, which penetrate into the complement of $\bar{\Omega}$ before returning to $\partial\Omega$ and passing through the wedge. This situation is quite different from Subcase A, where the trajectories emanating from $H$ penetrate immediately into the wedge. We expect therefore that in Subcase B and when $\mu > 1$ the true action-minimizing trajectory from $H$ to any point $x$ in the wedge will extend along $\partial\Omega$ before reentering the wedge. Moreover, since the putative Hessian matrix $\hat{Z}(H)$ corresponds to an unphysical (impossible) velocity field $\dot{x}(\cdot)$ for the most probable fluctuational paths, in Subcase B and when $\mu < 1$ we do not expect the action in the wedge to behave quadratically near $H$. The flow field of the action-minimizing trajectories within the wedge may differ from the predictions of the quadratic approximation.

We summarize our conclusions in Table 1. The behavior of $W$ near $H$, both inside and outside the wedge and for both $\mu < 1$ and $\mu > 1$, is quadratic and fully understood with the exception of the behavior inside the wedge in Subcase B, and when $\mu > 1$. Even these difficult cases lend themselves to a partial analysis. The precise behavior of $W$ inside the wedge in Subcase B and when $\mu > 1$, even in the linear approximation near $H$, can only be obtained by solving a difficult variational problem. But it is clear on physical grounds that $W(x)$ approaches $W(H)$ quadratically as $x$ approaches $H$ along $\partial\Omega$ from the wedge side of $H$. This is because on the wedge side of $H$, the cost function $W(\boldsymbol{x})$ on $\partial\Omega$ satisfies (in Subcase B, and when $\mu > 1$)

$$W(\boldsymbol{x}) - W(H) = \int_H^{\boldsymbol{x}} p_i \, dx^i, \tag{58}$$

the line integral being taken along $\partial\Omega$ from $H$ to $\boldsymbol{x}$. Here the momentum $\boldsymbol{p} = \boldsymbol{p}(\boldsymbol{x})$, at any point $\boldsymbol{x} \in \partial\Omega$, takes a value determined by (i) the zero-energy constraint, and (ii) the constraint that the corresponding velocity $\dot{\boldsymbol{x}}$ lie in $\partial\Omega$.

In Subcase A, in general $\boldsymbol{Z}(H) \ne \hat{\boldsymbol{Z}}(H)$; in Subcase B and when $\mu > 1$ there is no reason why the quadratic growth of $W$ along $\partial\Omega$, on the wedge side of $H$, should equal the quadratic growth on the other side (arising from $\boldsymbol{Z}(H)$, rather than from (58)). We conclude that *generically, $W$ fails to be twice continuously differentiable at $H$.* As $\boldsymbol{x} \to H$ from within $\Omega$, in Subcase A the Hessian matrix $W_{,ij}(H)$ will converge to $\boldsymbol{Z}(H)$ or $\hat{\boldsymbol{Z}}(H)$ depending on whether $\boldsymbol{x} \to H$ from outside or inside the wedge. The corresponding graph subspace $\mathcal{N}(\boldsymbol{x})$ will tend either to $\mathcal{N}(H)$, *i.e.*, to $T_{(H,0)} \mathcal{M}^u_{(S,0)}$, or to $\hat{\mathcal{N}}(H)$. The situation for Subcase B and $\mu > 1$ is similar, except that we lack a precise description of the behavior of $W$ inside the wedge. Notice that if $s$ denotes arc length along $\partial\Omega$ measured from $H$, with $s < 0$ on the wedge side, the value $\partial^2 W / \partial s^2 (s = 0+)$ will equal $e_s^t \boldsymbol{Z}(H) e_s$, and the value $\partial^2 W / \partial s^2 (s = 0-)$ will equal $e_s^t \hat{\boldsymbol{Z}}(H) e_s$ (in Subcase A), or a quantity derived from (58) (in Subcase B, or if $\mu > 1$). Generically these two one-sided second derivatives will be different; in fact if $\mu < 1$ the former, as we have noted in Table 1, will equal zero. This justifies the statements about $\partial^2 W / \partial s^2 (s = 0\pm)$ made in the Introduction.

It is worth noting that $W(\boldsymbol{x})$ will generically fail to be a twice continuously differentiable function of $\boldsymbol{x}$ not merely at $x = H$, but along the entire boundary of the wedge. This is strongly reminiscent of the Stokes phenomenon, which occurs in the asymptotic expansion of the solution (in the complex plane) of an ordinary differential equation with a small parameter multiplying the term with highest derivative. The exponent of the dominant exponential factor in the expansion is non-smooth along certain curves ('anti-Stokes lines') emanating from turning points. The implications of the Stokes phenomenon for the asymptotic expansion of Airy functions, etc., are well known. It is probably best to view the phenomenon of the wedge as a two-dimensional counterpart to the conventional Stokes phenomenon. For the forward Kolmogorov equation on $\Omega \subset \mathbb{R}^2$, which is a *partial* differential equation, the anti-Stokes line appears as a curve in $\Omega$ (the boundary of the wedge) rather than in the complex plane. (Cf. Berry and Howls [5].) In order to see it, no analytic continuation is required.

In the following three sections we approximate the principal eigenfunction $v_\epsilon^0$ and the pro-cess $\boldsymbol{x}_\epsilon(t)$ in the boundary region near $H$. Our results on the flow of classical (and piecewise classical) trajectories near $H$ will to a large extent dictate the choice of asymptotic approximations.

# 6 Matched Asymptotic Expansions, Continued

It is clear, in broad outline, how to apply the method of matched asymptotic expansions sketched in Section 3. In Section 4 we derived an outer approximation to the principal eigenfunction $v_\epsilon^0$ which is valid in the body of $\Omega$; it will necessarily match to the Gaussian approximation (53) in the stable region of size $O(\epsilon^{1/2})$ surrounding $S$. We lack only an inner approximation to $v_\epsilon^0$ valid in the boundary region near $H$. If it is to match to the outer approximation, the inner approximation must in the far field have leading asymptotics

$$v_\epsilon^0(\boldsymbol{x}) \sim \exp\left[-(x^i - H^i)W_{,ij}(H)(x^j - H^j)/2\epsilon\right]. \tag{59}$$

But as we have seen from the last section, this expression must be interpreted with care. By convention $W_{,ij}(H)$ signifies the limiting Hessian matrix obtained by taking $\boldsymbol{x} \to H$ along the most probable exit path (MPEP), or in general from within the complement of the classically forbidden 'wedge.' The asymptotic statement (59) is interpreted as holding when $(\boldsymbol{x} - H)/\epsilon^{1/2} \to \infty$ within the complement of the wedge. If the behavior of $W$ in the wedge is quadratic near $H$, the limiting Hessian matrix obtained by taking $\boldsymbol{x} \to H$ from within the wedge is denoted $\hat{W}_{,ij}(H)$. We have seen that in Subcase A of the case $\partial_i b^i(H) > 0$, the behavior of $W$ in the wedge is indeed quadratic near $H$, so the counterpart to (59) (involving $\hat{W}_{,ij}(H)$) should hold as $(\boldsymbol{x} - H)/\epsilon^{1/2} \to \infty$ within the wedge. In other words the far-field asymptotics of the inner approximation to $v_\epsilon^0$ must in this case be *piecewise bivariate Gaussian*: different decay (or growth) rates will be found in the wedge and its complement. In Subcase B or when $\partial_i b^i(H) < 0$ (*i.e.*, $\mu > 1$, in the notation of the last section), the behavior of $W$ in the wedge will not necessarily be quadratic near $H$. But the inner approximation to $v_\epsilon^0$ must still have Gaussian asymptotics as $(\boldsymbol{x} - H)/\epsilon^{1/2} \to \infty$ along $\partial\Omega$, with different decay rates on the two sides of $H$.

The inner approximation to $v_\epsilon^0$ can be found, at least in principle, by solving a linearized version of the (approximate) forward Kolmogorov equation $\mathcal{L}_\epsilon^* \rho = 0$ in the boundary region near $H$. In the linear approximation we take $b^i(\boldsymbol{x}) \approx B^i{}_j(H)(x^j - H^j)$ and $D^{ij}(\boldsymbol{x}) \approx D^{ij}(H)$, so the Kolmogorov

equation reduces to

$$(\epsilon/2)D^{ij}(H)\partial_i\partial_j\rho - \partial_i[B^i{}_j(H)(x^j - H^j)\rho] = 0. \tag{60}$$

Assume for the moment that the appropriate lengthscale on which the inner approximation should be defined is indeed the $O(\epsilon^{1/2})$ lengthscale. If so, we can employ the 'stretched' variable $X^i = (x^i - H^i)/\epsilon^{1/2}$, in terms of which (60) becomes

$$\frac{1}{2}D^{ij}(H)\frac{\partial^2\rho}{\partial X^i \partial X^j} - B^i{}_j(H)\frac{\partial}{\partial X^i}[X^j\rho] = 0. \tag{61}$$

We can change variables to reduce this covariant equation to a noncovariant, but more understandable form. Under a linear change of variables $(X')^i = L^i{}_j X^j$, i.e., $\boldsymbol{X}' = \boldsymbol{L}\boldsymbol{X}$, the matrices $\boldsymbol{B}(H)$ and $\boldsymbol{D}(H)$ transform to $\boldsymbol{L}\boldsymbol{B}(H)\boldsymbol{L}^{-1}$ and $\boldsymbol{L}\boldsymbol{D}(H)\boldsymbol{L}$ respectively. Choosing $\boldsymbol{L} = \boldsymbol{D}(H)^{-1/2}$ transforms $\boldsymbol{D}(H)$ to the identity matrix. But since $H$ is a saddle point, irrespective of coordinate transformations the linearized drift $\boldsymbol{B}(H)$ will have one positive eigenvalue ($\lambda_s(H)$) and one negative eigenvalue ($\lambda_u(H)$). By a further change of variables (a rotation) we can arrange matters so that $B^1{}_2(H) = 0$, and

$$\boldsymbol{B}(H) = \begin{pmatrix} \lambda_u(H) & 0 \\ c & \lambda_s(H) \end{pmatrix}, \tag{62}$$

for some real constant $c$. The constant $c$ is not determined by $\lambda_u(H)$ and $\lambda_s(H)$. Since $\boldsymbol{D}(H) = \boldsymbol{I}$ is preserved under rotations, with respect to the new system of coordinates equation (61) becomes

$$\frac{1}{2}\frac{\partial^2\rho}{\partial(X^1)^2} + \frac{1}{2}\frac{\partial^2\rho}{\partial(X^2)^2} - \lambda_u(H)\frac{\partial}{\partial X^1}[X^1\rho] - \lambda_s(H)\frac{\partial}{\partial X^2}[X^2\rho] - c\frac{\partial}{\partial X^2}[X^1\rho] = 0. \tag{63}$$

In terms of the transformed coordinates $(X^1, X^2)$ we may view the region $\Omega$ as the right-half plane $X^1 > 0$, and its boundary $\partial\Omega$ as the $X^2$-axis. This was the convention of Figures 2 and 3.

This system of coordinates is computationally easy to work with. Suppose for simplicity that $\lambda_u(H) = 1$; this is an innocuous normalization condition that can be absorbed into a redefinition of time $t$ (and noise strength $\epsilon$). Then

$$\boldsymbol{B}(H) = \begin{pmatrix} 1 & 0 \\ c & -\mu \end{pmatrix}, \tag{64}$$

where $\mu = |\lambda_s(H)|/\lambda_u(H)$ as in the last section. The forward Kolmogorov equation (63) reduces to

$$\frac{1}{2}\frac{\partial^2 \rho}{\partial (X^1)^2} + \frac{1}{2}\frac{\partial^2 \rho}{\partial (X^2)^2} - \frac{\partial}{\partial X^1}[X^1 \rho] + \mu \frac{\partial}{\partial X^2}[X^2 \rho] - c\frac{\partial}{\partial X^2}[X^1 \rho] = 0. \qquad (65)$$

Moreover a bit of matrix computation, using the form (64) for $B(H)$ and $D(H) = I$, yields

$$(e_s, 0) = (0, 1, 0, 0) \qquad (66)$$
$$(e_u, 0) = (\mu + 1, c, 0, 0) \qquad (67)$$
$$(\tilde{e}_s, \tilde{g}_s) = (\mu - 1, c, -2\mu + 2, 0) \qquad (68)$$
$$(\tilde{e}_u, \tilde{g}_u) = (-2\mu c, \mu^2 - 1 - c^2, -2\mu c(\mu - 1), 2\mu(\mu^2 - 1)) \qquad (69)$$

for the four eigenvectors of the linearized Hamiltonian flow $T(H)$ at the point $(H, 0)$ in phase space, as discussed in the last section. (Normalization is irrelevant here; the negatives of these vectors could equally well have been chosen.) The fact that $e_s = (1, 0)$ is in agreement with the convention of Figures 2 and 3.

The formulæ (66)–(69) explain the positioning of the rays $e_u$, $\tilde{e}_s$, and $\tilde{e}_u$ in Figures 3(a), 3(b), and 3(c). Recall that in those figures the MPEP is taken to approach from the first quadrant; if $\mu < 1$, it is generically tangent to $e_s$, and hence to the positive $X^2$-axis. By examination of (66)–(69), if $\mu < 1$ and $c < 0$ then $e_u$ will lie between the positive $X^2$-axis and $\tilde{e}_u$, while if $c > 0$ then $\tilde{e}_u$ (taken to point into the right half-plane) will lie between the positive $X^2$-axis and $e_u$. So the Subcases A and B of Figures 3(a) and 3(b) are simply the subcases $c < 0$ and $c > 0$. This correspondence assumes of course that the MPEP is tangent to the *positive* $X^2$-axis; if it approached $H$ from the fourth quadrant, and were tangent to the negative $X^2$-axis instead, then the interpretation in terms of sgn $c$ would be reversed.

Figure 3(c) is justified, as well. If $\mu > 1$ the approaching MPEP is known by the argument of the last section to be generically tangent to $\tilde{e}_s$. By (68), $\tilde{e}_s \propto (\mu - 1, c)$, and our convention that the MPEP approaches from the first quadrant mandates that $c \geq 0$. It is easy to verify, by examining (66)–(69), that in the right-half plane, when $\mu > 1$ and $c > 0$ the $e_u$ ray necessarily lies between the the $\tilde{e}_s$ ray and the the $\tilde{e}_u$ ray. This justifies the relative positioning of the rays in Figure 3(c).

Now that a canonical system of coordinates has been chosen, the rank-2 Hessian matrix

$\boldsymbol{Z}(H) = (W_{,ij}(H))$ may be computed; substituting (64) into (51) yields

$$\boldsymbol{Z}(H) = \frac{2(\mu - 1)}{(\mu - 1)^2 + c^2} \begin{pmatrix} c^2 + 1 - \mu & -c\mu \\ -c\mu & \mu^2 - \mu \end{pmatrix}. \tag{70}$$

Recall that generically this limiting Hessian matrix arises (when $\boldsymbol{x} \to H$ in the complement of the wedge) only when $\mu > 1$. If $\mu < 1$, $\boldsymbol{Z}(\boldsymbol{x}) \to \boldsymbol{0}$ as $\boldsymbol{x} \to H$ along the MPEP, and as noted in Table 1 we have $\boldsymbol{Z}(H) = \boldsymbol{0}$ instead. In Subcase A of $\mu < 1$, the limiting Hessian matrix $\hat{\boldsymbol{Z}}(H)$ in the wedge exists, and equals by Table 1 the rank-1 matrix corresponding to the graph subspace $\hat{\mathcal{N}}(H) = T_{(H,0)}\mathcal{M}^u_{(H,0)}$ in $\mathbb{R}^4$. This subspace is the linear span of $(\boldsymbol{e}_u, \boldsymbol{0})$ and $(\tilde{\boldsymbol{e}}_u, \tilde{\boldsymbol{g}}_u)$, and the explicit formula

$$\hat{\boldsymbol{Z}}(H) = \frac{2\mu}{c^2 + (\mu + 1)^2} \begin{pmatrix} c^2 & -c(\mu + 1) \\ -c(\mu + 1) & (\mu + 1)^2 \end{pmatrix}. \tag{71}$$

for the associated solution of the algebraic Riccati equation follows by some elementary manipulations. In Subcase B, and when $\mu > 1$, the behavior of $W(\boldsymbol{x})$ in the wedge as $\boldsymbol{x} \to H$ is not expected to be quadratic. However it follows from the formula (58) that its behavior as $\boldsymbol{x} \to H$ along $\partial\Omega$ (from the wedge side) is quadratic, with limiting second derivative

$$\hat{Z}_{\partial\Omega}(H) = 2\mu. \tag{72}$$

This limiting second derivative has a simple interpretation: on the wedge side of $H$, the cost (action) of the most probable trajectory leading to any point on $\partial\Omega$ arises from the drift $\boldsymbol{b}$ on $\partial\Omega$ itself.

The formulæ for the one-sided second derivatives $\boldsymbol{Z}(H)$, $\hat{\boldsymbol{Z}}(H)$, and $\hat{Z}_{\partial\Omega}(H)$ make quite precise the far-field ($\boldsymbol{X} \to \infty$) Gaussian (or inverted Gaussian) asymptotics which must be imposed on the solution $\rho(X^1, X^2)$ of the transformed Kolmogorov equation (65). First, we must have to leading order

$$\rho(X^1, X^2) \sim \exp\left(-\boldsymbol{X}^t \boldsymbol{Z}(H) \boldsymbol{X}/2\right), \qquad \boldsymbol{X} \to \infty \text{ outside the wedge.} \tag{73}$$

Also, in Subcase A of $\mu < 1$ we must have

$$\rho(X^1, X^2) \sim \exp\left(-\boldsymbol{X}^t \hat{\boldsymbol{Z}}(H) \boldsymbol{X}/2\right), \qquad \boldsymbol{X} \to \infty \text{ inside the wedge,} \tag{74}$$

and in Subcase B of $\mu < 1$, or when $\mu > 1$, we must have

$$\rho(X^1, X^2) \sim \exp\left(-\hat{Z}_{\partial\Omega}(H)(X^2)^2/2\right), \qquad \boldsymbol{X} = (0, X^2) \to \infty, \text{ on the wedge side,} \quad (75)$$

since the boundary $\partial\Omega$ is the $X^2$-axis. The location of the wedge here is as shown in Figure 2. If $\mu > 1$ the wedge is the sector bounded by the rays $\boldsymbol{e}_s$ and $\tilde{\boldsymbol{e}}_u$ which contains neither $\tilde{\boldsymbol{e}}_s$ nor $\boldsymbol{e}_u$; if $\mu < 1$ it is the sector bounded by the rays $\boldsymbol{e}_s$ and $\boldsymbol{e}_u$ which contains $\tilde{\boldsymbol{e}}_u$ but not $\tilde{\boldsymbol{e}}_s$. The Dirichlet (absorbing) boundary condition $\rho(0,\cdot) = 0$ must also be imposed. It expresses the fact that the quasistationary density is absorbed on the boundary.

In general it is not easy to solve the partial differential equation (65) on the half-plane $X^1 \geq 0$, subject to these boundary conditions. Our treatments of the generic $\mu < 1$ and generic $\mu > 1$ cases, in Sections 7 and 8 respectively, are designed to circumvent this problem. For the case $\mu < 1$ we shall expand on a larger lengthscale than $O(\epsilon^{1/2})$; for the case $\mu > 1$, on which we have less information, we shall use stochastic analysis. In advance of our detailed treatments, we observe that the *far-field* behavior of the exit location density $p_\epsilon$ (on the $O(\epsilon^{1/2})$ lengthscale near $H$) follows from (73), (74), and (75). By (18) the exit location density is simply proportional to the normal derivative of the inner approximation $v_\epsilon^0$. If we substitute the known expressions for $\boldsymbol{Z}(H)$, $\hat{\boldsymbol{Z}}(H)$, and $\hat{Z}_{\partial\Omega}(H)$, we obtain far-field asymptotics

$$p_\epsilon(s) \sim \begin{cases} \exp\left[0 \cdot (s^2/\epsilon)\right], & s/\epsilon^{1/2} \to +\infty, \\ \exp\left[-\dfrac{\mu(\mu+1)^2}{c^2 + (\mu+1)^2}(s^2/\epsilon)\right], & s/\epsilon^{1/2} \to -\infty \end{cases} \quad (76)$$

if $\mu < 1$ (Subcase A), and

$$p_\epsilon(s) \sim \begin{cases} \exp\left[0 \cdot (s^2/\epsilon)\right], & s/\epsilon^{1/2} \to +\infty, \\ \exp\left[-\mu(s^2/\epsilon)\right], & s/\epsilon^{1/2} \to -\infty \end{cases} \quad (77)$$

if $\mu < 1$ (Subcase B), and

$$p_\epsilon(s) \sim \begin{cases} \exp\left[-\dfrac{\mu(\mu-1)^2}{c^2 + (\mu-1)^2}(s^2/\epsilon)\right], & s/\epsilon^{1/2} \to +\infty, \\ \exp\left[-\mu(s^2/\epsilon)\right], & s/\epsilon^{1/2} \to -\infty \end{cases} \quad (78)$$

if $\mu > 1$. The zero coefficients in the exponents of (76) and (77) indicate that the corresponding asymptotics are sub-Gaussian. We have written $s$ for $\Delta x^2$ (the distance along $\partial\Omega$, measured from $H$), for consistency with the Introduction, and have taken the wedge side of $H$ to be the side on which $s < 0$. Notice that as $\mu \uparrow 1$ (resp. $\mu \downarrow 1$) the asymptotics of (77) and (78) come into agreement. Subcase A, however, has no $\mu > 1$ counterpart.

There is one degenerate case in which the partial differential equation (65), subject to the boundary conditions (73)–(75), can be solved explicitly. This is the case when $c = 0$, *i.e.*, when the linearized drift $B(H)$ at the saddle point has the property that its eigenvalues $e_s$ and $e_u$ are orthogonal with respect to the inner product specified by the local diffusivity tensor $D(H)$. If $c = 0$ and $\mu > 1$, the classically forbidden wedge *vanishes*: the rays $e_s$ and $\tilde{e}_u$, which form the boundary of the wedge, become identical. In picturesque language, as $c \to 0$ (when $\mu > 1$) the wedge disappears into the boundary $\partial\Omega$, and the action becomes twice continuously differentiable on a neighborhood of the saddle point. In this case it follows from (78), by setting $c = 0$, that the two Gaussian decay rates of the asymptotic exit location density are equal. This suggests that $p_\epsilon$ is asymptotic to a Gaussian. The confirmation of this, however, requires the solution of (65).

If there is no wedge, the imposed far-field asymptotics become Gaussian rather than piecewise Gaussian; they reduce to $\rho(X^1, X^2) \sim \exp\left(-X^t Z(H) X/2\right)$, since only (73) applies. But if $c = 0$ the expression (70) for the matrix $Z(H)$, which is valid when $\mu > 1$, simplifies to

$$Z(H) = \begin{pmatrix} -2 & 0 \\ 0 & 2\mu \end{pmatrix}. \tag{79}$$

So the far-field asymptotics which must be imposed on the solution of (65) simplify greatly; we have $\rho(X^1, X^2) \sim \exp[(X^1)^2 - \mu(X^2)^2]$. Moreover if $c = 0$ the final term in (65) vanishes, allowing a solution on $X^1 \geq 0$ with these asymptotics to be found by separation of variables. It is

$$\rho(X^1, X^2) = \exp\left[-\mu(X^2)^2\right] G(X^1) \tag{80}$$

in which

$$G(z) = e^{z^2} \operatorname{erf}(z) = \frac{2}{\sqrt{\pi}} e^{z^2} \int_{s=0}^{z} e^{-s^2}\, ds. \tag{81}$$

When $c = 0$ therefore (and $\mu > 1$) we have the inner approximation

$$v_\epsilon^0(x^1, x^2) \sim \{K(H) \exp\left(-W(H)/\epsilon\right)\} \exp\left[-\mu(x^2 - H^2)^2/\epsilon\right] G\left((x^1 - H^1)/\epsilon^{1/2}\right) \tag{82}$$

in which we have included the prefactor $K(H)\exp{(-W(H)/\epsilon)}$ to facilitate matching with the outer approximation.

Note that in (82) we must interpret $x^1 - H^1$, $x^2 - H^2$ as linearly transformed ('primed') versions of the original coordinates $x^1 - H^1$, $x^2 - H^2$. As remarked, $\boldsymbol{X} = (X^1, X^2)$ in (80) really means $\boldsymbol{X}' = \boldsymbol{O}\boldsymbol{D}(H)^{-1/2}\boldsymbol{X}$, for $\boldsymbol{O}$ a suitably chosen (orthogonal) rotation matrix; similarly, $\boldsymbol{x} - H$ in (82) must be interpreted as $\boldsymbol{O}\boldsymbol{D}(H)^{-1/2}(\boldsymbol{x} - H)$. Using (79), we can rewrite (82) in a partially covariant form as

$$v_\epsilon^0(x^1, x^2) \sim \{K(H)\exp{(-W(H)/\epsilon)}\} \exp\left[-(x^i - H^i)W_{,ij}(H)(x^j - H^j)/2\epsilon\right] \text{erf}\left[(x^1 - H^1)/\epsilon^{1/2}\right] \tag{83}$$

with the same proviso on the interpretation of the factor $x^1 - H^1$ in the argument of the error function. This way of writing the inner approximation to $v_\epsilon^0$ makes it clear that its leading asymptotics as $\boldsymbol{x}/\epsilon^{1/2} \to \infty$ are indeed those of (59). It also reveals that it is closely related to the Gaussian approximation (53), which is valid near $S$. The final factor in (83), which has no counterpart in (53), is attributable to the absorbing boundary condition at $X^1 = 0$. This boundary layer function may be written in fully covariant form as

$$\text{erf}\left(\lambda_u(H)^{1/2} n^i D_{ij}(H)(x^j - H^j)/\epsilon^{1/2}\right) \tag{84}$$

where $n^i$ is a contravariant unit normal vector to $\partial\Omega$ at $H$, satisfying $n^i D_{ij}(H)t^j = 0$ for any vector $t^i$ tangent to $\partial\Omega$ at $H$, and normalized so that $n^i D_{ij}(H)n^j = 1$. We have included in the argument of the error function a factor $\lambda_u(H)^{1/2}$, which will be present if $\lambda_u(H) \neq 1$.

We stress that we are able to find such a simple inner approximation as (83) in the case $\mu > 1$ *only* when $c = 0$, *i.e.*, only when the eigenvectors of $\boldsymbol{B}(H)$ are orthogonal with respect to the inner product specified by $D_{ij}(H)$. A covariant way of expressing this condition may be derived as follows. $e_s{}^i D_{ij} e_u{}^j = 0$, *i.e.*, $e_s^t \boldsymbol{D}(H)^{-1} \boldsymbol{e}_u = 0$, means that $\boldsymbol{D}(H)^{-1/2}\boldsymbol{e}_s$ and $\boldsymbol{D}(H)^{-1/2}\boldsymbol{e}_u$ are orthogonal in the conventional sense. Equivalently, $\boldsymbol{D}(H)^{-1/2}\boldsymbol{B}(H)\boldsymbol{D}(H)^{1/2}$ has orthogonal eigenvectors and is a symmetric matrix. But $\boldsymbol{D}(H)^{-1/2}\boldsymbol{B}(H)\boldsymbol{D}(H)^{1/2}$ if and only if $\boldsymbol{B}(H)\boldsymbol{D}(H)$ is symmetric, *i.e.*, if the tensor $b^i{}_{,j}D^{jk}$ is symmetric under the interchange of $i$ and $k$ at $x = H$. This is the covariant form of the condition. It is easily checked that if $b^i = -D^{ij}\partial_j \Phi$ for some scalar potential field $\Phi$ which has a saddle point at $H$, then the condition is necessarily satisfied. The $c = 0$ condition for the validity of the inner approximation (83), when $\mu > 1$, is really a condition that the drift $\boldsymbol{b}$ be *locally gradient* at $H$.

Stochastic exit models in which this local gradient condition holds are (sadly) nongeneric. But if $\mu > 1$, and the condition holds, it is an easy matter to apply the technique sketched in Section 3 to the inner approximation (83) to determine both the asymptotics of the exit location density on $\partial\Omega$, and the asymptotics of the principal eigenvalue $\lambda_\epsilon^{(0)}$ of $\mathcal{L}_\epsilon^*$. Equation (18) yields

$$p_\epsilon(\boldsymbol{x}) \propto \exp\left[-(x^i - H^i)W_{,ij}(H)(x^j - H^j)/2\epsilon\right], \qquad \boldsymbol{x} \in \partial\Omega \qquad (85)$$

for the density of the exit location distribution in the $\epsilon \to 0$ limit, on the $O(\epsilon^{1/2})$ lengthscale near $H$. So the density on $\partial\Omega$ is indeed asymptotically Gaussian, with the same falloff rate as $(\boldsymbol{x} - H)/\epsilon^{1/2} \to \infty$ to either side of the saddle point. Once again we stress that this Gaussian behavior can only occur in the absence of a classically forbidden wedge.

Equation (22), which expresses $\lambda_\epsilon^{(0)}$ in terms of the flux of probability into $\partial\Omega$ near $H$, when applied to the inner approximation (83) yields

$$(\mathsf{E}\tau_\epsilon)^{-1} \sim \lambda_\epsilon^{(0)} \sim \frac{1}{\pi}\sqrt{\det \boldsymbol{Z}(S)}\sqrt{\frac{\lambda_u(H)}{|\lambda_s(H)|}}K(H)\exp\left(-W(H)/\epsilon\right) \qquad \epsilon \to 0. \qquad (86)$$

Here the factor $\sqrt{\det \boldsymbol{Z}(S)}$ arises from the denominator of (22). We have approximated the integral in the denominator of (22) by an integral of the Gaussian approximation (53) to $v_\epsilon^0$ over the stable region of size $O(\epsilon^{1/2})$ near $S$; this equals $2\pi\sqrt{\epsilon/\det \boldsymbol{Z}(S)}$. The asymptotics of (86) are valid only if the coordinates $x^i$ near $H$ have been linearly transformed in such a way that $\boldsymbol{D}(H) = \boldsymbol{I}$; this was of course assumed in the derivation of the inner approximation (83).

$$(\mathsf{E}\tau_\epsilon)^{-1} \sim \lambda_\epsilon^{(0)} \sim \frac{\lambda_u(H)}{\pi}\sqrt{\frac{\det \boldsymbol{Z}(S)}{|\det \boldsymbol{Z}(H)|}}K(H)\exp\left(-W(H)/\epsilon\right), \qquad \epsilon \to 0 \qquad (87)$$

is the generalization to arbitrary coordinate systems.

Wentzell-Freidlin theory provides the leading order growth of the MFPT $\mathsf{E}\tau_\epsilon$ as $\epsilon \to 0$; it is exponential, with rate constant $W(H)$. The asymptotic expression (87) includes this exponential growth, and also a (constant) pre-exponential factor. The pre-exponential factor, like $W(H)$, is nonlocal: it is not determined by the behavior of the stochastic model in the vicinity of $S$ and $H$. This is because the nominal 'frequency factor' $K(H)$ can be computed only by integrating the system of ordinary differential equations (31), (32), (35), (36), (38) from $S$ to $H$ along the MPEP. The only exception to this is when $b^i = -D^{ij}\partial_j\Phi$ for some differentiable function $\Phi$, *i.e.*, when $b$ is globally as well as locally gradient. In this case it is easily checked that $W = 2\Phi$ and $K \equiv 1$, so the pre-exponential factor in (87) is locally determined.

The $\epsilon \to 0$ asymptotics which we have just derived for the exit location density $p_\epsilon$ and the MFPT $\mathsf{E}\tau_\epsilon$ are quite familiar from the literature [45, 65, 69, 74]. In fact in the context of chemical physics the MFPT formula can be traced back to Eyring [29]. But our derivation of these formulæ makes it clear that they are valid *only* when the classically forbidden wedge is absent. In general this will occur only in a single nongeneric case: when the drift field near $H$ satisfies the local gradient condition, and moreover the eigenvalue ratio $\mu > 1$ (*i.e.*, $\partial_i b^i(H) < 0$). In the generic case a wedge will be present, and the Hessian matrix $Z(x)$ will not have a unique limit as $x \to H$. As a consequence the inner approximation to the principal eigenfunction near $H$ will have different Gaussian falloff rates as $(x - H)/\epsilon^{1/2} \to \infty$ to either side of $H$, the asymptotic exit location density on $\partial\Omega$ will not be Gaussian, and equation (87) must at the very least be reinterpreted.

Most previous work has concentrated on globally gradient models, for which $b^i = -D^{ij}\partial_j\Phi$. Such models are extremely nongeneric in that irrespective of $\mu$, they have no forbidden wedge: $W = 2\Phi$ is always smooth at $H$, even when $\mu \geq 1$. It is perhaps this that has given rise to a general impression that the Hessian matrix $Z(x)$ always has a well-defined limit as $x \to H$, and that asymptotic exit location location distributions are always Gaussian. Generic models satisfy however neither $b^i = -D^{ij}\partial_j\Phi$ nor the local gradient condition $c = 0$, and we now see that their behavior is altogether different. In the next two sections we analyse the behavior of generic models with $\mu < 1$ and $\mu > 1$.

# 7 Skewing and MFPT Asymptotics When $\partial_i b^i(H) > 0$

We now consider models with $b^i{}_{,i}(H) > 0$, *i.e.*, models in which the eigenvalue ratio $\mu = |\lambda_s(H)|/\lambda_u(H) < 1$. We shall show that generically, the asymptotic exit location distribution near $H$ is a Weibull distribution on the $O(\epsilon^{\mu/2})$ lengthscale, as mentioned in the Introduction. It is non-Gaussian and asymmetric.

In this section we shall for simplicity take the linearized drift $B^i{}_j(H) = b^i{}_{,j}(H)$ to be

$$\boldsymbol{B}(H) = \begin{pmatrix} \lambda_u(H) & 0 \\ c & \lambda_s(H) \end{pmatrix} = \begin{pmatrix} 1 & 0 \\ c & -\mu \end{pmatrix}, \tag{88}$$

as in Section 6, and take $\boldsymbol{D}(H) = \boldsymbol{I}$. The lower triangular form for $\boldsymbol{B}(H)$ can be arranged by an appropriate linear change of coordinates near $H$, as can $\boldsymbol{D}(H) = \boldsymbol{I}$. $\lambda_u(H) = 1$ can be arranged by a rescaling of time and noise strength. With this choice of $\boldsymbol{B}(H)$, the boundary $\partial\Omega$ will (near $H$) be parallel to the $x^2$-axis. The most probable exit path (MPEP) from $S$ to $H$ will generically be tangent to $\partial\Omega$, and without loss of generality we assume that it approaches $H$ from the first quadrant, as in Figure 2(a).

The basic properties of $\mu < 1$ models were worked out in Sections 5 and 6. Generically there is a classically forbidden wedge emanating from $H$, on the boundary of which the Hessian matrix $W_{,ij}$ is discontinuous. If $c < 0$ ('Subcase A') then within the wedge $W_{,ij}(\boldsymbol{x})$ has rank-1 limit $\hat{W}_{,ij}(H)$ as $\boldsymbol{x} \to H$, though if $c > 0$ ('Subcase B') $W(\boldsymbol{x})$ in the wedge is not expected to be a quadratic function of $\boldsymbol{x}$ near $H$. If $\boldsymbol{x} \to H$ from outside the wedge, *e.g.*, along the MPEP, then $W_{,ij}(\boldsymbol{x}) \to 0$. Outside the wedge the action $W$ is 'flat' near $H$. Since the outer approximation to the quasistationary density $v_\epsilon^0(\boldsymbol{x})$ contains an exponential factor $\exp(-W(\boldsymbol{x})/\epsilon)$, this is a sign that on the classically allowed side of $H$, the quasistationary density falls off only slowly (more slowly than quadratically). As a consequence the appropriate lengthscale for an inner approximation near $H$ should be larger than $O(\epsilon^{1/2})$. On the $O(\epsilon^{1/2})$ lengthscale the asymptotics of $v_\epsilon^0$ and the exit location density $p_\epsilon$ are those of (73)–(77), but an approximation on that lengthscale is not particularly useful.

It was shown in Section 5 that when $\mu < 1$, the tangent space $T_{(H,0)}\mathcal{M}_{(S,0)}^u$ to the manifold $\mathcal{M}_{(S,0)}^u \subset \bar{\Omega} \times \mathbb{R}^2$ at $(H, 0)$ generically equals $\mathcal{N}_{(+,+)}$, the linear span of two eigenvectors of the linearized Hamiltonian flow at $(H, 0)$: the stable eigenvector $(\boldsymbol{e}_s, 0)$ and the unstable eigenvector $(\boldsymbol{e}_u, 0)$. Since the MPEP $\boldsymbol{q}^*$, regarded as a trajectory in the phase space $\bar{\Omega} \times \mathbb{R}^2$, terminates at $(H, 0)$, it must approach $(H, 0)$ along the tangent vector $(\boldsymbol{e}_s, 0)$; since $\boldsymbol{e}_s$ lies in $\partial\Omega$, that is why it is tangent to $\partial\Omega$. This tangency condition is of course an asymptotic statement, valid only as $t \to \infty$, *i.e.*, as $(H, 0)$ is approached. In fact the 4-by-4 matrix $\boldsymbol{T}(H)$ representing the linearization of the Hamiltonian flow has both $(\boldsymbol{e}_s, 0)$ and $(\tilde{\boldsymbol{e}}_s, \tilde{\boldsymbol{g}}_s)$ as stable eigenvectors. They have eigenvalues $\lambda_s(H) = -\mu$ and $-\lambda_u(H) = -1$ respectively, so a more precise description of the $t \to \infty$ asymptotics of the MPEP would be

$$\boldsymbol{q}^*(t) - H \sim C_s \exp(-\mu t)\,\boldsymbol{e}_s + \tilde{C}_s \exp(-t)\,\tilde{\boldsymbol{e}}_s, \qquad t \to \infty. \tag{89}$$

Since $\mu < 1$, the first term is dominant as $t \to \infty$, and gives rise to the approach along $\partial\Omega$. However the coefficient $\tilde{C}_s$, like $C_s$, is generically nonzero. The two coefficients can only be found by computing the MPEP explicitly: by integrating Hamilton's equations (31), (32) from $(\boldsymbol{x}, \boldsymbol{p}) = (S, 0)$ at $t = -\infty$ to $(H, 0)$ at $t = \infty$. The fact that $\tilde{C}_s$ is generically nonzero was taken into account when plotting Figures 2(a), 3(a), and 3(b); in those three figures, the slight deviation of the approaching MPEP from $\partial\Omega$ is due to the $\tilde{\boldsymbol{e}}_s$ term.

Explicit formulæ for the eigenvectors $\boldsymbol{e}_s$ and $\tilde{\boldsymbol{e}}_s$ appear in (66)–(69); we may take $\boldsymbol{e}_s = (0, 1)$ and $\tilde{\boldsymbol{e}}_s = (\mu - 1, c)$. So (89) may be rewritten, if $\Delta\boldsymbol{x}$ signifies $\boldsymbol{x} - H$, as

$$(\Delta x^1, \Delta x^2) \sim \left(\tilde{C}_s(\mu - 1)e^{-t}, C_s e^{-\mu t} + \tilde{C}_s c e^{-t}\right). \tag{90}$$

We must have $\tilde{C}_s < 0$ and $C_s > 0$, since it is our convention that the MPEP approaches $H$ from the first quadrant. As $t \to \infty$ and $H$ is approached, the MPEP will generically be asymptotic to the curve $\Delta x^2 = A(\Delta x^1)^\mu$, where

$$A = C_s \left(\frac{1}{\tilde{C}_s(\mu - 1)}\right)^\mu. \tag{91}$$

This asymptotic behavior occurs irrespective of the value of $c$; in this regard Subcases A and B are the same.

Since the boundary layer near the characteristic boundary $\partial\Omega$ will have thickness $O(\epsilon^{1/2})$, the inner approximation near $H$ should be valid when $\Delta x^1 = O(\epsilon^{1/2})$. But if the inner approximation is to be valid on a region containing a nontrivial (as $\epsilon \to 0$) portion of the MPEP, it should also be valid when $\Delta x^2 = O(\epsilon^{\mu/2})$. We see that $x^1 - H^1$ and $x^2 - H^2$ should be treated *asymmetrically*: when $\mu < 1$, the appropriate boundary region for the inner approximation is a strip near $\partial\Omega$ within which $\Delta x^1 \ll \Delta x^2$. This thin strip should extend from $H$ along $\partial\Omega$ in the direction of the approaching MPEP; equivalently, it should *not* extend in the direction of the forbidden wedge. As previously discussed, the appropriate lengthscale for an inner approximation on the wedge side of $H$ is $O(\epsilon^{1/2})$, not $O(\epsilon^{\mu/2})$.

Let us write $(x, z)$ for $\left(\Delta x^1, (\Delta x^2)^{1/\mu}\right)$; the change to noncovariant notation will emphasize the asymmetry. In the $(x, z)$-plane the boundary region will be the region where $x, z = O(\epsilon^{1/2})$. The MPEP will approach $H = (0, 0)$ as $t \to \infty$ along an asymptotically linear trajectory $z \sim A^{1/\mu} x$. It is an easy exercise to show, using the matrix Riccati equation (38) in the linear approximation near $H$, that irrespective of the asymptotic slope $A^{1/\mu}$ the matrix of second derivatives

$$\check{\boldsymbol{Z}}(x,z) = \begin{pmatrix} \dfrac{\partial^2 W}{\partial x^2} & \dfrac{\partial^2 W}{\partial x \partial z} \\ \dfrac{\partial^2 W}{\partial z \partial x} & \dfrac{\partial^2 W}{\partial z^2} \end{pmatrix} \tag{92}$$

will have a finite (nonzero) limit as $t \to \infty$, *i.e.*, as the MPEP approaches $H$. We accordingly expect that in the complement of the forbidden wedge, the action $W$ near $H$ behaves quadratically in $x$ and $z$.

This quadratic behavior will mandate Gaussian far-field asymptotics (as $(x,z)/\epsilon^{1/2} \to \infty$) for the inner approximation to $v_\epsilon^0$, much as in the last section. The inner approximation is best written in terms of the stretched variables $X = x/\epsilon^{1/2}$ and $Z = z/\epsilon^{1/2}$. It may be found by solving a linearized version of the approximate forward Kolmogorov equation $\mathcal{L}_\epsilon^* \rho = 0$, as follows. In the linear approximation we take $b^i(\boldsymbol{x}) \approx B^i{}_j(H)(x^j - H^j)$ and $D^{ij}(\boldsymbol{x}) \approx D^{ij}(H)$, and the Kolmogorov equation reduces to

$$(\epsilon/2)D^{ij}(H)\partial_i\partial_j \rho - \partial_i[B^i{}_j(H)(x^j - H^j)\rho] = 0. \tag{93}$$

Substituting both (88) and $\boldsymbol{D}(H) = \boldsymbol{I}$, and changing variables from $(x^1, x^2)$ to $X = (x^1 - H^1)/\epsilon^{1/2}$ and $Y = (x^2 - H^2)/\epsilon^{\mu/2}$ yields

$$\frac{1}{2}\frac{\partial^2 \rho}{\partial X^2} + \frac{\epsilon^{1-\mu}}{2}\frac{\partial^2 \rho}{\partial Y^2} - \frac{\partial}{\partial X}[X\rho] + \mu\frac{\partial}{\partial Y}[Y\rho] - c\epsilon^{(1-\mu)/2}\frac{\partial}{\partial Y}[X\rho] = 0. \tag{94}$$

In the $\epsilon \to 0$ limit this becomes

$$\frac{1}{2}\frac{\partial^2 \rho}{\partial X^2} - \frac{\partial}{\partial X}[X\rho] + \mu\frac{\partial}{\partial Y}[Y\rho] = 0 \tag{95}$$

and substituting $Y = Z^\mu$ yields

$$\frac{1}{2}\frac{\partial^2 \rho}{\partial X^2} - X\frac{\partial \rho}{\partial X} + Z\frac{\partial \rho}{\partial Z} - (1-\mu)\rho = 0. \tag{96}$$

This partial differential equation for $\rho = \rho(X, Z)$ must be solved in the boundary region.

In terms of the stretched variables $(X, Z)$ we may view the boundary region as the right-half plane $X \geq 0$, and the boundary $\partial \Omega$ as the $Z$-axis. The location of the classically forbidden wedge, in terms of $X$ and $Z$, is easily determined. As noted in Section 5 the wedge is bounded by the ray consisting of all multiples of $e_u$; since $e_u = (\mu + 1, c)$ by (67), its boundary has equation $\Delta x^2 = [c/(\mu + 1)]\Delta x^1$, or $Z = [c/(\mu+1)]^{1/\mu}\epsilon^{(1-\mu)/2\mu}X^{1/\mu}$. In the $\epsilon \to 0$ limit the boundary of the wedge therefore becomes the $X$-axis: of the first and fourth quadrants in the $(X, Z)$-plane one is classically allowed, and the other is forbidden. We are taking the MPEP to approach $(X, Z) = (0, 0)$ from the first quadrant, so it is the fourth quadrant that is forbidden. Equation (96) should therefore be solved only in the first quadrant. The extreme suppression of the quasistationary density in the forbidden wedge (on the $O(\epsilon^{\mu/2})$ lengthscale) allows us to set $\rho = 0$ when $Z \leq 0$, for both Subcase A and Subcase B.

A family of solutions of equation (96), each with far-field asymptotics that are Gaussian in $X$ and $Z$, can be found by inspection. Each solution in the family is of the form

$$\rho(X, Z) \propto \begin{cases} Z^{1-\mu} \exp(2BXZ - B^2 Z^2), & Z > 0 \\ 0, & Z \leq 0 \end{cases} \quad (97)$$

for some constant $B$. Actually we shall use antisymmetrized (odd) versions of these solutions, since $\rho$ must satisfy the Dirichlet boundary condition $\rho(0, \cdot) = 0$ on account of absorption of probability on $\partial\Omega$. Antisymmetrizing under $x \to -x$ yields

$$\rho(X, Z) \propto \begin{cases} Z^{1-\mu} \sinh(2BXZ) \exp(-B^2 Z^2), & Z > 0 \\ 0, & Z \leq 0 \end{cases} \quad (98)$$

Rewriting (98) in terms of $\Delta x^i = x^i - H^i$ gives

$$v_\epsilon^0(\boldsymbol{x}) \sim \begin{cases} C(\Delta x^2)^{(1/\mu)-1} \sinh\left[2B(\Delta x^1)(\Delta x^2)^{1/\mu}/\epsilon\right] \exp\left[-B^2(\Delta x^2)^{2/\mu}/\epsilon\right], & \Delta x^2 > 0 \\ 0, & \Delta x^2 \leq 0 \end{cases} \quad (99)$$

as the desired inner approximation for the case $\mu < 1$, with $C$ a constant to be found by matching to the outer approximation. Recall that in the far field, the inner approximation must match to an outer approximation of the form $K(\boldsymbol{x})\exp\left(-W(\boldsymbol{x})/\epsilon\right)$, where $W$ is generated by incoming classical trajectories (including the MPEP). From (99) we can read off the behavior of $W$ and $K$ as $\boldsymbol{x} \to H$ when $\Delta x^2 > 0$ (*i.e.*, in the complement of the forbidden wedge). Necessarily

$$W(\boldsymbol{x}) \sim -2B(\Delta x^1)(\Delta x^2)^{1/\mu} + B^2(\Delta x^2)^{2/\mu} \tag{100}$$

$$K(\boldsymbol{x}) \sim C(\Delta x^2)^{(1/\mu)-1}. \tag{101}$$

It is a useful exercise to verify that the formulæ (100) and (101), irrespective of the choice of $B$, are consistent with the system of ordinary differential equations (31), (32), (35), (36), (38) in the linear approximation near $H$. The existence of a limiting value for the matrix $\tilde{\boldsymbol{Z}}(x,z)$ of (92) as $\boldsymbol{x} \to H$ from outside the wedge is apparent: as $\boldsymbol{x} \to H$, $W$ is asymptotically quadratic in $x$ and $z$.

Notice that $K \to 0$ as $\boldsymbol{x} \to H$ from outside the wedge: in particular, along the MPEP. In other words, the nominal frequency factor $K(H)$ is *generically zero* when $\mu < 1$. This is an entirely new discovery, and has been confirmed in several models by numerical integration of the transport equation (36) from $(\boldsymbol{x}, \boldsymbol{p}) = (S, 0)$ to $(H, 0)$. The fact that $K(H) = 0$ is yet another reason why the classical formula (87) for the MFPT asymptotics cannot be generically applicable.

It turns out that the constant $B$ in the inner approximation is uniquely determined by the approach path taken by the MPEP. Since $p_i = \partial_i W$, differentiation of (100) yields

$$\boldsymbol{p}(H + \Delta\boldsymbol{x}) \sim \left( -2B(\Delta x^2)^{1/\mu}, -2B\mu^{-1}(\Delta x^1)(\Delta x^2)^{(1/\mu)-1} + 2B^2\mu^{-1}(\Delta x^2)^{(2/\mu)-1} \right) \tag{102}$$

near $H$, on the $\Delta x^2 = O\left((\Delta x^1)^\mu\right)$ lengthscale. But in the linear approximation near $H$

$$\boldsymbol{b}(H + \Delta\boldsymbol{x}) \approx (\Delta x^1, -\mu\,\Delta x^2 + c\,\Delta x^1) \tag{103}$$

by (88), and substitution into Hamilton's equation (31) for $\dot{\boldsymbol{x}}$ yields

$$\frac{d}{dt}(\Delta x^1, \Delta x^2) \sim \left( -2B(\Delta x^2)^{1/\mu} + \Delta x^1, -2B\mu^{-1}(\Delta x^1)(\Delta x^2)^{(1/\mu)-1} + 2B^2\mu^{-1}(\Delta x^2)^{(2/\mu)-1} - \mu\,\Delta x^2 + c\,\Delta x^1 \right) \tag{104}$$

as the equation of motion which must be followed by the classical trajectories near $H$ giving rise to $W$. This equation is less complicated than it looks. When $\Delta x^2 = O\left((\Delta x^1)^\mu\right)$, the dominant term in the second component on the right-hand side of (104) is the $-\mu\,\Delta x^2$ term. The other terms may simply be dropped, and the equation simplifies to

$$\frac{d}{dt}(\Delta x^1, \Delta x^2) \sim \left( -2B(\Delta x^2)^{1/\mu} + \Delta x^1, -\mu\,\Delta x^2 \right). \tag{105}$$

It is trivial to verify that this asymptotic equation of motion near $H$ is compatible with the asymptotic approach path $\Delta x^2 = A(\Delta x^1)^\mu$ if and only if $B = A^{-1/\mu}$. Since $A$ is the limit of the ratio $\Delta x^2/(\Delta x^1)^\mu$ along the MPEP as it approaches $H$, the constant $B$ in the inner approximation may be computed numerically.

Now that we have constructed an inner approximation to the quasistationary density $v_\epsilon^0$ which is asymptotically accurate as $\epsilon \to 0$, we may compute the asymptotic exit location distribution and MFPT asymptotics by the techniques sketched in Section 3. Since $\boldsymbol{D}(H) = \boldsymbol{I}$, equation (18) says that asymptotically, the density of the exit location measure on $\partial\Omega$ is proportional to the normal derivative of $v_\epsilon^0$. This is simply the rate at which probability is absorbed on $\partial\Omega$, as a function of position. By differentiating the expression (99) with respect to $\Delta x^1$ and setting $\Delta x^1$ to zero we obtain (since $B = A^{-1/\mu}$)

$$p_\epsilon(s) \sim \begin{cases} \left(\dfrac{2}{\mu A^{2/\mu}\epsilon}\right) s^{(2/\mu)-1} \exp[-(s/A)^{2/\mu}/\epsilon], & s > 0; \\ 0, & s \leq 0. \end{cases} \tag{106}$$

Here we have written $s$ for $\Delta x^2$ (the distance along $\partial\Omega$, measured from $H$) for consistency with the Introduction and Section 6. The overall normalization of (106) is fixed by the condition that $p_\epsilon$ have total unit mass.

The asymptotic exit location density (106), which is localized on the $s = O(\epsilon^{\mu/2})$ lengthscale, is of an unusual form. It is the density of a Weibull distribution [3], with shape parameter $2/\mu$ and scale parameter $1/A\epsilon^{\mu/2}$. Weibull-distributed random variables are simply powers of exponential random variables, and $p_\epsilon$ may be viewed as the density of an 'offset' random variable $\mathfrak{S}_\epsilon$ equal to $M^{\mu/2}$, where $M$ is an exponential variate of mean $A^{2/\mu}\epsilon$. The Weibull distribution is decidedly 'skewed'; in fact it is supported entirely on the $s > 0$ side of the saddle point. That $s > 0$, rather than $s < 0$, appears here is solely a matter of convention. By convention the $s > 0$ side of $H$ is the side from which the MPEP approaches, as in Figure 2(a). For later use we note that

$$\mathsf{E}\,\mathfrak{S}_\epsilon = \int_0^\infty s\, p_\epsilon(s)\, ds \sim A\Gamma(1 + \mu/2)\epsilon^{\mu/2}, \quad \epsilon \to 0 \tag{107}$$

is the expected offset from $H$ along $\partial\Omega$, when $\mu < 1$, at the time of exit.

The qualitative features of the asymptotic density (106) can be explained by reference to Figure 2(a). The quantity $s^{2/\mu}$ in the exponent is roughly proportional to the square of the distance between the point $(0, s)$ on $\partial\Omega$ and the closest point on the approaching MPEP. This is consistent with a picture developed elsewhere [56], according to which the MPEP is surrounded by a 'tube' of probability current, the tube having a Gaussian transverse profile. We have already discussed why the limiting $p_\epsilon(s)$, on the $O(\epsilon^{\mu/2})$ lengthscale, is zero for $s < 0$. The points on $\partial\Omega$ with $s < 0$ are classically forbidden, and $p_\epsilon(s)$ in the forbidden region falls to zero on the $s = O(\epsilon^{1/2})$ lengthscale, as summarized in (76), (77). On the $O(\epsilon^{1/2})$ lengthscale Subcase A and Subcase B differ significantly, but on the $O(\epsilon^{\mu/2})$ lengthscale they become identical.

It is clear how the $\epsilon \to 0$ MFPT asymptotics may be computed from the inner approximation (99). Asymptotically $\lambda_\epsilon^{(0)}$, *i.e.*, $(\mathsf{E}\tau_\epsilon)^{-1}$, is simply the flux of probability into $\partial\Omega$. As such it is proportional to the constant prefactor $C$ in (99). If the inner approximation is to match up with the outer approximation, by (101) we must have

$$C = L \exp\left(-W(H)/\epsilon\right), \tag{108}$$

where $L$ is the limit of the quantity $K(\boldsymbol{x})/(\Delta x^2)^{(1/\mu)-1}$ as $\boldsymbol{x} \to H$ along the MPEP. This formula for $C$ has novel consequences. The factor $\exp\left(-W(\boldsymbol{x})/\epsilon\right)$ gives rise to the familiar Wentzell-Freidlin growth factor in $\mathsf{E}\tau_\epsilon$. But the pre-exponential factor in the asymptotics of $(\mathsf{E}\tau_\epsilon)^{-1}$ is *not* proportional to the (generically zero) nominal frequency factor $K(H)$, as in the classical formula (87), but rather to $L$, the rate at which $K(\boldsymbol{x})$ approaches zero as the MPEP approaches $H$. In fact substituting the inner approximation into (22), and using $B = A^{-1/\mu}$, yields

$$(\mathsf{E}\tau_\epsilon)^{-1} \sim \lambda_\epsilon^{(0)} \sim \frac{\mu A^{1/\mu} L}{4\pi} \sqrt{\det \boldsymbol{Z}(S)} \exp\left(-W(H)/\epsilon\right), \qquad \epsilon \to 0, \tag{109}$$

which is the generic replacement for the classical formula (87) when $\mu < 1$. The $\sqrt{\det \boldsymbol{Z}(S)}$ factor arises from the denominator of (22), as in (87). We remind the reader that we are assuming $\lambda_u(H) = 1$ and $\boldsymbol{D}(H) = \boldsymbol{I}$ here.

The generic applicability (when $\mu < 1$) of this formula for the MFPT asymptotics, and the generic *inapplicability* of the classical formula (87), have not previously been recognized. It is remarkable that despite the nominal frequency factor $K(H)$ equalling zero, the pre-exponential factor in (109) fails to be $\epsilon$-dependent. Naively one would have expected it to contain a positive power of $\epsilon$. A positive power of $\epsilon$ is known to occur in the weak-noise reciprocal MFPT asymptotics of stochastic models where the nominal frequency factor equals zero on account of the exit location on $\partial\Omega$ converging to an unstable fixed point [54].

# 8 Skewing When $\partial_i b^i(H) < 0$

We now consider models with $b^i{}_{,i}(H) < 0$, *i.e.*, models in which the eigenvalue ratio $\mu = |\lambda_s(H)|/\lambda_u(H) > 1$. The asymptotic exit location distribution near $H$ is localized on the $O(\epsilon^{1/2})$ lengthscale, and as shown in Section 6 is generically non-Gaussian: it has different Gaussian falloff rates to either side of $H$. Although we shall not compute an explicit expression for its density, we shall work out a scheme for computing its moments of any desired order. Our treatment will be based on a probabilistic analysis, rather than on the construction of an inner approximation to the quasistationary density.

In this section, as in Sections 6 and 7, we shall without loss of generality take the linearized drift $B^i{}_j(H) = b^i{}_{,j}(H)$ to be

$$\boldsymbol{B}(H) = \begin{pmatrix} \lambda_u(H) & 0 \\ c & \lambda_s(H) \end{pmatrix} = \begin{pmatrix} 1 & 0 \\ c & -\mu \end{pmatrix}, \tag{110}$$

and take $\boldsymbol{D}(H) = \boldsymbol{I}$. With this choice of $\boldsymbol{B}(H)$ the boundary $\partial\Omega$ near $H$ will be parallel to the $x^2$-axis, as in Figures 2 and 3. We also translate coordinates so that $H = (0,0)$. With these normalizations the stochastic differential equation (1) becomes, in the linear approximation near $H$,

$$dx^1_\epsilon(t) = x^1_\epsilon(t)\,dt + \epsilon^{1/2}dw_1(t) \tag{111}$$

$$dx^2_\epsilon(t) = -\mu x^2_\epsilon\,dt + cx^1_\epsilon(t)\,dt + \epsilon^{1/2}dw_2(t). \tag{112}$$

If the 'stretched' process $(X(t), Y(t))$ in the $(X,Y)$-plane is defined to equal $\left(x^1_\epsilon(t), x^2_\epsilon(t)\right)/\epsilon^{1/2}$, we have

$$dX(t) = X\,dt + dw_1(t) \tag{113}$$

$$dY(t) = -\mu Y\,dt + cX\,dt + dw_2(t). \tag{114}$$

We see that $t \mapsto X(t)$ is an inverted (repelling) Ornstein-Uhlenbeck process. On the $O(\epsilon^{1/2})$ lengthscale near $H$ the region $\bar{\Omega}$ becomes the right-half plane $X \geq 0$, and the fact that equation (113) does not involve $Y$ indicates that with these normalizations the exit problem becomes essentially one-dimensional.

Our interest is in the final approach to the boundary, which as $\epsilon \to 0$ will take place along the MPEP (most probable exit path) determined in Sections 5 and 6. Generically the MPEP, as shown in Figure 2(b), is tangent to the stable ray $\tilde{e}_s$ emanating from $H$. As computed in (68), $\tilde{e}_s$ equals $(\mu - 1, c)$. So as $\epsilon \to 0$ the final approach to $\partial\Omega$, in the linear approximation near $H$, will be increasingly concentrated on the line $x^2/x^1 = c/(\mu - 1)$. On the $O(\epsilon^{1/2})$ lengthscale the approach path will not have a deterministic limit as $\epsilon \to 0$. However the straight-line deterministic asymptotics should appear in the far field of the $O(\epsilon^{1/2})$ lengthscale; going backward in time from the boundary hitting time, the approach path should be asymptotic to the line $Y/X = c/(\mu - 1)$.

The random variable $\mathfrak{S}_\epsilon = \epsilon^{1/2} Y(\tau_\epsilon)$, the displacement from $H$ along $\partial\Omega$ at the boundary hitting time $\tau_\epsilon$, is the quantity whose distribution we wish to compute. We define the time-reversed process $\left(\tilde{X}(u), \tilde{Y}(u)\right)$, $u \geq 0$, to equal $(X(\tau_\epsilon - u), Y(\tau_\epsilon - u))$; $\tau_\epsilon$ is of course a random variable, which depends on the sample path. With this convention, $\tilde{X}(0) = 0$ and $\tilde{Y}(0) = \mathfrak{S}_\epsilon/\epsilon^{1/2}$. A straightforward integration of (114) from $u = 0$ to $u = T$, *i.e.*, from $t = \tau_\epsilon$ to $t = \tau_\epsilon - T$, yields that for any $T > 0$,

$$\tilde{Y}(0) = \tilde{Y}(T) e^{-\mu T} + \int_0^T e^{-\mu u}[c\tilde{X}(u)\,du - dw_2(u)]. \tag{115}$$

As will shortly be seen, as $\epsilon \to 0$ the expected transit time of the final approach path tends to infinity. This justifies the taking of the $T \to \infty$ limit when computing $\epsilon \to 0$ asymptotics. Taking the $T \to \infty$ limit yields

$$\tilde{Y}(0) \sim \int_0^\infty e^{-\mu u}[c\tilde{X}(u)\,du - dw_2(u)], \tag{116}$$

which is to be interpreted as a statement that the left and right-hand sides are distributed identically in the $\epsilon \to 0$ limit. But $\int_0^\infty e^{-\mu u} dw_2(u)\,du$ is a Gaussian random variable of mean zero and variance $1/2\mu$. It follows that

$$\mathfrak{S}_\epsilon/\epsilon^{1/2} \sim c\,\mathfrak{I}(\mu) + \mathfrak{Z}/\sqrt{2\mu}, \tag{117}$$

where $\mathfrak{Z}$ is standard normal and the integral

$$\mathfrak{I}(\mu) \stackrel{\text{def}}{=} \int_0^\infty e^{-\mu u} \tilde{X}(u)\,du \tag{118}$$

is a weighted area under the graph of the time-reversed process $\tilde{X}(u)$, $u \geq 0$. The two terms in (117) are independent.

The asymptotic exit location density $p_\epsilon(s)$ equals $(d/ds)\mathsf{P}\{\mathfrak{S}_\epsilon \leq s\}$, the density of $\mathfrak{S}_\epsilon$. In Sec-tion 6 we deduced that when $\mu > 1$, $p_\epsilon(s)$ has different Gaussian decay rates as $s/\epsilon^{1/2} \to \pm\infty$; from (78),

$$p_\epsilon(s) \sim \begin{cases} \exp\left[-\dfrac{\mu(\mu-1)^2}{c^2 + (\mu-1)^2}(s^2/\epsilon)\right], & s/\epsilon^{1/2} \to +\infty, \\ \exp\left[-\mu(s^2/\epsilon)\right], & s/\epsilon^{1/2} \to -\infty. \end{cases} \quad (119)$$

In deriving (119) the convention was adopted that the MPEP should approach $H$ from the $s \geq 0$ side; this amounts to assuming that $c \geq 0$. This being the case, it is interesting to compare (119) with the asymptotic equality in distribution (117). They are perfectly consistent: the Gaussian falloff of the density of the $\mathfrak{Z}/\sqrt{2\mu}$ term in (117) may be viewed as the cause of the comparatively rapid Gaussian decay of $p_\epsilon(s)$ as $s/\epsilon^{1/2} \to -\infty$, since the random variable $\mathfrak{I}(\mu)$ is non-negative. In fact by independence, the density $p_\epsilon(\cdot)$ will be the convolution of the densities of $\epsilon^{1/2}\mathfrak{I}(\mu)$ and $\epsilon^{1/2}\mathfrak{Z}/\sqrt{2\mu}$. Equivalently, the generic asymptotic exit location density on the $O(\epsilon^{1/2})$ lengthscale will be the convolution of the density of $c\,\mathfrak{I}(\mu)$ with a Gaussian (the density of $\mathfrak{Z}/\sqrt{2\mu}$). A striking conclusion follows: *The skewing of the exit location distribution when $\mu > 1$ is attributable to the asymmetry of the distribution of $c\,\mathfrak{I}(\mu)$*. Only if $c = 0$ is this asymmetry absent. When $c = 0$, as we have already seen in Section 6, there is no skewing: the exit location distribution is asymptotically Gaussian, with variance $1/(2\mu)$.

To determine the distribution of $\mathfrak{I}(\mu)$, or at least its moments, we need to study the one-dimensional process $t \mapsto x_\epsilon^1(t)$, conditioned on its exit at time $\tau_\epsilon$. By the 'final approach path' we shall mean that segment of the trajectory $t \mapsto \boldsymbol{x}_\epsilon(t)$ which leaves some specified neighborhood of $S$ and terminates on $\partial\Omega$ at time $\tau_\epsilon$. Let $a > 0$ be specified, and suppose that along the final approach path, $x_\epsilon^1(t)$ first reaches the point $x^1 = a\epsilon^{1/2}$ at time $t = \tau_\epsilon - \mathfrak{u}$; equivalently, that $X(t)$ first reaches the point $X = a$ at time $t = \tau_\epsilon - \mathfrak{u}$. The process $X(t)$, $\tau_\epsilon - \mathfrak{u} \leq t \leq \tau_\epsilon$, is an inverted Ornstein-Uhlenbeck process conditioned to satisfy $X(t) > 0$ for all $t \in (\tau_\epsilon - \mathfrak{u}, \tau_\epsilon)$, and $X(\tau_\epsilon) = 0$.

It is not difficult to compute the asymptotics of the distribution of $\mathfrak{u}$, the 'additional time to absorption' variable, in the large-$a$ limit. The transition density $p(X_0, t_0; X_1, t_1)$ of an inverted Ornstein-Uhlenbeck process is of the form

$$p(X_0, t_0; X_1, t_1) = [2\pi\sigma_{t_1-t_0}^2]^{-1/2} \exp\left[-(X_1 - e^{t_1-t_0}X_0)^2/2\sigma_{t_1-t_0}^2\right] \quad (120)$$

where $\sigma_z^2 \stackrel{\text{def}}{=} (e^{2z} - 1)/2$ is the variance at elapsed time $z$. If the process $X(t)$ is conditioned to begin at $a > 0$ at some specified time $t_0$, the probability of its having reached $X = 0$ by time $t_0 + \tilde{u}$ will by the method of images equal [16]

$$2\left\{1 - \int_{X_1 \geq 0} p(a, t_0; X_1, t_0 + \tilde{u}) \, dX_1\right\}. \tag{121}$$

This absorption probability equals $\mathsf{P}\{\mathfrak{u} \leq \tilde{u}\}$, the probability that the additional time to absorption is no greater than $\tilde{u}$. Substitution of (120) into (121), and some elementary manipulations, yield that

$$\mathsf{P}\{\mathfrak{u} \leq \tilde{u}\} \sim \exp\left(-e^{-2(\tilde{u} - \log a)}\right), \quad a \to \infty. \tag{122}$$

It follows that one may write

$$\mathfrak{u} \sim \log a + \mathfrak{G}, \quad a \to \infty, \tag{123}$$

where $\mathfrak{G}$ satisfies

$$\mathsf{P}\{\mathfrak{G} \leq \tilde{u}\} = \exp\left(-e^{-2\tilde{u}}\right). \tag{124}$$

Equation (123) is an asymptotic equality in distribution, and the distribution of $\mathfrak{G}$ is a so-called Gumbel (or double exponential) distribution of the sort that arises in extreme value theory [67].

It is noteworthy that if $a$ is taken to equal $C\epsilon^{-1/2}$ for some $C > 0$, so that $\mathfrak{u}$ is the amount of time that elapses between the moment the final approach path reaches $x^1 = C$ and the moment of final absorption at $x^1 = 0$, the formula (123) implies that to leading order as $\epsilon \to 0$,

$$\mathsf{E}\mathfrak{u} \sim \log\left(C\epsilon^{-1/2}\right) \sim (1/2)\log(\epsilon^{-1}) \tag{125}$$

irrespective of $C$. We have commented elsewhere on the implications of this logarithmic growth [54, 56]; see also Ludwig [50]. Its interpretation is as follows: the time needed for the process to make its final approach to the characteristic boundary grows logarithmically in $\epsilon$. This logarithmic growth is to be contrasted with the *exponential* growth of the MFPT $\mathsf{E}\tau_\epsilon$ as $\epsilon \to 0$. The exponential timescale is the timescale on which a successful exit is expected to occur; when it occurs, however, it takes place on a much shorter (logarithmic) timescale. $(1/2)\log(\epsilon^{-1})$ is best viewed as the time needed for the deterministic MPEP to approach within an $O(\epsilon^{1/2})$ distance of the characteristic boundary. On that lengthscale (the width of the boundary layer) the MPEP ceases to be well-defined; equivalently, the limiting approach path process ceases to be deterministic. The Gumbel random variable $\mathfrak{G}$ is the additional amount of time needed for the process to reach the boundary.

Here we are interested primarily in the implications of the asymptotic representation (123) for the time-reversed process $\tilde{X}(u)$, $u \geq 0$, and the weighted area $\mathfrak{I}(\mu)$. The time-reversed process is an Ornstein-Uhlenbeck process satisfying

$$d\tilde{X}(u) = -\tilde{X}(u)\,du + dw(u), \tag{126}$$

conditioned to satisfy $\tilde{X}(0) = 0$ and $\tilde{X}(u) > 0$ for all $u > 0$. Moreover, $\tilde{X}(\mathfrak{u}) = a$ by definition. In the large-$a$ limit we may write this condition as $\tilde{X}(\log a + \mathfrak{G}) = a$, so conditioning on the event $\mathfrak{G} = \tilde{u}$ simply imposes an additional condition

$$\tilde{X}(\log a + \tilde{u}) = a \tag{127}$$

on $u \mapsto \tilde{X}(u)$. In the language of random processes, once a value $\tilde{u}$ for the random variable $\mathfrak{G}$ is specified, the process $\tilde{X}(u)$, $u \geq 0$, becomes a *conditioned Ornstein-Uhlenbeck meander process*. 'Meander' refers to the fact that $\tilde{X}(u) > 0$ for all $u > 0$, *i.e.*, the fact that return to zero (*i.e.*, to $\partial\Omega$) is not allowed [23].

We shall shortly see that imposing the additional condition (127), and taking the $a \to \infty$ limit, yields a well-defined process which we may denote $\tilde{X}_{\tilde{u}}(u)$, $u \geq 0$. This being the case, define

$$\mathfrak{I}_{\tilde{u}}(\mu) = \int_0^\infty e^{-\mu u}\tilde{X}_{\tilde{u}}(u)\,du \tag{128}$$

to be the conditioned version of $\mathfrak{I}(\mu)$. By (117) the moments of $\mathfrak{S}_\epsilon/\epsilon^{1/2}$, the normalized displacement along $\partial\Omega$ at the time of hitting, satisfy

$$\begin{aligned}
\mathsf{E}(\mathfrak{S}_\epsilon/\epsilon^{1/2})^k &\sim \sum_{l=0}^k \binom{k}{l} c^l (2\mu)^{-(k-l)/2}\, \mathsf{E}\,\mathfrak{I}(\mu)^l\, \mathsf{E}\,\mathfrak{Z}^{k-l} \\
&= \sum_{l=0}^k \binom{k}{l} c^l (2\mu)^{-(k-l)/2} [(k-l)!!]\, \mathsf{E}\,\mathfrak{I}(\mu)^l \\
&= \sum_{l=0}^k \binom{k}{l} c^l (2\mu)^{-(k-l)/2} [(k-l)!!] \int_{\tilde{u}=0}^\infty \mathsf{E}\mathfrak{I}_{\tilde{u}}(\mu)^l\, d[\exp(-e^{-2\tilde{u}})] \quad (129)
\end{aligned}$$

where the integral arises from removing the conditioning $\mathfrak{G} = \tilde{u}$. This formula expresses the moments of the asymptotic exit location distribution $p_\epsilon(s)\,ds$, when $\mu > 1$, in terms of those of the random variables $\mathfrak{I}_{\tilde{u}}(\mu)$. And by (128) the moments of $\mathfrak{I}_{\tilde{u}}(\mu)$ can be computed, by repeated integration, from the correlation functions (finite-dimensional distributions) of the process $\tilde{X}_{\tilde{u}}(u)$, $u \geq 0$.

For any $\tilde{u}$, $\tilde{X}_{\tilde{u}}(u)$, $u \geq 0$, is a Markov process whose transition probabilities may be computed by taking the above $a \to \infty$ limit. However, it turns out to be time-inhomogeneous. It is preferable to express $\mathfrak{I}_{\tilde{u}}(\mu)$ in terms of a closely related time-homogeneous process, which is based on Brownian rather than Ornstein-Uhlenbeck motion. This process is introduced as follows. Recall that a standard Ornstein-Uhlenbeck process $o(u)$, $u \geq 0$ satisfies

$$o(u) = e^{-u} w\left((e^{2u} - 1)/2\right) \tag{130}$$

in the sense of equality in distribution; here $w(t)$, $t \geq 0$, is a standard Wiener process. Formally, imposing the condition $o(u) > 0$ for all $u > 0$ is equivalent to imposing the condition $w(t) > 0$ for all $t > 0$; in other words, the transformation (130) relates Ornstein-Uhlenbeck meander $o^+$ to Brownian meander $w^+$, just as it relates unconstrained Ornstein-Uhlenbeck motion $o$ to Brownian motion $w$. Similarly, imposing the condition $o^+(u) = a$ on Ornstein-Uhlenbeck meander $o^+$ at $u = \log a + \tilde{u}$ is by (130) equivalent to imposing the condition

$$e^{-u} w^+\left((e^{2u} - 1)/2\right) = a \tag{131}$$

at $u = \log a + \tilde{u}$ on the corresponding Brownian meander $w^+$, *i.e.*, imposing the condition

$$w^+\left([a^2 e^{2\tilde{u}} - 1]/2\right) = a^2 e^{\tilde{u}}. \tag{132}$$

By defining $T = a^2 e^{2\tilde{u}}/2$ we see that for any fixed $\tilde{u}$, in the large-$a$ limit this condition is to leading order a requirement that

$$w^+(T) = 2e^{-\tilde{u}} T. \tag{133}$$

Informally, as $a \to \infty$ and $T \to \infty$ the conditioned Brownian meander corresponding to the original conditioned Ornstein-Uhlenbeck meander is forced by the condition (133) to drift upward at a mean speed $2e^{-\tilde{u}}$.

We define $w^+_{\tilde{u}}(t)$, $t \geq 0$, to be the weak limit as $T \to \infty$ of the standard Brownian meander process $w^+(t)$, $t \geq 0$, when constrained by the condition (133). We necessarily have, as an equality in distribution,

$$\tilde{X}_{\tilde{u}}(u) = e^{-u} w^+_{\tilde{u}}\left((e^{2u} - 1)/2\right), \tag{134}$$

so that

$$\begin{aligned}
\mathfrak{I}_{\tilde{u}}(\mu) &= \int_0^\infty e^{-\mu u} \tilde{X}_{\tilde{u}}(u) \, du \\
&= \int_0^\infty e^{-(\mu+1)u} w^+_{\tilde{u}}\left((e^{2u} - 1)/2\right) \, du \\
&= \int_0^\infty (1 + 2t)^{-(\mu+3)/2} w^+_{\tilde{u}}(t) \, dt.
\end{aligned} \tag{135}$$

The last equality follows by a change of variables $t = (e^{2u} - 1)/2$. The formula (135) permits the computation of the moments of $\mathfrak{I}_{\tilde{u}}(\mu)$, as required by (129), from the correlation functions (*i.e.*, finite-dimensional distributions) of the process $w^+_{\tilde{u}}$. We shall shortly see that $w^+_{\tilde{u}}$ is time-homogeneous, making this representation particularly useful.

The $n$-point correlation functions of $w^+_{\tilde{u}}$ may be computed by taking the $T \to \infty$ limit of the $n$-point correlation functions of Brownian meander $w^+$, conditioned by (133). The evaluation of this limit is facilitated by the following fact. Recall that a *three-dimensional Bessel process* $B(t)$, $t \geq 0$, conventionally taken to satisfy $B(0) = 0$, is the radial coordinate in $\mathbb{R}^3$ of a diffusing particle. That is, $B(t)$ equals $[w_1^2(t) + w_2^2(t) + w_3^2(t)]^{1/2}$, where the $w_i(t)$ are independent Wiener processes. It is a standard result [41] that the Bessel process $B$, when conditioned to satisfy $B(t') = x'$ for any specified $t' > 0$ and $x' > 0$, and Brownian meander $w^+$, when conditioned to satisfy $w^+(t') = x'$, become identical in distribution on the time interval $0 \leq t \leq t'$. This allows us to substitute the Bessel process for Brownian meander, and to compute instead the $T \to \infty$ limit of the $n$-point correlation functions of $B$, conditioned on $B(T) = 2e^{-\tilde{u}} T$. The substitution of $B$ for $w^+$ simplifies the computation, for the Bessel process is (unlike Brownian meander) time-homogeneous.

Denote by $q(w_1, t_1; w_2, t_2)$ the transition density of the standard Wiener process, *i.e.*,

$$q(w_1, t_1; w_2, t_2) \stackrel{\text{def}}{=} [2\pi(t_2 - t_1)]^{-1/2} \exp\left[-(w_2 - w_1)^2/2(t_2 - t_1)\right]. \tag{136}$$

If instants $0 < t_1 < \cdots < t_n$ are specified, the $n$-point correlation function $p^{(n)}(\cdot, t_1; \ldots ; \cdot, t_n)$ for the Bessel process $B$, which is defined by

$$p^{(n)}(w_1, t_1; \ldots ; w_n, t_n) \stackrel{\text{def}}{=} \mathsf{P}\{B(t_i) \in w_i + dw_i,\ 1 \leq i \leq n\} \bigg/ \prod_{i=1}^{n} dw_i\,, \tag{137}$$

satisfies

$$p^{(n)}(w_1, t_1; \ldots ; w_n, t_n) = p^{(1)}(w_1, t_1) \prod_{i=1}^{n-1} t(w_i, t_i; w_{i+1}, t_{i+1}) \tag{138}$$

where

$$p^{(1)}(w_1, t_1) = \sqrt{\frac{2}{\pi t_1^3}} w_1^2 \exp\left(-w_1^2/2t_1\right) \tag{139}$$

and

$$t(w_1, t_1; w_2, t_2) = (w_2/w_1)[q(w_1, t_1; w_2, t_2) - q(-w_1, t_1; w_2, t_2)] \tag{140}$$

are the probability density and transition density for the Bessel process. Conditioning on the event $B(T) = 2e^{-\tilde{u}}T$, where $T > t_n$, yields a process with $n$-point correlation function

$$p^{(n)}_{\tilde{u},T}(w_1, t_1; \ldots ; w_n, t_n) = \frac{p^{(n+1)}(w_1, t_1; \ldots ; w_n, t_n; 2e^{-\tilde{u}}T, T)}{p^{(1)}(2e^{-\tilde{u}}T, T)}. \tag{141}$$

In particular, the conditioned density $p^{(1)}_{\tilde{u},T}(w_1, t_1)$ satisfies

$$p^{(1)}_{\tilde{u},T}(w_1, t_1) = p^{(1)}(w_1, t_1) \left[\frac{t(w_1, t_1; 2e^{-\tilde{u}}T, T)}{p^{(1)}(2e^{-\tilde{u}}T, T)}\right], \tag{142}$$

and the conditioned transition density $t_{\tilde{u},T}(w_1, t_1; w_2, t_2)$ satisfies

$$t_{\tilde{u},T}(w_1, t_1; w_2, t_2) = \frac{p^{(2)}_{\tilde{u},T}(w_1, t_1; w_2, t_2)}{p^{(1)}_{\tilde{u},T}(w_1, t_1)} = t(w_1, t_1; w_2, t_2) \left[\frac{t(w_2, t_2; 2e^{-\tilde{u}}T, T)}{t(w_1, t_1; 2e^{-\tilde{u}}T, T)}\right]. \tag{143}$$

Let $\tilde{p}^{(1)}_{\tilde{u}}(w_1, t_1)$ and $\tilde{t}_{\tilde{u}}(w_1, t_1; w_2, t_2)$ be the $T \to \infty$ limits of $p^{(1)}_{\tilde{u},T}(w_1, t_1)$ and $t_{\tilde{u},T}(w_1, t_1; w_2, t_2)$

respectively. It follows by taking the $T \to \infty$ limit of the two factors in brackets that

$$\begin{aligned}
\tilde{p}_{\tilde{u}}^{(1)}(w_1, t_1) &= p^{(1)}(w_1, t_1) e^{-v^2 t_1/2} \left( \frac{\sinh v w_1}{v w_1} \right) \\
&= \frac{1}{\sqrt{2\pi t_1^3}} (w_1/v) [e^{-(w_1 - v t_1)^2/2 t_1} - e^{-(w_1 + v t_1)^2/2 t_1}]
\end{aligned} \quad (144)$$

and

$$\tilde{t}_{\tilde{u}}(w_1, t_1; w_2, t_2) = e^{-v^2(t_2 - t_1)/2} \left( \frac{\sinh v w_2}{\sinh v w_1} \right) [q(w_1, t_1; w_2, t_2) - q(-w_1, t_1; w_2, t_2)]. \quad (145)$$

Here $v \stackrel{\text{def}}{=} 2e^{-\tilde{u}}$. $\tilde{p}_{\tilde{u}}^{(1)}(w_1, t_1)$ and $\tilde{t}_{\tilde{u}}(w_1, t_1; w_2, t_2)$ are the density and transition density of the limiting process $w_{\tilde{u}}^+(t)$, $t \geq 0$.

The transition density $\tilde{t}_{\tilde{u}}(w_1, t_1; w_2, t_2)$ is invariant under time translation, so the limiting process is a time-homogeneous Markov process as promised. It is well known [62], and follows by examination of the transition density (140), that the three-dimensional Bessel process has generator $-(1/2)d^2/dw^2 - (1/w)d/dw$. By examination of the formula (145) for $\tilde{t}_{\tilde{u}}(w_1, t_1; w_2, t_2)$, the process $w_{\tilde{u}}^+(t)$, $t \geq 0$, has generator $-(1/2)d^2/dw^2 - (v \coth vw)d/dw$. The function $v \coth vw$ is asymptotic to $1/w$ as $w \to 0$, and to $v$ as $w \to \infty$. This confirms that $v = 2e^{-\tilde{u}}$ has an interpretation as the strength of a superimposed upward drift.

Now that the probability density and transition density of the process $w_{\tilde{u}}^+$ are known, its $n$-point correlation functions $p_{\tilde{u}}^{(n)}(w_1, t_1; \ldots; w_n, t_n)$ follow from

$$\tilde{p}_{\tilde{u}}^{(n)}(w_1, t_1; \ldots; w_n, t_n) = \tilde{p}_{\tilde{u}}^{(1)}(w_1, t_1) \prod_{i=1}^{n-1} \tilde{t}_{\tilde{u}}(w_i, t_i; w_{i+1}, t_{i+1}). \quad (146)$$

Since $\mathfrak{I}_{\tilde{u}}(\mu)$ is by (135) a weighted area under the process $w_{\tilde{u}}^+$, its moments can be expressed in terms of these $n$-point correlation functions by repeated integration. It is not clear whether a closed-form expression for the distribution of $\mathfrak{I}_{\tilde{u}}(\mu)$ exists. If it does it is likely to be quite intricate, as is suggested by the results of other authors. The problem of computing the distribution of the (unweighted) area under a Brownian bridge (*i.e.*, a pinned Wiener process) was solved by Cifarelli [13] and Shepp [72], and the corresponding problem for a Brownian excursion by Louchard [47, 48]. More recently, Takács [73] has computed the distribution of the integral of the absolute value of a Wiener process. All these distributions have closed-form expressions that are surprisingly complicated; they involve, for example, double Laplace transforms of the logarithmic derivative of an Airy function.

It is unfortunate that one cannot go directly from (144) and (145), or from the explicit expression for the generator, to a closed-form expression for the distribution of $\mathfrak{I}_{\tilde{u}}(\mu)$. If such an expression were known, it would be possible to remove (by integration) the conditioning on $\mathfrak{G} = \tilde{u}$, and obtain a closed-form expression for the distribution of $\mathfrak{I}(\mu)$. This in turn would yield a closed-form expression for the limiting exit location distribution on the $O(\epsilon^{1/2})$ lengthscale. But in the absence of such an expression one can at least compute the moments of the limiting distribution to any desired order, by using (129), (135), (144), and (145).

The case of the first moment (the expected offset from $H$ along $\partial\Omega$, at the time $\partial\Omega$ is reached) is particularly straightforward. By (129),

$$\mathsf{E}\,(\mathfrak{S}_\epsilon/\epsilon^{1/2}) \sim c\,\mathsf{E}\,\mathfrak{I}(\mu) \tag{147}$$

where

$$\mathsf{E}\,\mathfrak{I}(\mu) = \int_{\tilde{u}=0}^{\infty} \mathsf{E}\mathfrak{I}_{\tilde{u}}(\mu)\,d[\exp(-e^{-2\tilde{u}})]. \tag{148}$$

Moreover, by (135)

$$\mathsf{E}\,\mathfrak{I}_{\tilde{u}}(\mu) = \int_0^\infty (1+2t)^{-(\mu+3)/2}\,\mathsf{E}\,w_{\tilde{u}}^+(t)\,dt, \tag{149}$$

in which

$$\mathsf{E}\,w_{\tilde{u}}^+(t) = \int_0^\infty w\,p_{\tilde{u}}^{(1)}(w,t)\,dw, \tag{150}$$

with the density $p_{\tilde{u}}^{(1)}(w,t)$ given by (144). Evaluating the integral (150) yields

$$\mathsf{E}\,w_{\tilde{u}}^+(t) = (v^{-1}+vt)\,\mathrm{erf}\left(\sqrt{v^2 t/2}\right) + \sqrt{\frac{2t}{\pi}}e^{-v^2 t/2}, \tag{151}$$

which is a result of independent interest; this quantity is the expected distance from $\partial\Omega$ (on the $O(\epsilon^{1/2})$ lengthscale) at $t$ time units before exit, if one conditions on the Gumbel random variable $\mathfrak{G}$ equalling $\tilde{u}$. Substitution of (151) into (149) and (148) yields, after various manipulations,

$$\mathsf{E}\,\mathfrak{I}(\mu) = \sqrt{\frac{2}{\pi}}\frac{1}{\mu^2 - 1} + \frac{B(1/2, \mu/2)}{4\sqrt{\pi}(\mu - 1)} \tag{152}$$

where $B(\cdot, \cdot)$ is the Euler beta function. We conclude, by (147), that the expected offset from $H$ at the time of exit, *i.e.*, $\mathsf{E}\mathfrak{S}_\epsilon = \int_0^\infty s\, p_\epsilon(s)\, ds$, is when $\mu > 1$ asymptotically equal to $c\epsilon^{1/2}$ times this function of the eigenvalue ratio $\mu$.

This result on the first moment provides additional information on the degree of skewing present in generic $\mu > 1$ models, over and above the differing $s/\epsilon^{1/2} \to \pm\infty$ asymptotics of $p_\epsilon(s)$ given in (119). It is in fact possible to speculate, on the basis of the Gaussian falloff rates and the first moment, on the functional form of the generic asymptotic exit location distribution when $\mu > 1$. But we shall resist the temptation.

We note briefly, in conclusion, that the probabilistic analysis of this section may be extended to models with $\mu < 1$ as well. If $\mu < 1$, it was shown in Section 5 that generically the MPEP is tangent to $\partial\Omega$. If $\boldsymbol{B}(H)$ is normalized as in (88), and $\boldsymbol{D}(H) = \boldsymbol{I}$, the MPEP approaches $H = (0,0)$ along a curve $x^2 \sim A(x^1)^\mu$, where $A$ can only be computed by integrating Hamilton's equations from $S$ to $H$. In this case the conditioning on $x^1 = a\epsilon^{1/2}$ at time $t = \tau_\epsilon - \mathfrak{u}$, as $\epsilon \to 0$, must be supplemented by a conditioning on $x^2 = A(a\epsilon^{1/2})^\mu$. It follows from the stochastic differential equation (112) that asymptotically, $x_\epsilon^2(\tau_\epsilon) \sim x_\epsilon^2(\tau_\epsilon - \mathfrak{u})\exp(-\mu\mathfrak{u})$, irrespective of the value taken by the coefficient $c$. Since $\mathfrak{u} \sim \log a + \mathfrak{G}$, this implies that when $\mu < 1$,

$$\mathfrak{S}_\epsilon \sim A(a\epsilon^{1/2})^\mu \exp[-\mu(\log a + \mathfrak{G})], \tag{153}$$

*i.e.*,

$$\mathfrak{S}_\epsilon \sim A\epsilon^{\mu/2}\exp(-\mu\mathfrak{G}). \tag{154}$$

By examination, this offset random variable $\mathfrak{S}_\epsilon$ has the Weibull distribution previously computed in (106) by the method of matched asymptotic expansions. For models with $\mu < 1$, the probabilistic analysis and the method of asymptotic expansions are in agreement.

Even though a probabilistic analysis can be performed when $\mu < 1$, the method of matched asymptotic expansions is superior in that it yields an asymptotic approximation to the quasi-stationary density in the boundary layer. Although our probabilistic results on the case $\mu > 1$ are fairly strong, when $\mu > 1$ we have as yet no analogous boundary layer approximation.

## 9  Two-Dimensional Ackerberg-O'Malley Resonance

Now that we have to a large extent determined the generic asymptotic exit location distributions, we can examine the implications of our results for the singularly perturbed boundary value problem introduced in Section 2. Recall that if

$$\mathcal{L}_\epsilon = -(\epsilon/2) D^{ij} \partial_i \partial_j - b^i \partial_i \tag{155}$$

is the generator of the process $\boldsymbol{x}_\epsilon(t)$, $t \geq 0$, then the solution $u_\epsilon$ of the boundary value problem

$$\mathcal{L}_\epsilon u_\epsilon = 0 \text{ in } \Omega, \, u_\epsilon = f \text{ on } \partial\Omega \tag{156}$$

will satisfy

$$u_\epsilon(\boldsymbol{y}) = \mathsf{E}_{\boldsymbol{y}} f(\boldsymbol{x}_\epsilon(\tau_\epsilon)) \tag{157}$$

where the subscript on $\mathsf{E}$ signifies the starting point $\boldsymbol{x}_\epsilon(0)$ for the process. As $\epsilon \to 0$ the function $u_\epsilon$ is expected to 'level,' or tend to a constant, exponentially rapidly; this has been verified rigorously for the case of a non-characteristic boundary by Eizenberg [25]. The levelling is in agreement with the probabilistic picture that for any $\boldsymbol{y} \in \Omega$, as $\epsilon \to 0$ it becomes overwhelmingly (exponentially) likely that a sample path for the process $\boldsymbol{x}_\epsilon(t)$ will first flow toward $S$, and approach $S$ to within an $O(\epsilon^{1/2})$ distance, before experiencing further fluctuations. As a consequence the expectation $\mathsf{E}_{\boldsymbol{y}}$ in (157) may up to exponentially small errors be replaced by $\mathsf{E}_S$, and $u_\epsilon(\boldsymbol{y})$ may be approximated by a $\boldsymbol{y}$-independent constant. Since (if $\boldsymbol{x}_\epsilon(0) = S$) the exit location on $\partial\Omega$ converges in probability to $H$, if $f$ is continuous at $H$ then $u_\epsilon(\boldsymbol{y}) \to f(H)$ for all $\boldsymbol{y} \in \Omega$.

Even though $u_\epsilon$ will level exponentially rapidly and may be approximated by a constant, the constant itself will be $\epsilon$-dependent. It is possible to apply our results to determine its asymptotics, and the speed of its convergence to $f(H)$, as $\epsilon \to 0$. The key results are (107) and (152), in which we determined the expected offset from $H$ along $\partial\Omega$, at the time the process $\boldsymbol{x}_\epsilon(t)$ exits $\Omega$. In the standardization of the last two sections (the unstable eigenvalue $\lambda_u(H)$ taken to equal unity, and $D(H)$ taken to equal $I$), we found that the expected offset $\mathsf{E}\mathfrak{S}_\epsilon = \int_0^\infty s\, p_\epsilon(s)\, ds$ has asymptotics

$$\mathsf{E}\,\mathfrak{S}_\epsilon \sim [A\Gamma(1+\mu/2)]\,\epsilon^{\mu/2}, \quad \epsilon \to 0 \tag{158}$$

if $\mu < 1$, i.e., $\partial_i b^i(H) > 0$, and

$$\mathsf{E}\,\mathfrak{S}_\epsilon \sim c\left[\sqrt{\frac{2}{\pi}}\frac{1}{\mu^2-1} + \frac{B(1/2,\mu/2)}{4\sqrt{\pi}(\mu-1)}\right]\epsilon^{1/2}, \quad \epsilon \to 0 \tag{159}$$

if $\mu > 1$, i.e., $\partial_i b^i(H) < 0$. Here $\mu = |\lambda_s(H)|/\lambda_u(H)$ as always, $A$ is a quantity that may be computed from the way in which the generically unique Wentzell-Freidlin trajectory from $S$ to $H$ (the MPEP) approaches $H$, and $c$ is the off-diagonal element of the linearization of $\boldsymbol{b}$ at $H$ (see (88) and (110)). $B(\cdot,\cdot)$ is the Euler beta function.

It follows from (157), (158), and (159) that if $f$ on $\partial\Omega$ is continuously differentiable at $H$, then for all $\boldsymbol{y} \in \Omega$, $u_\epsilon(\boldsymbol{y})$ has leading $\epsilon \to 0$ asymptotics

$$u_\epsilon(\boldsymbol{y}) \sim \begin{cases} f(H) + [A\Gamma(1+\mu/2)]\,f'(H)\epsilon^{\mu/2}, & \text{if } \mu < 1; \\ f(H) + c\left[\sqrt{\dfrac{2}{\pi}}\dfrac{1}{\mu^2-1} + \dfrac{B(1/2,\mu/2)}{4\sqrt{\pi}(\mu-1)}\right]f'(H)\epsilon^{1/2}, & \text{if } \mu > 1, \end{cases} \tag{160}$$

which are independent of $\boldsymbol{y}$. If $\mu > 1$ and $c = 0$ (a *local gradient* condition at $H$, as discussed in Section 6) then the $O(\epsilon^{1/2})$ correction to $f(H)$ will have zero coefficient. The same will occur, irrespective of $\mu$, if $b^i = -D^{ij}\Phi_{,j}$, i.e., if the drift field $\boldsymbol{b}$ is globally gradient. In these nongeneric cases the asymptotic exit location distribution will be a Gaussian centered on $H$, on the $O(\epsilon^{1/2})$ lengthscale, and the leading correction to $f(H)$ in $u_\epsilon$ will necessarily be $o(\epsilon^{1/2})$. The same is true, incidentally, when $\Omega$ has non-characteristic boundary. When $\Omega$ is attracted to $S$ but $\partial\Omega$ is non-characteristic, it can be shown by the method of matched asymptotic expansions that the exit location distribution will generically be a Gaussian on the $O(\epsilon^{1/2})$ lengthscale, centered on the point on $\partial\Omega$ (generically unique) at which $W$ attains its minimum.

The asymptotics of (160) are striking, especially in the case $\mu < 1$. Since $\mu$ need not be rational, the presence of a leading correction term proportional to $\epsilon^{\mu/2}$ implies that $u_\epsilon$ on $\Omega$ *cannot in general be expanded in an asymptotic series in integral powers of $\epsilon$, or even in fractional powers*. This has not previously been realized. We expect that a similar phenomenon will occur if the boundary value problem (156) is generalized to read

$$\mathcal{L}_\epsilon u_\epsilon = \lambda_0^{(n)} \text{ in } \Omega, \ u_\epsilon = f \text{ on } \partial\Omega, \tag{161}$$

where $\lambda_0^{(n)}$ (as defined in Section 2) is the limit as $\epsilon \to 0$ of the eigenvalue $\lambda_\epsilon^{(n)}$ of the $n$'th eigenfunction $u_\epsilon^n$ of $\mathcal{L}_\epsilon$. In general we should have $u_\epsilon \sim C_n(\epsilon) u_0^n$, $\epsilon \to 0$, where $u_0^n$ is the limit of $u_\epsilon^n$ as $\epsilon \to 0$, *i.e.*, a solution of the degenerate first-order problem $\mathcal{L}_0 u_0^n = \lambda_0^{(n)} u_0^n$. Since $\lambda_0^{(0)} = 0$ and $u_0^0$ is constant, this is a generalization from $n = 0$ to arbitary $n$ of the behavior we have just found. The small-$\epsilon$ behavior of $C_n(\epsilon)$ for general $n$ is likely to be at least as unusual as that of $C_0(\epsilon)$.

To place these results in context, we remind the reader that the analogue for one-dimensional problems of the phenomenon we are investigating (the existence of a nontrivial $\epsilon \to 0$ limit on $\Omega$ for the solution $u_\epsilon$ of the singularly perturbed boundary value problem (156), or its generalization (161)), is known as *Ackerberg-O'Malley resonance* [1, 66]. In the one-dimensional case the partial differential equation $\mathcal{L}_\epsilon u_\epsilon = \lambda u_\epsilon$ reduces to an ordinary differential equation, namely

$$-(\epsilon/2) D(x) u_\epsilon'' - b(x) u_\epsilon' = \lambda u_\epsilon. \tag{162}$$

If an interval $\Omega = (x_0, x_1)$ is to have 'characteristic boundary' $\partial\Omega = \{x_0, x_1\}$ in the sense of this paper, both $x_0$ and $x_1$ must be linearly unstable turning points; they must be zeroes of $b(\cdot)$ with $b'(x_0)$ and $b'(x_1)$ both positive. Moreover $S \in (x_0, x_1)$ must be a linearly stable turning point: the only zero of $b(\cdot)$ in $\Omega$, with $b'(S) < 0$. If $\lambda$ is arbitrary, irrespective of the choice of boundary values $u_\epsilon(x_0)$ and $u_\epsilon(x_1)$ the function $u_\epsilon$ will normally converge to zero exponentially as $\epsilon \to 0$, uniformly on any compact subset of $\Omega$. The only exception to this occurs when $\lambda = \lambda_0^{(n)}$ for some $n$, *i.e.*, when the boundary value problem is (asymptotically) 'in resonance' with the $n$'th eigenmode of the operator $\mathcal{L}_\epsilon$, equipped with Dirichlet boundary conditions. In this case $u_\epsilon$ converges to a nontrivial solution of the degenerate problem $\mathcal{L}_0 u = \lambda_0^{(n)} u$. Since this convergence is uniform on compact subsets of $(x_0, x_1)$, there are normally boundary layers near the endpoints $x_0$ and $x_1$, of $O(\epsilon^{1/2})$ width as $\epsilon \to 0$.

The explicit solution of the $n = 0$ one-dimensional problem is instructive. In one dimension the drift field $b(\cdot)$ is necessarily a gradient; $b(x) = -D(x) \partial_x \Phi(x)$, where

$$\Phi(x) = \int_S^x [-b(y)/D(y)]\, dy, \tag{163}$$

and the action $W$ equals $2\Phi$ as usual. When $n = 0$ we set $\lambda = \lambda_0^{(0)} = 0$, and the differential equation (162) simplifies. The closed-form solution of (162) is (cf. O'Malley [66])

$$u_\epsilon(x) = \left[\frac{\int_x^{x_1} \exp\left(W(y)/\epsilon\right) dy}{\int_{x_0}^{x_1} \exp\left(W(y)/\epsilon\right) dy}\right] u_\epsilon(x_0) + \left[\frac{\int_{x_0}^{x} \exp\left(W(y)/\epsilon\right) dy}{\int_{x_0}^{x_1} \exp\left(W(y)/\epsilon\right) dy}\right] u_\epsilon(x_1). \tag{164}$$

If $W(x_0) < W(x_1)$ (resp. $W(x_1) < W(x_0)$), then by examination $u_\epsilon(x) \to u_\epsilon(x_0)$ (resp. $u_\epsilon(x) \to u_\epsilon(x_1)$), uniformly on compact subsets of $\Omega$ as $\epsilon \to 0$. Moreover, $u_\epsilon$ both levels and converges exponentially rapidly. The exponential levelling, and the constant limit, have the same interpretation in terms of the stochastic exit problem as they do in two-dimensional models.

The one-dimensional boundary value problem cannot usually be solved in closed form when $n \geq 1$, but a method of matched asymptotic expansions [66], involving the construction of both inner and outer approximations to $u_\epsilon$, may be employed to show that for any $x \in \Omega$, $u_\epsilon(x)$ has an asymptotic expansion in fractional powers of $\epsilon$. Although this is more complicated than the asymptotic behavior for the one-dimensional $n = 0$ problem, it is still simpler than the behavior we have discovered in its two-dimensional analogue. In two-dimensional Ackerberg-O'Malley resonance (at least, when $\Omega$ has characteristic boundary) the outer expansion of $u_\epsilon$ in the body of $\Omega$ may involve non-fractional powers.

The presence of irrational powers of $\epsilon$ in the outer expansion of $u_\epsilon$, as in (160), suggests that they may also be present in the outer expansions of the eigenfunctions $u_\epsilon^n$ and $v_\epsilon^n$ of $\mathcal{L}_\epsilon$ and $\mathcal{L}_\epsilon^*$. For this reason, unlike many authors we have refrained from approximating the quasistationary density $v_\epsilon^0$ in the body of $\Omega$ by a formal asymptotic series, since it is unclear what powers of $\epsilon$ should be present. Instead, we have worked only to leading order. As we noted at the beginning of Section 4, for an outer expansion to be useful it must match to an inner expansion. And the expansion beyond leading order of the quasistationary density $v_\epsilon^0$ in the boundary layer, if $\partial\Omega$ is characteristic, remains an unsolved problem.

## 10 Conclusions

The generic features of the two-dimensional stochastic exit problem, when exit from the region $\Omega$ occurs near a saddle point $H$, are now clear. As the noise strength $\epsilon \to 0$, the exit location near $H$ will be concentrated on the $O(\epsilon^{\mu/2})$ lengthscale near $H$ (if the eigenvalue ratio $\mu < 1$) or the $O(\epsilon^{1/2})$ lengthscale (if $\mu > 1$). In the $\mu < 1$ case the exit location distribution is asymptotic to the Weibull distribution (106), which includes a scale factor that can only be computed from the approach path taken by the MPEP (the optimal, or most probable trajectory) from $S$ to $H$. In the $\mu > 1$ case the limiting exit location distribution, whose moments are computable (see, *e.g.*, (152)), contains no free parameters: it is determined by the behavior of the stochastic model in the vicinity of $H$.

In both cases the limiting distribution will be 'skewed': non-Gaussian and asymmetric. Normally, it is Gaussian only when the deterministic drift $\boldsymbol{b}$ satisfies $b^i = -D^{ij}\partial_j\Phi$ for some potential function $\Phi$, or when $\mu > 1$ and a local version of this equality (the $c = 0$ condition of Section 6) holds near $H$. These cases, which are characterized by the absence of a classically forbidden 'wedge' emanating from $H$, are nongeneric. Although our two-dimensional stochastic model differs from the barrier crossing models employed in chemical physics, we believe that the generic skewing phenomenon is related to the phenomenon of 'saddle point avoidance' [2, 4, 46]. A number of authors have in fact already noted the presence (in particular models) of a classically forbidden region. In the literature the boundary of the forbidden region is sometimes called the 'stochastic separatrix,' or a 'switching line' [2, 4, 24, 40, 65].

It is clear from our treatment that the generic features of models with $\mu < 1$ are particularly interesting. In such models the nominal frequency factor $K(H)$ (the value of the WKB prefactor at $\boldsymbol{x} = H$, which would normally be interpreted as a factor by which the frequency of excursions to the vicinity of $H$ is multiplied) equals *zero*. This feature, like the anomalously large lengthscale over which the exit location distribution is spread [$O(\epsilon^{\mu/2})$ rather than $O(\epsilon^{1/2})$] can be traced to the unusual approach path taken by the MPEP. When $\mu < 1$ the MPEP is generically tangent to the separatrix $\partial\Omega$ at $H$, as in Figure 2(a). This grazing behavior causes the exit location distribution to be anomalously wide. It also causes the WKB prefactor $K(\boldsymbol{x})$ to tend to zero as $\boldsymbol{x} \to H$ along the MPEP, as can be shown by integrating the system of ordinary differential equations (31), (32), (35), (36), (38) from $S$ to $H$. Another unusual feature of generic models with $\mu < 1$ is that the classical formula (87) for the mean exit time asymptotics must be replaced by (109). The formula (109) is unaffected by $K(H)$ equalling zero, and by the fact that generically, the Hessian matrix of the Wentzell-Freidlin action $W$ does not exist at $H$.

The system (31), (32), (35), (36), (38), which generates the WKB (outer) approximation to the quasistationary density, is particularly useful for numerical work. We commented briefly on the interpretation of the matrix Riccati equation (38) for $\partial_i\partial_j W$ in terms of symplectic geometry. It turns out, as we may explain elsewhere, that the transport equation (36) for $K(\cdot)$ also has a geometric interpretation. The natural setting for the outer approximation $K(\boldsymbol{x})\exp\left(-W(\boldsymbol{x})/\epsilon\right)$ is the theory of semiclassical expansions for partial differential equations [22, 59], which has deep geometric underpinnings.

It should be possible to extend the approach of this paper in several different directions. It has recently become clear that caustics (folds in the unstable manifold $\mathcal{M}^u_{(S,0)}$ in phase space comprising the most probable fluctuational trajectories, as in Section 4) occur very frequently [24, 55]. Their effect on first passage and exit time phenomena is now under investigation. Another extension, of particular value in applications, would be to the case of singular diffusion. The results of this paper apply to what is known in physics as the *overdamped limit* of barrier crossing models; the analysis of exit location distributions in models with underdamped dynamics will require an extension to the case when the diffusion tensor is allowed to become singular [65]. An extension to higher dimensionality would also prove useful, and is now under study.

As regards the connections with the phenomenon of Ackerberg-O'Malley resonance discussed in Section 9, we cannot resist quoting O'Malley [66]:

> We note that resonance is sometimes related to certain exit-time problems for stochastic equations, but regret to report that the mathematical phenomenon under discussion (despite much attention in the literature) has not yet substantially helped us understand much new physics.

It now appears that the desired connections to physical models can indeed be found, by going from one-dimensional resonance (for ordinary differential equations) to multidimensional resonance (for partial differential equations). Skewed exit location distributions and saddle point avoidance have implications for the asymptotic expansions used in analysing multidimensional resonance, and vice versa.